\documentclass[UKenglish, acmsmall]{acmart}

\acmJournal{PACMPL}
\acmVolume{}
\acmNumber{CONF} 
\acmArticle{}
\acmYear{}
\acmMonth{}
\acmDOI{} 
\startPage{}

\setcopyright{none}

\usepackage{
  ragged2e,
  relsize,
  stackengine,
}

\usepackage{booktabs}   
\usepackage{subcaption} 

\usepackage{wrapfig}
\usepackage{pifont}

\usepackage{ifthen}
\usepackage{xargs}

\newcommandx{\yaHelper}[2][1=\empty]{%
\ifthenelse{\equal{#1}{\empty}}%
  { \ensuremath{ \scriptstyle{ #2 } } } 
  { \raisebox{ #1 }[0pt][0pt]{ \ensuremath{ \scriptstyle{ #2 } } } }  
}

\newcommandx{\yrightarrow}[4][1=\empty, 2=\empty, 4=\empty, usedefault=@]{%
  \ifthenelse{\equal{#2}{\empty}}
  { \xrightarrow{ \protect{ \yaHelper[ #4 ]{ #3 } } } } 
  { \xrightarrow[ \protect{ \yaHelper[ #2 ]{ #1 } } ]{ \protect{ \yaHelper[ #4 ]{ #3 } } } } 
}

\usepackage[inline]{enumitem}
\setlist{left=0pt}
\newlist{contributions}{itemize}{1}
\setlist[contributions]{label=$\blacktriangleright$}
\newlist{hypotheses}{enumerate}{1}
\setlist[hypotheses]{label=H\arabic*.}
\newlist{goals}{enumerate}{1}
\setlist[goals]{label=(G\arabic*)}
\newlist{concepts}{enumerate}{1}
\setlist[concepts]{label=(C\arabic*)}

\usepackage[capitalise]{cleveref}
\Crefname{section}{\S$\!$}{\S\S$\!$}
\crefname{figure}{Fig.}{fig.}
\crefname{hypotheses}{Hypothesis}{Hypotheses}
\crefalias{hypothesesi}{hypotheses}
\crefname{goals}{Goal}{Goals}
\crefalias{goalsi}{goals}
\crefname{concepts}{Design Concept}{Design Concepts}
\crefalias{conceptsi}{concepts}
\crefalias{line}{line}
\crefname{lstlisting}{Listing}{Listings}
\Crefname{lstlisting}{Listing}{Listings}

\newcommand{\NA}{\ensuremath{\bot}}

\newcommand{\FigureScaleFactor}{0.24}

\definecolor{RED}{RGB}{255, 0, 0}
\definecolor{BLUE}{RGB}{0, 0, 255}
\newcommand{\RED}[1]{\textcolor{RED}{#1}}
\newcommand{\BLUE}[1]{\textcolor{BLUE!66}{#1}}
\definecolor{POORCHOICEOFWORD}{RGB}{127, 255, 63}

\newcommand{\Verona}{Verona}
\newcommand{\Venice}{Reggio}

\newcommand{\TODO}[1]{\RED{[TODO: #1]}}
\renewcommand{\TODO}[1]{}

\newcommand{\ie}{\emph{i.e.},}
\newcommand{\eg}{\emph{e.g.},}
\newcommand{\cf}{\emph{c.f.},}

\usepackage{pifont}

\newcommand{\CapabilityFormatting}[1]{\textsf{#1}}

\newcommand{\Mut}{\CapabilityFormatting{mut}}
\newcommand{\Tmp}{\CapabilityFormatting{tmp}}
\newcommand{\Imm}{\CapabilityFormatting{imm}}
\newcommand{\Iso}{\CapabilityFormatting{iso}}
\newcommand{\Paused}{\CapabilityFormatting{paused}}
\newcommand{\Var}{\CapabilityFormatting{var}}

\newcommand{\VPA}[2]{#1\mathop{\odot}#2}

\newcommand{\Ptr}[2]{\ensuremath{#1}\,\ensuremath{\rightarrow}\,\ensuremath{#2}}

\newcommand{\Use}{\ensuremath{\mathit{use}}}
\newcommand{\lval}{\ensuremath{\mathit{lval}}}
\newcommand{\CL}{\ensuremath{~\mathit{CL}}}
\newcommand{\CLTag}{\ensuremath{\mathsf{\#}\mathit{CL}}}
\newcommand{\Undef}{\ensuremath{\mathbf{undef}}}
\newcommand{\fnc}{\ensuremath{\mathit{fnc}}}

\newcommand{\Cell}{\texttt{Cell}}
\newcommand{\CellTag}{\texttt{\#Cell}}
\newcommand{\Valfield}{\texttt{val}}

\newcommand{\subtag}{\mathop{\text{<\#:}}}

\newcommand{\Graph}{\ensuremath{\mathcal{G}}}

\newcommand{\rn}[1]{\textnormal{\textls{\uppercase{\scriptsize #1}}}}
\newcommand{\many}[1]{\ensuremath{\overline{#1}}}

\newenvironment{DisplayRule}{%
\vspace*{1ex}%
\begin{center}}{%
\end{center}%
\vspace*{1ex}}

\usepackage{listings}

\lstdefinelanguage{verona}{
  language=C++,
  morekeywords={enter,explore,var,ref,iso,mut,tmp,indy,read,imm,paused,def,let,freeze,merge,stack,heap,self,drop,match,continue,type_alias,drop,in,exit,load,swap,if,typetest},
  deletekeywords={delete},
  sensitive=false,
  morecomment=[l]{//},
  morecomment=[s]{/*}{*/},
  morestring=[b]",
  breakatwhitespace=false,
  breaklines=false,
  escapeinside={(*@}{@*)},
  mathescape=true,
  captionpos=b,
  numbers=none,
  numbersep=5pt,
  showspaces=false,
  showstringspaces=false,
  showtabs=false,
  tabsize=2,
  columns=fullflexible,
}

\lstdefinestyle{tinyverona}{
  basicstyle=\fontsize{8}{8}\selectfont\tt\color{black},
  keywordstyle=\fontsize{8}{8}\selectfont\tt\bfseries\color{black},
  numberstyle=\fontsize{5}{8}\selectfont\tt\color{black},
  commentstyle=\color{blue!66}\it,
  language=verona
}
\lstdefinestyle{piccoloverona}{
  basicstyle=\fontsize{7}{8}\selectfont\tt\color{black},
  keywordstyle=\fontsize{7}{8}\selectfont\tt\bfseries\color{black},
  numberstyle=\fontsize{5}{8}\selectfont\tt\color{black},
  commentstyle=\color{blue!66}\it,
  language=verona
}
\lstnewenvironment{tightcode}{\lstset{style=tinyverona}}{}

\lstdefinestyle{verona}{
  basicstyle=\fontsize{8.5}{9}\selectfont\tt\color{black},
  keywordstyle=\fontsize{8.5}{9}\selectfont\tt\bfseries\color{black},
  numberstyle=\fontsize{5}{9}\selectfont\tt\color{black},
  commentstyle=\color{blue!66}\it,
  language=verona
}
\lstnewenvironment{code}[1][]{\lstset{style=verona,numbers=left,#1}}{}
\renewcommand{\c}[1]{\lstinline[basicstyle=\fontsize{9}{11}\selectfont\tt,keywordstyle=\fontsize{9}{11}\selectfont\tt\bfseries\color{black},language=verona]!#1!}
\newcommand{\cc}[1]{\lstinline[basicstyle=\fontsize{7.5}{11}\selectfont\tt,keywordstyle=\fontsize{7.5}{11}\selectfont\tt\bfseries\color{black},language=verona]!#1!}

\usepackage[supertabular,lineBreakHack]{ottalt}

  \input{anc/semantics.ott.tex}
  \renewottcommands[ott]

\newcommand{\SB}{\startblock}
\newcommand{\FB}{\finishblock}
\newcommand{\Cfg}{\ensuremath{\mathit{cfg}}}
\newcommand{\RCfg}{\ensuremath{\mathit{rcfg}}}
\newcommand{\CCfg}{\ensuremath{\mathit{ccfg}}}
\renewcommand{\SB}{\startblock}
\renewcommand{\FB}{\finishblock}

\newcommand{\HelperFormat}{\mathsf}

\newcommand{\Hcap}[1]{\ensuremath{\mathsf{cap}(#1)}}
\newcommand{\Hopen}[1]{\ensuremath{\mathsf{open}(#1)}}

\newcommand{\Hmakemut}[1]{\ensuremath{\mathit{make\_mut}(#1)}}
\newcommand{\Hmakeiso}[1]{\ensuremath{\mathit{make\_iso}(#1)}}
\newcommand{\Hmakeimm}{\ensuremath{\mathit{make\_imm}}}

\newcommand{\reachableregions}{\mathsf{reachable\_regions}}

\newcommand{\Kap}{\ensuremath{\mathsf{cap}}}

\newcommand{\Hop}{\ensuremath{H_{op}}}
\newcommand{\Hcl}{\ensuremath{H_{cl}}}
\newcommand{\Hfr}{\ensuremath{H_{fr}}}

\newcommand{\RCfglong}[1]{\ensuremath{\langle #1 \rangle}}

\newcommand{\cupdisj}{\ensuremath{\uplus}}

\usepackage{scalerel}

\newcommand{\FUndef}{~\downarrow~}

\newcommand{\Reg}{\ensuremath{\mathit{Reg}}}
\newcommand{\Oid}{\ensuremath{\mathit{Oid}}}
\newcommand{\X}{\ensuremath{\mathit{Var}}}

\newcommand{\Eff}{\ensuremath{\mathit{Eff}}}

\newcommand{\Cls}{\ensuremath{\mathit{Cls}}}
\newcommand{\RF}{\ensuremath{\mathit{RF}}}
\newcommand{\RS}{\ensuremath{\mathit{RS}}}

\newcommand{\src}[1]{\ensuremath{\HelperFormat{src}(#1)}}
\newcommand{\dst}[1]{\ensuremath{\HelperFormat{dst}(#1)}}
\newcommand{\reg}[1]{\ensuremath{\HelperFormat{reg}(#1)}}
\newcommand{\regions}[1]{\ensuremath{\HelperFormat{regions}(#1)}}
\newcommand{\references}[1]{\ensuremath{\HelperFormat{references}(#1)}}
\newcommand{\REF}[1]{\ensuremath{\mathit{ref\!}_{#1}}}

\renewcommand{\REF}{\ensuremath{\mathit{ref}}}

\newcommand{\iso}{\Iso{}}
\newcommand{\var}{\Var{}}
\newcommand{\tmp}{\Tmp{}}
\newcommand{\mut}{\Mut{}}
\newcommand{\paused}{\Paused{}}
\newcommand{\imm}{\Imm{}}

\newcommand{\enter}[1]{\textit{enter}(#1)}
\newcommand{\badenter}[1]{\textit{badenter}(#1)}
\newcommand{\exit}[1]{\textit{exit}(#1)}
\newcommand{\load}[1]{\textit{load}(#1)}
\newcommand{\swap}[1]{\textit{swap}(#1)}
\newcommand{\halloc}[1]{\textit{halloc}(#1)}
\newcommand{\salloc}[1]{\textit{salloc}(#1)}
\newcommand{\bind}[1]{\textit{bind}(#1)}
\newcommand{\freeze}[1]{\textit{freeze}(#1)}
\renewcommand{\merge}[1]{\textit{merge}(#1)}
\newcommand{\cast}[1]{\textit{cast}(#1)}
\newcommand{\nocast}[1]{\textit{nocast}(#1)}

\newcommand{\Drop}{\kw{drop}~}

\newcommand{\Get}{\mathbf{get}}
\newcommand{\CfgLoad}{\mathsf{cfg\_load}}
\newcommand{\store}{\mathsf{store}}
\newcommand{\sload}{\mathsf{load}}

\newcommand{\heapstore}{\mathsf{heap\_store}}
\newcommand{\stackstore}{\mathsf{stack\_store}}
\newcommand{\stackload}{\mathsf{stack\_load}}
\newcommand{\heapload}{\mathsf{heap\_load}}

\newcommand{\FTypes}[1]{\mathbf{ftypes}(#1)}

\newcommand{\FType}{\mathbf{ftype}}

\newcommand{\FrozenRegs}{\ensuremath{H_{\it fr}}}
\newcommand{\ClosedRegs}{\ensuremath{H_{\it cl}}}
\newcommand{\OpenRegs}{\ensuremath{H_{\it op}}}
\newcommand{\sees}{\mathop{\odot}}

\newcommand{\AN}{~\land~}
\newcommand{\OR}{~\lor~}

\newcommand{\CapOK}{\ensuremath{\mathsf{capability\_ok}}}
\newcommand{\RegionOrder}{\ensuremath{\mathsf{region\_order}}}
\newcommand{\LocationOK}{\ensuremath{\mathsf{location\_ok}}}
\newcommand{\VarUnique}{\ensuremath{\mathsf{var\_unique}}}
\newcommand{\VPU}{\ensuremath{\mathsf{var\_pw\_unique}}}
\newcommand{\DeepFreeze}{\ensuremath{\mathsf{deep\_freeze}}}
\newcommand{\TopOK}{\ensuremath{\mathsf{topology\_ok}}}
\newcommand{\topokgraph}{\ensuremath{\mathsf{topology\_ok\_graph}}}
\newcommand{\entrypoints}{\ensuremath{\mathsf{entrypoints\_ok\_graph}}}
\newcommand{\entrypointsok}{\ensuremath{\mathsf{entrypoints\_ok}}}
\newcommand{\ToppOK}{\ensuremath{\mathsf{topology\_pw\_ok}}}
\newcommand{\toppok}{\ToppOK}

\newcommand{\Open}[0]{\ensuremath{\mathit{Open}}}
\newcommand{\Closed}[0]{\ensuremath{\mathit{Closed}}}
\newcommand{\Frozen}[0]{\ensuremath{\mathit{Frozen}}}

\newcommand{\open}{\kw{open}}

\newcommand{\Fr}{\ensuremath{\mathit{Fr}}}
\newcommand{\Cl}{\ensuremath{\mathit{Cl}}}

\newcommand{\Heap}[0]{\ensuremath{\mathtt{heap}}}
\newcommand{\Temp}[0]{\ensuremath{\mathtt{temp}}}
\newcommand{\Root}[0]{\ensuremath{\mathtt{root}}}

\newcommand{\heap}{\Heap}
\newcommand{\temp}{\Temp}
\newcommand{\ruut}{\Root}

\newcommand{\dom}{\ensuremath{\mathbf{dom}}}
\newcommand{\defined}{\ensuremath{\mathbf{defined}}}

\newcommand{\RefTo}[1]{\ensuremath{\xrightarrow{#1}}}
\newcommand{\Rid}{\ensuremath{\mathit{rid}}}
\newcommand{\rid}{\Rid}
\newcommand{\rids}{\ensuremath{\mathit{rids}}}
\newcommand{\LocRS}{\ensuremath{\mathit{loc}_{RS}}}
\newcommand{\LocH}{\ensuremath{\mathit{loc}_{H}}}
\newcommand{\LocS}{\ensuremath{\mathit{loc}_{S}}}

\newcommand{\Loc}{\ensuremath{\mathit{loc}}}
\newcommand{\loc}{\Loc}
\newcommand{\refto}[1]{\RefTo{#1}}

\newcommand{\topok}{\TopOK}
\newcommand{\capok}{\CapOK}
\newcommand{\regionorder}{\RegionOrder}
\newcommand{\locationok}{\LocationOK}
\newcommand{\varunique}{\VarUnique}
\newcommand{\deepfreeze}{\DeepFreeze}

\newcommand{\ObjSep}{\ensuremath{\mathtt{obj}}}

\newcommand{\SepLoc}{\ensuremath{\mathit{seploc}}}
\newcommand{\SepRef}{\ensuremath{\mathit{sepref}}}
\newcommand{\RefRS}{\ensuremath{\mathit{ref}_{RS}}}
\newcommand{\RefH}{\ensuremath{\mathit{ref}_{H}}}
\newcommand{\RefS}{\ensuremath{\mathit{ref}_{S}}}
\newcommand{\RefF}{\ensuremath{\mathit{ref}_{F}}}

\newcommand{\WFG}{\ensuremath{\mathit{WFG}}}
\newcommand{\WFSrcDst}{\ensuremath{\mathit{WFSrcDst}}}
\newcommand{\WFRefs}{\ensuremath{\mathit{WFRefs}}}

\newcommand{\GraphSet}{\ensuremath{\mathit{GRAPH}}}
\newcommand{\RCfgSet}{\ensuremath{\mathit{RCFG}}}
\newcommand{\AllLocs}{\ensuremath{\mathit{LOCS}}}
\newcommand{\AllRefs}{\ensuremath{\mathit{REFS}}}
\newcommand{\RefSet}{\ensuremath{\mathcal{R}}}
\newcommand{\LocSet}{\ensuremath{\mathcal{L}}}

\newcommand{\partialto}{\rightharpoonup}

\newcommand{\NilF}{\ensuremath{\varepsilon}}
\newcommand{\NilS}{\ensuremath{\varepsilon}}
\newcommand{\NilRS}{\ensuremath{\varepsilon}}
\newcommand{\NilGamma}{\ensuremath{\varepsilon}}
\newcommand{\NilGammas}{\ensuremath{\varepsilon}}
\newcommand{\NilDelta}{\ensuremath{\varepsilon}}
\newcommand{\Nilrho}{\ensuremath{\varepsilon}}
\newcommand{\Nildelta}{\ensuremath{\varepsilon}}

\newcommand{\gammadefincl}{\mathop{\overset{\mathsf{pw}}{\subseteq}}}
\newcommand{\rhosep}[1]{\underset{#1}{::}}
\newcommand{\rhosepe}{::}

\newcommand{\Pow}{\mathcal{P}}
\newcommand{\Len}{\mathit{len}}
\newcommand{\Fnames}{\mathbf{fnames}}

\newcommand{\Regions}{\ensuremath{\mathit{Regions}}}
\newcommand{\If}{\text{ if }}
\newcommand{\Ow}{\text{ otherwise}}

\newcommand{\RegStep}[1]{\xrightarrow{#1}}
\newcommand{\ComStep}[1]{\xrightarrow{#1}}

\newcommand{\Gammas}{\many{\Gamma}}
\newcommand{\Suchthat}{~\middle|~}
\newcommand{\suchthat}{~|~}

\usepackage{enumitem}
\usepackage{xifthen}
\usepackage{pgfkeys}

\setlist[itemize]{label=--}

\newcommand{\lrefcounterprefix}{lrefcounter}
\newcommand{\lrefcounter}[1]{\newcounter{\lrefcounterprefix#1}}

\newcommand{\lastcounter}{pf}
\newcommand{\lastcountername}{\lrefcounterprefix\lastcounter}
\newcommand{\lastcounterget}{\arabic{\lastcountername}}

\newcommand{\lrefstep}[1]{
  \stepcounter{\lrefcounterprefix#1}
  \renewcommand{\lastcounter}{#1}
}

\newcommand{\llabel}[2][]{
  \ifthenelse{\isempty{#1}}
  {\label{local:\lastcountername:\lastcounterget:#2}}
  {\label{local:\lrefcounterprefix#1:\arabic{\lrefcounterprefix#1}:#2}}
}

\newcommand{\lref}[2][]{
  \ifthenelse{\isempty{#1}}
  {\ref{local:\lastcountername:\lastcounterget:#2}}
  {\ref{local:\lrefcounterprefix#1:\arabic{\lrefcounterprefix#1}:#2}}
}

\lrefcounter{pf}
\lrefcounter{case}
\lrefcounter{scase}

\newcommand{\pf}{\lrefstep{pf}}
\newcommand{\pfcase}[1]{\lrefstep{case}\item #1:}
\newcommand{\pfsubcase}[1]{\lrefstep{scase}\item #1:}

\newcommand{\pfsetscope}[1]{
  \renewcommand{\lastcounter}{#1}
}

\newenvironment{pfcases}{\begin{enumerate}[label=\textbf{Case \arabic*}]}{\end{enumerate}}
\newenvironment{pfsubcases}{\begin{enumerate}[label*=\textbf{.\alph*}]}{\end{enumerate}}

\begin{document}

\title[Reference Capabilities for Flexible Memory Management: Extended Version]{Reference Capabilities for Flexible Memory Management: Extended Version} 
\titlenote{This extended version of the paper contains an appendix with the
full formal development and proofs. Citations should use the
original version of the paper~\cite{reggio}.}


\author[E. Arvidsson]{Ellen Arvidsson}
\orcid{0009-0004-0878-3641}             
\affiliation{
  \department{Information Technology}
  \institution{Uppsala University}
  \country{Sweden}                    
}
\email{ellen.arvidsson@it.uu.se}          

\author[E. Castegren]{Elias Castegren}
\orcid{0000-0003-4918-6582}             
\affiliation{
  \department{Information Technology}             
  \institution{Uppsala University}           
  \country{Sweden}                   
}
\email{elias.castegren@it.uu.se}         

\author[S. Clebsch]{Sylvan Clebsch}
\orcid{0009-0004-4049-134X}             
\affiliation{
  \department{Azure Research}           
  \institution{Microsoft}               
  \country{UK}                     
}
\email{sylvan.clebsch@microsoft.com}         

\author[S. Drossopoulou]{Sophia Drossopoulou}
\orcid{0000-0002-1993-1142}
\affiliation{
  \department{Department of Computing}
  \institution{Imperial College London}
  \country{England}
}
\email{s.drossopoulou@imperial.ac.uk}

\author[J. Noble]{James Noble}
\orcid{0000-0001-9036-5692}
\affiliation{
 \institution{Creative Research \& Programming}
 \country{New Zeeland}
}
\email{kjx@acm.org}

\author[M. Parkinson]{Matthew J. Parkinson}
\orcid{0009-0004-3937-1260}             
\affiliation{
  \department{Azure Research}           
  \institution{Microsoft}               
  \country{UK}                     
}
\email{mattpark@microsoft.com}         

\author[T. Wrigstad]{Tobias Wrigstad}
\orcid{0000-0003-4918-6582}             
\affiliation{
  \department{Information Technology}             
  \institution{Uppsala University}           
  \country{Sweden}                   
}
\email{tobias.wrigstad@it.uu.se}         


\begin{abstract}
  \Verona{} is a concurrent object-oriented programming language
  that organises
  all the objects in a program into a forest of isolated regions.
  Memory is managed
  locally for each
  region, so programmers can control a program's memory use
  by adjusting objects' partition into regions, and by
  setting each region's memory management strategy.
  A thread can only mutate (allocate, deallocate) objects
  within one active region --- its ``window of mutability''.
  Memory management costs are localised to the active
  region, ensuring overheads can be predicted and controlled.
  Moving the mutability window between regions
  is explicit, so code can be executed wherever it
  is required,  yet programs remain in control of memory use.
  An ownership type system based
  on reference capabilities enforces region isolation,
  controlling aliasing within and between regions,
  yet supporting objects moving between regions and threads.
  Data accesses
  never need expensive atomic operations, and
  are always thread-safe.
\end{abstract}



\maketitle

\section{Introduction}

Memory management has always been challenging, and programmers and
programming language designers have developed a wide range of
techniques and patterns to deal with it~\cite{smallmemory}. Most early
languages like FORTRAN and COBOL supported only fixed memory
allocation, where memory was allocated before a program began
to execute. Algol-60 popularised stack allocation, enabling
recursive procedures to be expressed clearly, and then languages like
C and Pascal popularised heap allocation, where programmers
could manually request memory from the runtime system, and
manually return that memory when it was no longer required.
LISP introduced the first automatic memory management
system---a garbage collector---which relieved programmers from the
need to explicitly free memory, rather memory will be automatically
reclaimed once it is no longer needed~\cite{jones_garbage_2016}.  As
well as reducing the amount of bookkeeping code programmers have to
write, garbage collection typically provides ``memory safety'' which
prevents a number of characteristic errors common to manual memory
management, such as failing to free objects that are no longer needed, or
accidentally freeing objects that are still in use.

While there is now a 60+ year history of research in garbage
collection algorithms and implementations, many programmers seem
resistant to using garbage collection, despite the
pitfalls of manual memory management. According to the TIOBE index
of programming languages~\cite{TIOBE}, about half out of the top
twenty programming languages eschew garbage collection, and rely
on various forms of manual memory management.
The 1st and 3rd of the top 25 Common Weaknesses in
CWE 2020--2022 are writing and reading outside of allocated memory and
using memory after freeing it is 7th. Memory leaks come in at 32nd and race
conditions 33rd~\cite{CWE_top_25}.

In short, manual memory management (\eg{} C/C++) is unsafe and
prone to errors but allows programmers to leverage domain
knowledge to optimise memory management. Some compile-time memory
management (\eg{} Rust) and automatic GC (\eg{} Java/C\#)
avoids the memory unsafety, but instead leads programmers to write
unsafe code for a variety of reasons.
In Rust, programmers must use unsafe code to construct well-known
data structures, and object topologies without clear domination.
In Java, programmers use unsafe code~\cite{10.1145/2814270.2814313} to leverage domain knowledge
to optimise memory management and to make GC performance more
predictable. In general, reasoning about the performance of
automatic GC is made difficult by the systemic effects of GC on
program performance and the heuristics which control when and how
GC is run.

\subsubsection*{Contributions}

This paper introduces \Verona{}'s region system---\Venice{}---and its
accompanying type system. \Venice{} gives
programmers control over memory management costs by dividing a
program's heap into a flexible forest of independent regions.
Programmers can pick a suitable strategy for how memory in each
region is managed, irrespective of what other regions do. Within
each thread, the programmer explicitly moves a single ``window of
mutability'' through the region forest. The single window of
mutability makes clear which region each part of a program is
working within, and how the program affects that region, in
particular with respect to object liveness, and
also permits a flexible aliasing. As a region can only be made accessible through a single pointer,
programs become free of data-races by design, and cheap ownership transfer to
support reconfiguring the region topology is easy.
Memory management overheads (\eg{} tracing, and reference count
manipulations) are likewise localised to just the mutable region.

\section{Background}

The continuing appeal of manual memory management
highlights the research problem we aim to solve:
how can languages give programmers the level of
control offered by manual memory management, while maintaining
memory safety? Two broad research streams tackle
this problem, one based on \textit{regions} and one based on
\textit{ownership}.

\subsubsection*{Regions}

\citet{explicitRegions1998} introduced explicit
regions for managing memory in C-like languages: objects are
allocated in regions; and entire regions of objects are deleted in
one operation, rather than deleting objects individually. A later
version of this scheme added annotations to indicate that a
pointer refers to an object in the same region, in an enclosing
region, or is not allocated via the region
system~\cite{gay-regions-pldi2001}. \citet{utting1995} had previously shown
how regions could help local reasoning, based on
the ``collections'' or ``local stores'' used to differentiate
pointers in Euclid~\cite{euclid1977}.

Rather than using explicit, programmer-visible regions, Tofte and
Talpin~\citeyearpar{tofte_region-based_1997,
  tofte-retrospective-2004} demonstrated how Milner-style
inference~\cite{baker-unify1990} could be extended to implicitly
allocate objects to regions, and allocate and deallocate those
regions, without either explicit first-class regions, or
additional annotations on programs or types. Their
MLKit~\cite{mlkit-4.6.0} runs ML programs using stack allocation,
as the regions are allocated last-in, first-out. Because these
inferred regions are implicit, the region structure does not
capture a programmer's intent about how and where memory should be
allocated and freed. MLKit remains under continuous
development, in particular showing how regions can be supported in
a straightforward monadic style~\cite{fluet-monadic-jfp2006}, and
integrating generational-style GC within
regions~\cite{elsman-generational-regions-jfp2021}.

Safe region allocation was then tested at scale in
Cyclone~\cite{grossman_region-based_2002} an extension of C
with an explicit region construct, rather than using
inference. Like the MLKit, Cyclone regions were originally stack
based; later versions also adopted support for unique pointers and
reference counted objects to permit deallocation of individual
objects inside a region, at the cost or introducing memory leaks
due to cycles or failure to deallocate a dropped unique
pointer~\cite{cyclone-ismm2004}. Both MLKit style implicit /
inferred regions, and Cyclone explicit / annotated regions can be
modelled by a common core calculus based on linear references to
explicit, first-class
regions~\cite{fluet-linear-regions-esop2006}.


\subsubsection*{Ownership}

Work on object ownership effectively begins with Hogg's
Islands~\citeyearpar{hogg_islands_1991} and a general
recognition of the need to control topologies of programs~\cite{geneva91} in
languages where object identity (\ie{} dynamic allocation, mutable
state), encapsulation, and even ``automatic storage management''
are taken as essential design principles, rather than accidental
optimisations~\cite{lehrmann_madsen_object-oriented_1993,
  ingalls1981, demeter1989}.

Based on ``Flexible Alias Protection''
\cite{noble_flexible_1998}, \textit{ownership
  types}~\cite{clarke_ownership_1998,clarke_2001}
offer
compile-time enforcement of pointer encapsulation, including the
property that, considering paths through the object graph, an
``owner'' object should dominate all other objects that programmers
intend to encapsulate ``inside'' it \cite{potter:insandouts}.
Leveraging ``owners-as-dominators'', extensions to ownership types
have been applied to encapsulate object
invariants~\cite{muller_universes_1999}, record conformance to
software architecture~\cite{aldrich_ownership_2004}, localise
program effects~\cite{clarke_ownership_2002}, scope object
cloning~\cite{paley_li_sheep_2012}, ensure actor
isolation~\cite{gruber_ownership-based_2013, clarke_minimal_2008, srinivasan_kilim_2008},
prevent data races~\cite{boyapati_parameterized_2001,
  parallel-compsurv2022}, support safe
parallelisation~\cite{francis-landau_fine-grained_2016,
  bocchino_deterministic_2011} or ensure data is only accessed
under a mediating lock~\cite{flanagan_type-based_2000}---the first
fifteen years of these efforts are surveyed
in~\cite{clarke_aliasing_2013}. Owners-as-dominators leads
directly to applications in memory management, as deleting a
dominating node from a graph, by definition, must also delete
every node it dominates. This was first demonstrated in
SafeJava~\cite{boyapati-rtsj} using ownership types to compile
straightforwardly annotated programs to the Real-Time
Specification for Java~\cite{rtsjmem}, which supports fine-grained
control over memory via explicit dynamically-scoped regions.

Distinguishing between the inside and outside of an encapsulated
object lets languages generalise traditional pointer-based
uniqueness to \emph{external uniqueness}
\cite{clarke_external_2003}, where an object may have only one
pointer from the outside, but any number of pointers from its
inside. As with regions for unique objects, an externally unique
object can be represented as the sole object in a region; however
for external uniqueness, the object's region can also contain one
or more enclosed subregions in turn containing the object's
insides. External uniqueness is almost as strong as regular
uniqueness~\cite{wrigstad_ownership-based_2006}, and in particular
makes it easier to change objects' ownership, or dually, to
transfer objects between actors or independent threads
\cite{clarke_minimal_2008, haller_capabilities_2010,
  gordon_uniqueness_2012, haller_lacasa_2016}.

\subsubsection*{Rust}

Regions and ownership have been brought together recently in the
design of Rust, combining control of memory use, safe concurrency,
and excellent compiler error messages~\cite{RustPL, RustPopular,
  MSRust, turon_fearless_2015}. Rust essentially inherits
Cyclone's and MLKit's regions, but strongly integrated with
uniqueness. In particular, only values which are uniquely
referenced or passed by copy are mutable. Nested uniqueness brings
nested ownership: a unique value is owned by its storage
location, ensuring that when an owner is deallocated, all the
memory owned by that owner can also be deallocated.

To make programming in Rust practical, Rust allows unique values
to be borrowed without nullifying their source: a unique mutable
reference can be passed up the stack without losing
uniqueness~\cite{boyland_alias_2001} or traded for multiple
read-only borrowed references~\cite{wadler_linear_1990}. To
ensure absence of dangling pointers, Rust tracks lifetimes of
borrowed references and ensure that a ``longer-lived'' (or
enclosing) object can never point to a ``shorter-lived'' (or
encapsulated) object. This borrowing semantics is reminiscent of
fractional permissions~\cite{boyland_fractional_2013}. Through
uniqueness, Rust imposes a multiple-reader/single-writer
concurrency model~\cite{lea98}.

The strict rules surrounding ownership and borrowing, and
compilers' inability to accept safe programs that they ``cannot
understand'', make Rust hard to learn and to use
correctly~\cite{LearnRust, VizRust, HardRust, SafeRust, FightRust,
  rustGC}---to the point where the difficulty of implementing
first-year data structures (such as doubly-linked lists) has now
become an Internet trope~\cite{RustDLL1, RustDLL2, RustDLL3,
  RustDLL4}. When faced with these problems in practice, Rust
programmers either escape into unsafe Rust or revert to the
birthplace of aliasing, using integer array indices, FORTRAN
style~\cite{bendersky-blog2021}.

\section{\Venice{} Regions}
\label{sec:regions}

In this section we describe
the central concepts of the \Venice{} region system:
regions,
the region topology,
operations on regions,
the single window of mutability,
and the properties of the region system.
But first, let us overview the goals of this work.

\subsection{Motivation}

The design decisions in this paper are motivated by our primary goal:

\begin{goals}
\item \emph{Controllable and Predictable Memory Management
    Costs}.%
  \label{goal:local}
  \label{goal:first}
  It should be possible for a programmer to reason about and
  control the impact of memory management on performance.
\end{goals}

Our approach is to divide a program's heap into \emph{isolated regions} and
make each region an isolated unit of memory management. Concretely, we
set the following five sub-goals:

\begin{goals}\setcounter{goalsi}{1}
\item \textit{Mix-and-Match Memory Management}.%
  \label{goal:mix-match}
  A region is free to manage its own memory however it likes, irrespective of any other regions in the program.
  Thus, a programmer is free to pick a memory management strategy suited to the needs of particular operations.

\item \textit{Incremental Memory Management}.%
  \label{goal:incremental}
  Performance of memory management in one region should not be affected by activities in another region.
  Thus, fine-grained partitioning gives finer cost-control.

\item \textit{Zero-Copy Ownership Transfer}.%
  \label{goal:ownership-transfer}
  Ownership transfer between regions must be possible without copying objects.
  Thus, fine-grained partitioning does not have a hidden expensive cost, and heap topology can be modified cheaply.

\item \textit{Concurrent Memory Management}.%
  \label{goal:concurrent}
  Timing of memory management in one region should not be contingent by activities in another region.
  Thus, a programmer can initiate an operation---memory management or not---without having to wait, or forcing a wait upon any other part of the program.

\item \emph{Safe Concurrency}.%
  \label{goal:drf}%
  \label{goal:last}
  A thread that has access to a datum may access it freely without any need for synchronisation, and with a guarantee of data-race freedom.
%
\end{goals}

Because this paper does not deal explicitly with concurrency, we
will refrain from discussing the last two goals until
\cref{sec:concurrency}.

\subsection{High-Level Overview}

We distinguish between mutable and immutable objects. In this
paper, we are mostly concerned with the former. Immutable objects
do not live in regions, and can be accessed freely in a program.
In contrast, every mutable object belongs to a particular region.
In certain circumstances, mutable objects may be made
\emph{temporarily immutable}. To avoid confusion, we will
sometimes use the phrase \emph{permanently immutable} to denote
objects whose mutability is irretrievably lost.

\subsubsection{Regions and Region Topology}

A region is a set of objects whose memory is managed together. At
any moment, one of these objects is designated as the
\emph{bridge} object. A region can be \emph{opened} or
\emph{closed}. Closed regions are isolated from the rest of the
program which means that with the exception of the bridge object,
objects in a closed region are only referenced from within the
region. Bridge objects are externally
unique~\cite{clarke_external_2003} so they may have an additional,
single external incoming reference.

The only outgoing references from objects in a region are either
to immutable objects or to bridge objects of other (nested) regions.
Thus, a program's region topology forms a forest,
and moving the external reference to the bridge object changes the
topology. The topology of references \emph{within} regions
is unrestricted: any object can point to any other within the
same region.

\begin{wrapfigure}{r}{\FigureScaleFactor\textwidth}
  \vspace*{-.5em}
  \includegraphics[page=1,scale=\FigureScaleFactor,trim=0 620 1530 0,clip=true]{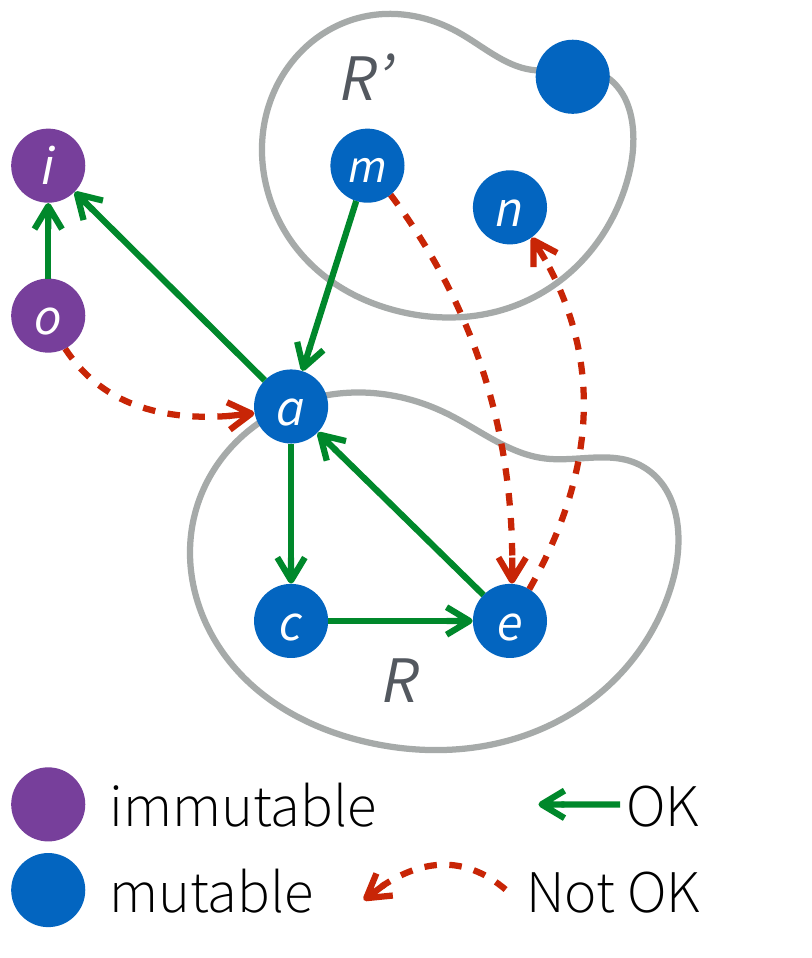}
  \caption{Region isolation of two closed regions.}
  \label{fig:isolation}
  \vspace*{-1em}
\end{wrapfigure}

\cref{fig:isolation} illustrates the isolation of the region $R$
(bean-shaped boundary). Object $a$ is the current bridge object of
the region, denoted by drawing it on the boundary. Also $c$ and
$e$ could serve as the bridge object. References \Ptr{e}{n} and \Ptr{m}{e} are not
permitted as they break isolation. The reference from \Ptr{a}{i} is permitted as $i$ is
immutable. The reference \Ptr{m}{a} is permitted because $a$ is the current
bridge object. Immutable objects do not live in regions and may
only refer to other immutable objects. Therefore, the reference
\Ptr{o}{a} is not allowed. Last, bridge objects may only have one
incoming reference from outside the region, so no more references
to $a$ are allowed from outside of $R$, regardless of their
origin.

\subsubsection{Regions and Memory Management}

Every region manages the memory of its objects in isolation, and
according to a strategy picked by the programmer specifically for
that region at the time of its creation. Code inside of a region
is agnostic to how memory is managed, meaning that a library can
leave such decisions to its users.

When an external reference to a bridge object is dropped, the
entire corresponding region can be free'd along with any nested
regions. In \cref{fig:isolation}, dropping \Ptr{m}{a} makes all
objects in region $R$ unreachable as external references to its
objects (\eg{} \Ptr{m}{e}) are not permitted. Thus, they can be
collected immediately.

\subsubsection{Single Window of Mutability}

A region must be explicitly opened to be accessed, and must be
closed before it can be opened again. The open regions form a LIFO
stack. The top region is \emph{active} and the remaining regions
on the stack are \emph{suspended}. An active region permits
allocation, deallocation and mutation of its objects. When a
region is suspended, neither allocation, deallocation nor mutation
is permitted in it.

Making a region active temporarily weakens its \emph{outgoing}
encapsulation: its objects are permitted to reference objects in
the suspended regions further down the stack. Suspending a region
conversely weakens its \emph{incoming} encapsulation: its objects
can be referenced by the regions further up the stack.
\Cref{tab:overview} overviews the allowed actions depending on a
whether a region is active, suspended or closed. When the active
region is closed, it is popped from the stack, and the new top
region goes from suspended to active. Because closed regions are
not permitted outgoing refereces, any references to objects in
open regions must be invalidated.

To the active region, the suspended regions appear as a single
\emph{temporarily immutable} region whose objects can be
referenced as long as the active region remains on the stack. Programmers
can thus \emph{trade mutability for access, and any object can
be temporarily accessed from anywhere, provided the containing
  regions are opened on the stack in a permitting order.}

\begin{table}[t]
  \centering
  \caption{Allowed actions depending on a region's state. Incoming
    and outgoing denote references to mutable objects from and to other regions respectively.
    Bridge means only to bridge objects; any$^*$ means to any object
    of a previously (outgoing) or subsequently (incoming) opened
    region. Free object denotes ability to free individual objects
    inside of a region. Free region denotes the ability to free an
    entire region.}%
  \label{tab:overview}
  \small\setlength{\tabcolsep}{1ex}
  \begin{tabular}{rcccccccc}
    \toprule
            & \multicolumn{2}{c}{Encapsulation} & \multicolumn{2}{c}{Effects} & \multicolumn{3}{c}{Memory Management} & \sf Nested                                                       \\
	\sf State & \sf Incoming                       & \sf Outgoing                 & \sf Mutate                             & \sf Read & \sf Alloc\,object & \sf Free\,object & \sf Free\,region & Regions \\
    \cmidrule(l{3pt}r{3pt}){1-1}\cmidrule(l{3pt}r{3pt}){2-3}\cmidrule(l{3pt}r{3pt}){4-5}\cmidrule(l{3pt}r{3pt}){6-8}\cmidrule(l{3pt}r{3pt}){9-9}
  active	  & bridge                             & any$^*$                          & yes                                    & yes      & yes           & yes          & no           & yes     \\
  suspended	& any$^*$                                & any$^*$                          & no                                     & yes      & no            & no           & no           & yes     \\
  closed	  & bridge                              & bridge                       & no                                     & no       & no            & no           & yes          & yes     \\
    \bottomrule
  \end{tabular}
\end{table}

In \cref{fig:isolation}, to open $R$ we must first
open $R'$ to access the reference \Ptr{m}{a}. Opening $R$
through this reference suspends $R'$, making $m$ and $n$
temporarily immutable, $a$, $c$, and $e$ mutable, and permitting
\Ptr{e}{n}. With the topology of \cref{fig:isolation}, we cannot
open $R$ and $R'$ in a way that permits the creation of \Ptr{m}{e}
as $m$ is immutable when $R'$ is suspended, and $e$ is not
accessible when $R$ is closed. To do so, we must change the
topology by moving \Ptr{m}{a} out of $R'$, \eg{} to a stack
variable or other region.

In combination, the design decision to only permit mutation in one
region at a time, the LIFO order of the region stack and the
inaccessibility of closed regions facilitates reasoning about
side-effects. The main motivation, however, is to control the
costs of memory management. As we shall see, direct overheads
related to memory management such as maintaining reference counts
or tracing object structures are only applicable to active
regions.

\subsubsection{Navigating Through the Region Forest}

\begin{wrapfigure}{r}{0.45\textwidth}
  \vspace*{-1.5em}
  \begin{subfigure}{.49\linewidth}
    \centering
    \includegraphics[page=2,scale=\FigureScaleFactor,trim=0 660 1590 0,clip=true]{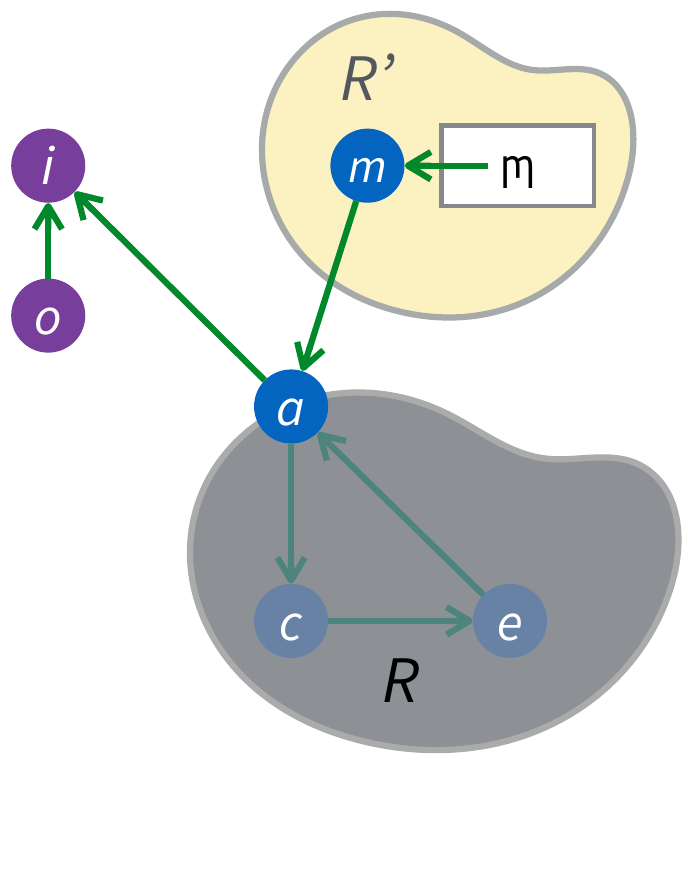}
  \end{subfigure}
  \begin{subfigure}{.49\linewidth}
    \centering
    \includegraphics[page=3,scale=\FigureScaleFactor,trim=0 660 1590 0,clip=true]{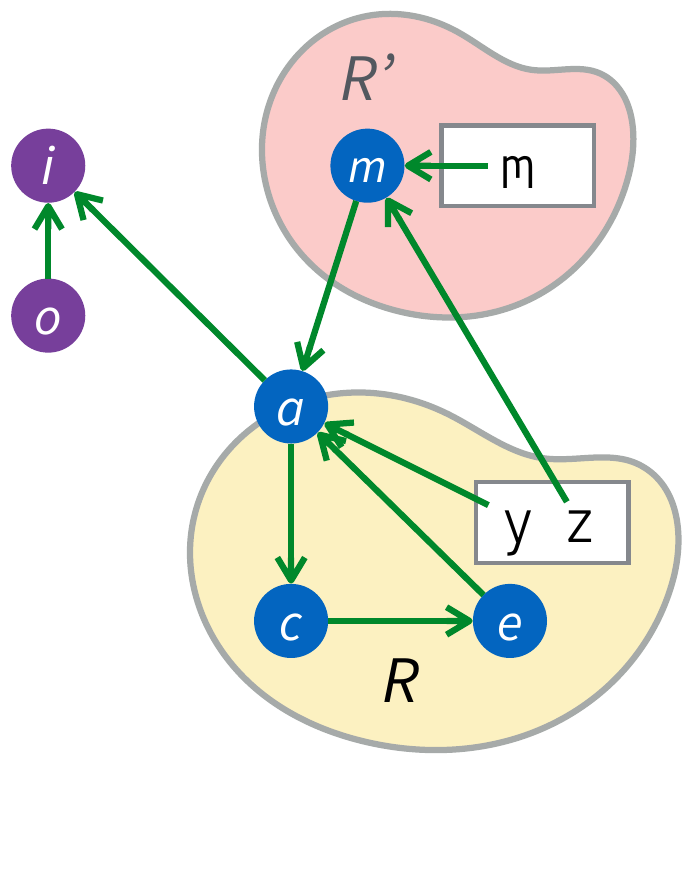}
  \end{subfigure}
  \vspace*{-1.5em}
  \caption{
    Left: $R'$ is active and $R$ is closed.
    Right: we opened $R$ making it active and suspending $R'$.
  }
    \label{fig:isolation:first}
    \label{fig:isolation:second}
  \label{fig:region}
\end{wrapfigure}

Due to the single window of mutability, programs require explicit
navigation through the region forest. The left subfigure of
\Cref{fig:isolation:first} shows
\cref{fig:isolation}, denoting regions' states by colour when $R'$
is active and $R$ is closed. The white box denotes the stack frame
of $R'$ with its local variable(s).
If we proceed by opening $R$ we
arrive at the right subfigure of \Cref{fig:isolation:second}: a
new top frame is created inside $R$ containing its own local
variables, with \c{y} holding a reference to $a$; $R'$
is suspended and $R$ active, and the window of mutability is moved
from $R'$ to $R$. The reference \Ptr{z}{m} shows the weakened
isolation allowing outgoing references from $R$ and incoming
references to $R'$.
To continue, we may close $R$ or open any reachable region $R''$. The former
will invalidate any references from $R$ to $R'$ since these would violate
isolation. The latter gives mutable access to $R''$ and suspends $R$, and
permits references from $R''$ to both $R$ and $R'$.


\begin{figure}[b]
  \includegraphics[page=5,scale=\FigureScaleFactor,trim=0 820 300 50,clip=true]{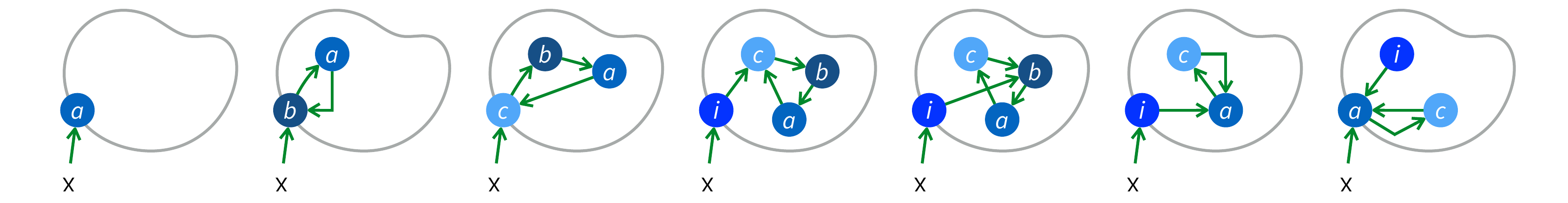}
\vspace*{-1em}
  \caption{Examples of bridge swapping. Time moves left; we use different shades of blue for clarity. Subfigures 1--3 construct the cyclic
    linked list $[a,b,c]$, using the most recently added link as the bridge object.
    Subfigures 4--6 construct an iterator, for the list, and make it the bridge object. We
    subsequently use the iterator to iterate to the $b$ link and unlink it, before dropping the
    iterator and making $a$ the bridge object (and with a garbage iterator whose
    removal is determined by the region's memory management strategy).}%
  \label{fig:swap}
\end{figure}

\subsubsection{Swapping the Bridge Object}%
\label{sec:swapp-bridge-object}

Any object in a region can serve as the bridge object. While a
region is open, any object can be designated as the new bridge
object, and we make this choice visible when the region closes.
Thus, it is possible to \eg{} create a region with a stack where
the bridge object is always the top link. If we wish to create an
external iterator to the stack, we can create an iterator inside
the region, and make the iterator the bridge object for the
duration of the iteration, and then switch back again.
\cref{fig:swap} shows the situation pictorially.

\subsubsection{Merging and Freezing Regions}%
\label{sec:immutable}%
\label{sec:freeze}

\begin{figure}[ht]
  \includegraphics[page=6,scale=\FigureScaleFactor,trim=0 835 350 2,clip=true]{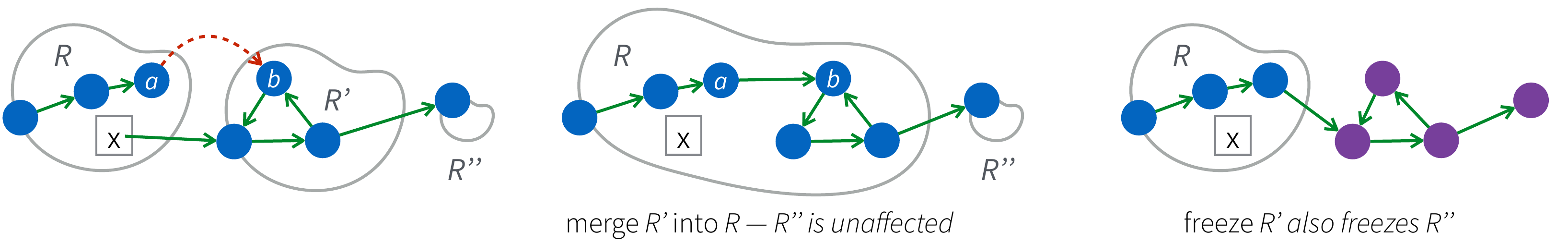}
  \caption{Left: Three nested regions $R$, $R'\!$, and $R''$. Centre: Merging region $R'$ into $R$. Right: Freezing $R'$.}%
  \label{fig:merge}%
  \label{fig:freeze}
  \vspace*{-1em}
\end{figure}

A closed region's contents may be merged into another region by dropping
the uniqueness of the bridge object. For clarity, we use an
explicit merge operation.
\Cref{fig:merge} shows merging $R'$ into $R$, which moves the objects in $R'$
into $R$ and creating the (now legal) \Ptr{a}{b} reference from the variable
\c{x}, after which $R'$ ceases to exist. Merging a region (source) into
another (sink) \emph{moves} all regions directly nested in the source to the sink, but
does not \emph{merge} those regions into the sink (see $R''$ in \Cref{fig:merge}).

Permanently immutable structures are created by constructing a mutable region
and then turning its entire contents immutable through
an explicit freeze operation that operates on closed regions.
In contrast to merging, freezing a region also freezes its nested
regions (see~\cref{fig:freeze}). This preserves the property that immutable objects may
only reference other immutable objects. Freezing dissolves region
boundaries, making the frozen objects freely accessible to objects
in all regions.

\section{Reference Capabilities for Statically Enforcing Region Isolation}
\label{sec:typing-regions}

We now introduce a type system that statically enforces region
isolation according to the \emph{encapsulation} and \emph{effects}
columns of \cref{tab:overview}. A region is opened through an
\c{enter} operation that takes a reference to bridge object and a
lambda, executes the lambda inside the region passing it the
bridge object as argument, and then closes the region. Its
companion operation \c{explore} opens a region in a suspended state
(while still suspending the former active region).

Our type system uses reference
capabilities which decorate all types $\tau$ in a program:

\begin{itemize}[label=--]
\item \Mut{} $\tau$ denotes an intra-regional reference to an object of type $\tau$ (\c{r}, \c{c}, \c{e},
  and \c{m} in \cref{lst:isolation});
\item \Tmp{} $\tau$ is like \Mut{} $\tau$, but the lifetime of
  the object is bound to the \c{enter}/\c{explore} scope;
\item \Imm{} $\tau$ denotes a reference to a permanently immutable
    object of type $\tau$ (\c{i} and \c{o});
\item \Iso{} $\tau$ denotes an externally unique reference
  to a bridge object of type $\tau$ of a closed
    region (\c{a});
\item \Paused{} $\tau$ denotes a reference to an immutable
  object in a suspended region (\c{z} in
  \cref{fig:isolation:second}).
\end{itemize}

\noindent
On assignment, the \textsc{lhs} capability and the \textsc{rhs}
capability must be the same. \c{merge} has the signature
\Iso{}$\to$\Mut{} and \c{freeze} has the signature \Iso{}$\to$\Imm{}.
To \c{enter} or \c{explore}, an \Iso{} is needed.




All expressions are typed from the point of view of the currently
active region. A field \c{f} declared with a \Mut{} capability
will only appear as such when the object $o$ containing \c{f} is
in the active region. If $o$'s containing region is suspended,
\c{f} will also appear as suspended; if $o$'s containing region is
closed, \c{f} is not even visible to the program. This is handled
by viewpoint adaptation (\cf{} \cref{sec:vpa}).

An invariant in our system is that aliases to an object that are
\emph{accessible simultaneously} have the same capability. This
design is motivated in-part by region isolation and in-part by a desire
to keep complexity down in this presentation. For example: If a
\Mut{} and \Paused{} could alias, the immutability of the latter
would be weakened to read-only. If \Paused{} and \Imm{} could
alias, it would violate the temporary vs. permanent nature of
their immutability. (This restriction could be relaxed to permit
some aliasing across the mutable capabilities and aliasing across
immutable capabilities.)

\begin{wrapfigure}{r}{0.38\textwidth}
\vspace*{-1ex}
\begin{code}[style=tinyverona,numbers=right,caption={Code creating the region $R$ from $R'$ as depicted in \cref{fig:isolation}.},label=lst:isolation]
// Freeze creates immutable objects, c.f. (*@\BLUE{\it\cref{sec:freeze}}@*)
let i = freeze new iso Cell(42) (*@\label{imm}\label{first_ok}@*)
let o = freeze new iso Cell(i) (*@\label{o_constructor}@*)
let a = new iso Link // creates R (*@\label{r_created}@*)
// { r $\textcolor{blue}{\Rightarrow}$ ... } is a lambda with argument r
enter a { r (*@$\Rightarrow$\hspace{1pt}@*) let c = new mut Link (*@\label{enter}@*)
                 let e = new mut Link (*@\label{e_not_tmp}@*)
                 r.elem := i
                 r.next := c
                 c.next := e
                 e.next := r } (*@\label{enterends}@*)
let m = new mut Cell(a) // buries a (*@\label{bury}\label{massigned}\label{last_ok}@*)
\end{code}
\vspace*{-1em}
\end{wrapfigure}

\paragraph{Constructing~\cref{fig:isolation}}

\cref{lst:isolation} shows code that
creates the regions, objects, and (permitted) references of
\cref{fig:isolation}. Line~\ref{r_created} creates the $R$ region.
On Line~\ref{imm},
the immutable object $i$ is constructed by creating a region and
freezing it. The object $o$ is created similarly. Its reference to
$i$ does not break region isolation as $i$ is immutable. We could
get rid of the \c{freeze} on Line~\ref{imm} since
Line~\ref{o_constructor} moves $i$ into $o$'s region and then
freezes the entire structure.

When a region is created, it is closed and empty, except for its
bridge object. To populate $R$ as in \cref{fig:isolation}, it must
first be made active. The \c{enter} keyword is used to open a
region and making it active. It takes a unique reference to a
bridge object as its argument. Lines~\ref{enter}--\ref{enterends}
of \cref{lst:isolation} show the use of an \c{enter} block to open
the $R$ region to allocate and mutate its objects. (The code executes with region $R$ active and region $R'$
suspended.)

Entering a region moves control inside it and places a \Mut{}
reference to its bridge object in a variable on the stack (\c{r})
that can be used to call methods, or obtain and store references to other
objects in the region. Exiting the \c{enter} block (after
Line~\ref{enterends}) closes the region, and moves control back to
the previous region. While a region is open, the external
reference to its bridge object is
buried~\cite{boyland_alias_2001}, meaning it is not accessible to
the program.

Upon exiting the \c{enter} block, all variables referencing
objects in the now-closed region (\c{c}, \c{e}, and \c{r}) are
invalidated, save for the reference to the bridge object in \c{a}.
Any temporarily permitted references to objects in suspended regions,
such as \Ptr{e}{n} in \cref{fig:isolation}, will be invalidated as well. (We will show how
this is enforced statically in \cref{sec:tmp}.)

\subsection{Controlling Effects through Viewpoint Adaptation}
\label{sec:vpa}

\begin{wraptable}{r}{0.5\textwidth}\small
  \vspace*{-3ex}
  \caption{Viewpoint adaptation.
    If the capabilities of \c{x} and \c{f} are $\alpha$ and $\beta$,
    then the capability of \c{x.f} is $\VPA{\alpha}{\beta} = \gamma$,
    which we read as ``$\alpha$ sees $\beta$ as $\gamma$.''
    For $\dag$, \cf{}~\cref{sec:tmp}.
    The meaning of $\bot$ is inaccessible;
    For $\bot$/\Iso{} see text.
  }%
  \label{tab:vpa}
  \small\centering\setlength{\tabcolsep}{4pt}
  \begin{tabular}{rccccc}
    \toprule
       Capabil- & \multicolumn{5}{c}{Capability on \c{f}}             \\ %
       ity on \c{x}   & \Mut{}    & \Tmp{}    & \Imm{} & \Iso{} & \Paused{} \\ 
    \cmidrule(l{2pt}r{2pt}){1-1}\cmidrule(l{2pt}r{2pt}){2-6}
    \Mut{}        & \Mut{}    & \NA$\dag$ & \Imm{} & $\bot$/\Iso{} & \NA$\dag$ \\ 
    \Tmp{}        & \Mut{}    & \Tmp{}    & \Imm{} & $\bot$/\Iso{} & \Paused{} \\ 
    \Imm{}        & \Imm{}    & \Imm{}    & \Imm{} & \Imm{} & \Imm{}    \\ 
    \Iso{}        & \NA{}     & \NA{}     & \NA{}  & \NA{}  & \NA{}     \\ 
    \Paused{}     & \Paused{} & \Paused{} & \Imm{} & $\bot$/\Iso{} & \Paused{} \\ 
    \bottomrule
  \end{tabular}
  \vspace*{-1em}
\end{wraptable}

We rely on viewpoint adaptation~\cite{dietl_generic_2007} to
capture how a reference's type changes depending on its enclosing
region's relation to the active region. Viewpoint adaptation
means that the type of an object may appear differently depending on
from where it is accessed. For example, when accessed through a
variable of type \Imm{}, a field declared with type \Mut{} appears
as \Imm{}. (This particular case  ensures
that turning a unique reference to a bridge object immutable will
propagate the immutability to the entire region and nested
regions.)

Viewpoint adaptation also changes the types of variables captured
by an \c{enter} or \c{explore} block to propagate suspension.
Captured \Iso{}s retain their \Iso{}-ness rather than become
\Paused{}. \cref{tab:vpa} shows the viewpoint adaptation rules.
The meaning of $\bot/\Iso{}$ is that an \Iso{} location is
inaccessible through a \Mut{}, \Tmp{} or \Paused{},
unless it is \emph{swapped}, \emph{buried} or \emph{borrowed}
(\cf{}~\cref{sec:uniqueness}).



\newcommand{\ScopeA}{\textcolor{blue!66}{\ensuremath{\mathcal{A}}}}
\newcommand{\ScopeB}{\textcolor{blue!66}{\ensuremath{\mathcal{B}}}}

To illustrate viewpoint adaptation,
consider \c{$\ScopeA\ $ enter x \{ y $\ \Rightarrow $    $\ \ScopeB\ $   \}}.
In scope \ScopeA{}, let the
variables \c{x} and \c{v} have the types $\Iso{}~\tau_1$ and
$k~\tau_2$ respectively.
In scope \ScopeB{}, \c{x} is undefined and \c{y} is introduced
with type $\Mut{}~\tau_1$. This reflects the region pointed to by
\c{x} going from closed (denoted by \c{x} being \Iso{}) to active (denoted by \c{y} being \Mut{}).
%
Moving control from \ScopeA{} to \ScopeB{} suspends the region active in \ScopeA{}
(denoted there by \Mut{} and \Tmp{}). Thus, in scope \ScopeB{}, the type of \c{v} is
$(\VPA{\Paused{}}{k})~\tau_2$. Through a \Paused{} reference,
objects in the same region are \Paused{}. \Paused{}
and \Imm{} references stay \Paused{} and \Imm{} respectively
(permanently immutable is stronger than temporarily immutable).
\Iso{} references stay \Iso{}. This
permits opening nested regions of a suspended region.

If $k$ = \Iso{}, then \c{v.f} is not typeable in neither \ScopeA{}
nor \ScopeB{} as $\VPA{\Iso}{k'}=\bot$, regardless of what $k'$
(the capability of \c{f}) is. This is as expected, as \Iso{}'s
cannot be dereferenced (they must be \c{enter}'d).

\subsection{Region Isolation and Bridge Object External Uniqueness}
\label{sec:uniqueness}

Regions which are closed or active only have a single incoming
reference, which goes to the bridge object. Thus, when a region is
closed, it can be moved in and out of other regions by moving its
single incoming reference. When a region is opened, its containing
region is suspended, which means that the object containing field
holding the reference to the bridge object is \Paused{} so the
field cannot be reassigned. Thus, regions cannot move while open.
Finally, when a region is suspended, incoming references are
permitted from the stack and heap of subsequently opened regions
(\cf{} \cref{sec:tmp}). As regions are opened using a lexically
scoped construct (\c{enter} or \c{explore}), regions are opened
and closed in LIFO order. This means that when a suspended region
becomes active again, the permitted incoming aliases that could be
declared in the block have gone out of scope, and the region's
bridge object is again the only incoming alias.

As shown in \cref{tab:sbb} uniqueness of bridge object references
is maintained by a combination of swapping, burying, and borrowing.

\begin{table}[h]
  \caption{Maintaining uniqueness of bridge object references.}
  \label{tab:sbb}
  \vspace*{-1ex}
\small\setlength{\tabcolsep}{.75ex}
\begin{tabular}{p{.14\linewidth}p{.83\linewidth}}
\toprule
\raggedright Swap\newline\cite{harms_copying_1991} & Reading a mutable variable containing an \Iso{} requires that its contents is replaced. For example, \c{y = x} is not permitted when \c{x} is \Iso. However, \c{y = x =}~~$v$ is, which replaces the value of \c{x} by the value $v$ and moves the \emph{previous} value of \c{x} into \c{y}. \\
\midrule
Bury\newline\cite{boyland_alias_2001} & Reading a let-bound variable with an \Iso{} invalidates the variable. For example, \c{foo(x, x)} is not permitted when \c{x} is \Iso{} as the second use of \c{x} cannot be typed. \\
\midrule
Borrow\newline\cite{wadler_linear_1990} & Dereferencing an \Iso{} requires opening its region, where aliasing of the bridge object is unrestricted, and region isolation protects aliases to the bridge object from escaping.\\
\bottomrule
\end{tabular}
\end{table}


Entering a region borrows and/or buries the variable or field
referencing the bridge object. In the case of a stack variable,
the variable is buried to prevent the region from being multiply
opened. In the case of a field, we instead resort to a dynamic
check of the region's state. If the region is closed, it may be
opened. If the region is already open, an exception is thrown.

\subsection{Temporary Objects Allow References to Suspended Regions on the Heap}
\label{sec:tmp}

As exemplified already, we permit local variables to store references to
objects in a suspended region (\eg{} \c{z} in
\cref{fig:isolation:second}). This is sound as objects in
suspended regions are immutable, and because local variables in
the active region are guaranteed to go out of scope before a
region opened by an enclosing \c{enter} or \c{explore} becomes
active (and thus mutable), as explained above.

Because objects with \Tmp{} capability are created in, and bounded
by, a lexical scope, they have the same lifetimes as local
variables declared therein. Therefore, we can grant the same
permissions to store references to suspended regions to \Tmp{}
objects. We permit accessing \Paused{} and \Tmp{} fields through a
\Tmp{}, but not through a \Mut{} (as shown in \cref{tab:vpa}).
Permitting a \Mut{} object to store a suspended reference could
lead to a breach of region isolation (see
\cref{sec:types-enforce-isolation} for an example). Thus from
$\VPA{k_1}{k_2}=\Tmp{}$ it follows that $k_1=\Tmp{}$.

In terms of \cref{fig:isolation}, making $e$ a temporary object
allows its fields to store references with \Tmp{} capability. This
permits the reference \Ptr{e}{n} when $R$ is active and $R'$ is
suspended. However, \Ptr{c}{e} would no longer be permitted unless
$c$ is also \Tmp, etc.

\subsection{Storage Locations, Strong Updates and Bridge Object Swapping}

We unify the treatment of mutable variables (denoted \c{var} as
opposed to \c{let}) and fields through a \emph{storage location}
abstraction (similar to a pointer to a variable or field in C).

Storage locations are typed $k$~\texttt{Store[$k'\,\tau$]} where
$k$ is the capability of the frame or object containing the
location and $k'$ is the capability of the value stored at the
location.

We add a new capability that we call \Var{} for use in mutable
local variables. \Var{} differs from \Tmp{} in that it supports
strong updates. Its viewpoint adaptation rule is
$\VPA{\Var}{k}=k$ (``\Var{} sees $k$ as $k$'').

As shown in \cref{lst:store}, a mutable local variable \c{x}
holding a $\tau$-typed value has the type \c{var}
\c{Store[$\tau$]}. Storage locations are subject to the normal
rules for viewpoint adaptation, so opening another region when
\c{x} is already in scope will change the type of \c{x} to
\c{paused} \c{Store[$\tau$]}. We introduce a dereference operator
\c{*} and an update operator \c{:=} to access the contents of a
storage location.
A storage location must be \Mut{}, \Tmp{}, or \Var{} to be
updated.
We apply viewpoint adaptation to type the result of dereferencing.
On Line~\ref{deref}, \c{*x} has type
($\VPA{\Paused}{\Mut}$)~\c{Cell}, \ie{} \c{paused Cell}.

\begin{wrapfigure}{r}{0.48\textwidth}
\vspace*{-1em}
\begin{code}[numbers=right,firstnumber=1,caption={Storage locations example.},label=lst:store]
var u = new iso Cell(42) // (*@\BLUE{\footnotesize var Store[iso Cell]}@*)
var x = new Cell(4711) // (*@\BLUE{\footnotesize var Store[mut Cell]}@*)
enter u { y $\Rightarrow$
  // y has type (*@\BLUE{\footnotesize var$\,$Store[mut Cell]}@*)
  // x has type (*@\BLUE{\footnotesize paused$\,$Store[mut Cell]}@*)
  y := new Foo(*y) // changes bridge object('s type) (*@\label{type-change}@*)
  y := *x // (*@\RED{\it rejected:}@*) *x is (*@\BLUE{\footnotesize paused$\,$Cell}@*), not (*@\BLUE{\footnotesize mut$\,$Cell}@*) (*@\label{deref}@*)
  x := ... // (*@\RED{\it rejected:}@*) the x storage location is (*@\BLUE{\footnotesize paused}@*)
} // change to u becomes visible
\end{code}
\vspace*{-1em}
\end{wrapfigure}

We support changing the bridge object of a region---including
changing it for an object of a different type---by presenting the
borrowed bridge object reference internally in the \c{enter} block
as a \Var{} storage location. (Note that it is not possible to
update the bridge object in an \c{explore} as it opens the region as
suspended.)

Line~\ref{type-change} shows that changing the bridge
object to an object of another type is possible by simply
assigning to \c{y}. Strong updates of fields are not possible,
and this is handled by using \c{tmp} \c{Store[$\ldots$]} instead of
\c{var} \c{Store[$\ldots$]} to type a bridge object reference
borrowed from a field as opposed to a stack variable. 

\subsection{Types Enforce Region Isolation and the Single Window of Mutability}
\label{sec:types-enforce-isolation}

Region isolation means \emph{no references into \underline{active}
  or \underline{closed} regions from other regions (modulo unique
  references to bridge objects) or from \underline{immutable}
  objects, and no outgoing references from \underline{closed}
  regions to open regions}. Let's see how our types enforce this, by
looking at $R$ in~\cref{lst:isolation}.

The newly created region $R$ (Line \ref{r_created}) is isolated as \Iso{}
constructors only accept \Iso{}'s and \Imm{}'s as arguments. 
Right after creation, $R$ is \underline{closed}, and its only
external reference is \Iso{}. Since \Iso{}'s cannot be the
receivers of method calls or field accesses, we cannot read or
write internal objects in $R$, so we cannot create the illegal
references \Ptr{m}{e} or \Ptr{e}{n}.

When $R$ is \underline{active} (Line
\ref{enter}--\ref{enterends}), all previously suspended regions
stay suspended (none in~\cref{lst:isolation}), and are joined by
the current region ($R'$). This is captured by viewpoint
adaptation which changes all variables which are \Mut{}, \Tmp{},
or \Var{} in the enclosing region to \Paused{}. This prevents
these variables from being updated, and reading them yields
\Paused{} references. Field updates via \Paused{} or
\underline{\Imm{}} references are not allowed, and method calls on
such references require that the method's self type matches the
external view, meaning any callable method cannot perform a field
update on \c{self} (or call such a method). Thus, we cannot create
\Ptr{m}{e} or \Ptr{o}{a}.

As permitted by our definition of region isolation, we may store
\Paused{} references in the fields of \Tmp{} objects in $R$ while
$R$ is active (\eg{} \Ptr{e}{n} if $e$ is created as \Tmp{} on
Line~\ref{e_not_tmp}). These references will be invalidated when
$R$ closes as \Tmp{} references can only be stored in variables
local to the \c{enter} block (since mutable variables in the
enclosing scope have been suspended), or in other \Tmp{} objects
(\ie{} $a \to c \to e$ is an impossible path if $e$ is \Tmp{}, as
the $a$ object is \Mut{} by definition).

If we did not invalidate references into suspended regions such as
\Ptr{e}{n}, we could circumvent region encapsulation. For example,
imagine closing $R$ (without invalidating \Ptr{e}{n}), then
closing $R'$ while moving $R$ out of $R'$. Then reopen $R$ without
first opening $R'$. Now \Ptr{e}{n} would constitute a reference into
the internals of a closed region, thereby breaking region
isolation.

Inside an \c{enter} or \c{explore} block the enclosing scope is
immutable. Together with region isolation this gives that only one
region at a time is mutable, \ie{} a ``single window of
mutability''.

Last, region isolation means that reassigning the pointer to the
externally unique bridge object effectively changes the region
topology of the heap~\ref{goal:ownership-transfer}.

\newcommand{\ScopeR}[1]{\textcolor{blue!66}{\ensuremath{\mathcal{R}_{#1}}}}

\section{The Use of Regions for Managing Liveness}
\label{sec:gc-example}

We have sketched how our type system enforces region isolation and the single window of mutability.
In this section, we will show how this translates to costs for managing liveness when considering
memory management in isolated regions.

\begin{wrapfigure}{r}{0.48\textwidth}
  \centering
  \includegraphics[page=4,scale=\FigureScaleFactor,trim=0 530 1130 0,clip=true]{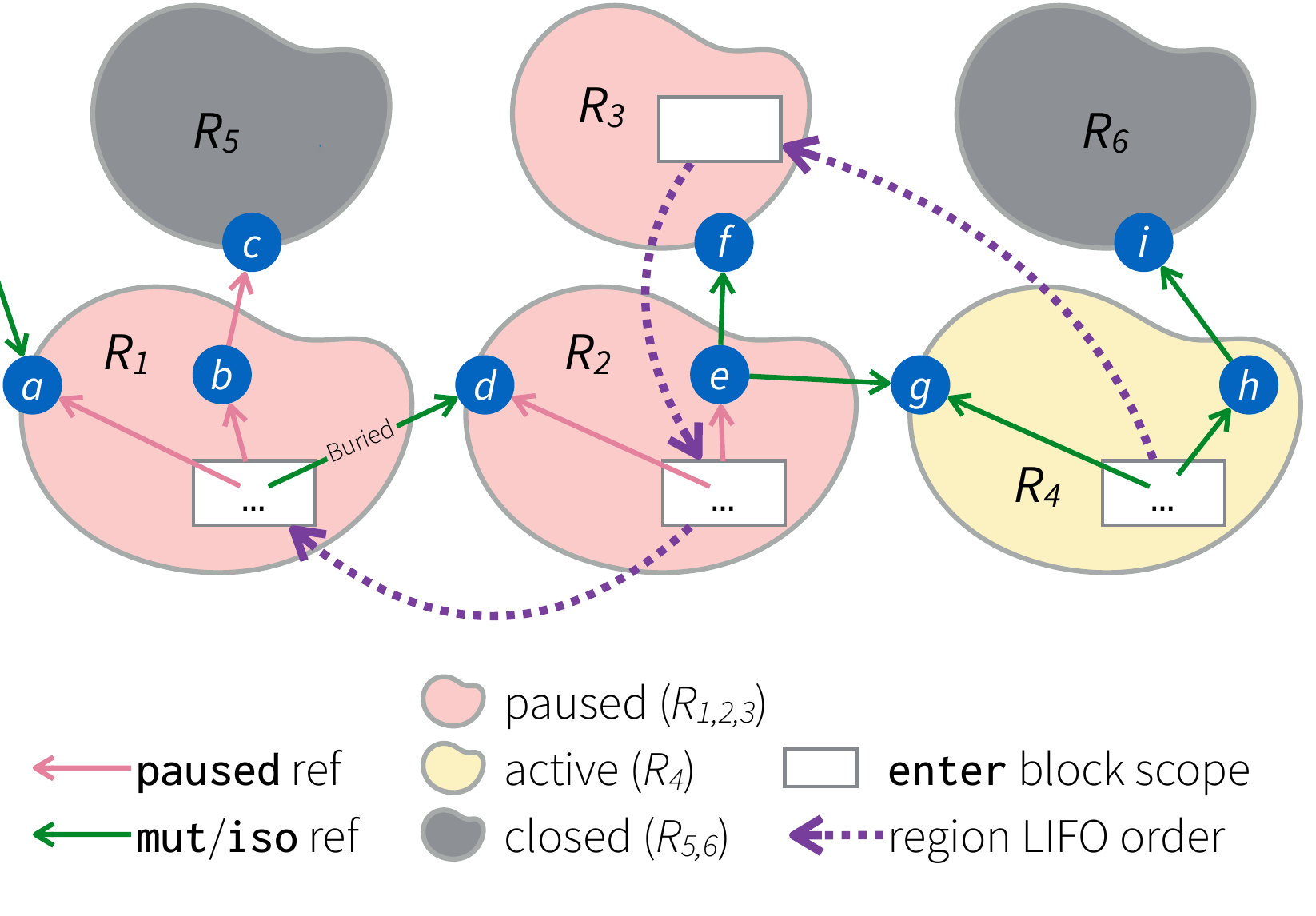}
\vspace*{-1em}
\caption{Program with 4 open and 2 closed regions.}
\label{fig:explanation}
\vspace*{-1em}
\end{wrapfigure}

\cref{fig:explanation} shows a heap consisting of regions
$R_1$ to $R_6$.
Presently, the program has entered $R_1$, $R_2$, $R_3$ and $R_4$
in that order. Ignoring method call
indirections, we can imagine a corresponding program shape starting in $R_1$:
\c{$\ScopeR{1}\ $ enter x \{ d $\ \Rightarrow\ \ScopeR{2}\ $ enter e.f \{ f $\ \Rightarrow\ \ScopeR{3}\ $ enter e.g \{ g $\ \Rightarrow\ $    $\ \ScopeR{4}\ $   \}  \}  \}}.

The region stack is thus [$R_4$, $R_3$, $R_2$, $R_1$] where
$R_4$ is active. Region isolation prevents references from objects in region $i$ to objects
in region $j$ if $i < j$ with the exception of the bridge object
references: \Ptr{\texttt{x}}{d}, \Ptr{e}{f}, and \Ptr{e}{g}. The
first of these is made inaccessible right after the first \c{enter}
by making \c{x} undefined in \ScopeR{2} and nested scopes. While
$R_4$ is open, $R_2$ and $R_3$ are suspended, meaning the fields
holding \Ptr{e}{f} and \Ptr{e}{g} cannot be reassigned (static
check). Furthermore, they cannot be dereferenced (\ie{} opened)
since the regions are already open (dynamic check). Thus, while
$R_4$ is open, the incoming references to the bridge objects will
remain the same, \ie{} the path that holds the region alive is
\emph{stable}.\footnote{Since an \cc{enter} block can change the
  bridge object on exit, the incoming reference does not affect
  liveness. We can think of the incoming bridge object reference
  as being \emph{invisibly nullified} during the \cc{enter}, and
  reinstated at the exit.}

Because of this, as long as $R_4$ remains active, liveness of objects in regions $R_1$--$R_3$
and $R_5$ is \emph{invariant} as no activity in $R_4$ can cause
objects in these regions to become garbage.\footnote{Every
  reference in $R_4$ to an object in $R_1$--$R_3$ is a copy of a
  reference inside one of those (immutable) regions that still
  exists.} This does not hold for $R_6$, as $R_4$ could drop
\Ptr{h}{i} to make the entire region $R_6$ garbage. It does
hold for \Ptr{b}{c} however, since $b$ is in a suspended region ($b$ is
temporarily immutable).

Furthermore, because of the absence of references from $R_1$--$R_3$
into $R_4$---with the exceptions of the stable bridge object
references that are either buried or cannot be re-opened, objects in $R_1$--$R_3$ \emph{cannot affect the
  liveness of objects in $R_4$} (this is also true for $R_5$ and
$R_6$ as they are closed). Thus, we can safely ignore references
to objects in suspended regions when managing liveness. For
example, if objects in $R_1$, $R_2$ or $R_3$ are managed by
reference counting, we do not need to increment or decrement
reference counts when manipulating \Paused{} references
in $R_4$. Similarly, if objects in $R_1$, $R_2$ or $R_3$ are
managed by a tracing GC, tracing in $R_4$ does not need to follow
\Paused{} references. Thus, when managing memory in the
active region, we can safely ignore any outgoing references, and
so it is possible for $R_1$--$R_3$ to use different strategies,
unbeknownst to $R_4$ and irrespective of how $R_4$ manages its
memory~\ref{goal:mix-match}. The only references to objects in
$R_5$ and $R_6$ that are possible (given that the regions are
closed) are to the bridge objects $c$ and $i$. Aliases of $c$ are
not possible in $R_4$ as \Iso{} references are unique, and we
cannot transfer references out of an immutable $b$. However, we do
need to track liveness of the reference to $i$, which is possible
to do statically \eg{} as in Rust.

From the reasoning above it follows that liveness of objects in $R_4$
can be determined by looking at objects in $R_4$ alone, meaning
that the costs of memory management are determined by the contents
of and activity in $R_4$~\ref{goal:local}. This makes it possible
to collect garbage in just $R_4$~\ref{goal:incremental}.

\section{Programming with \Venice{} Regions}%
\label{sec:examples}

As an example of how regions enable predictable memory
management performance, consider the server application ``\emph{Po}'' with
the following key characteristics (see \cref{lst:skeleton} for skeleton code):

\begin{enumerate}
\item The server serves incoming requests. Tasks that process
  requests are short-lived and their side-effects are typically in
  the form of data stored in a database.
\item Responses to requests are served from data in an in-memory
  key--value store implemented as a skip list. This storage will
  shrink and grow continuously during execution.
\item Values in the store can have a complicated graph-like
  structure (\eg{} they may contain cycles)\@.
\end{enumerate}

We now explain how we can express this, and compare with
Cyclone, MLKit, Pony, and Rust.

\subsubsection*{Processing Requests}

To manage allocations necessary to process a request, each request
is wrapped in a region (Line 16, Line 49). These regions use
\emph{arena allocation}: allocations in the region persist until
the region itself is deallocated. This gives cheap bump-pointer
allocation and fast deallocation of the entire region once the
response has been computed. If a request must be processed by
different threads, this can be done cheaply due to the
transferability of \Iso{}'s. If some requests turn out to require
considerable processing time, their corresponding regions might
switch away from arena allocation at the small cost of changing one
annotation at the creation site of the region(s).

\paragraph{Comparison}

Cyclone's LIFO regions are perfect for this purpose (as are
MLKit's, provided that the inference engine infers according to
intentions, or better). While Cyclone does not support switching
from arena allocation, it does permit manually managed reference
counts or unique objects that can be manually deallocated during
the arena's lifetime. Pony only supports memory management on a
per-actor level, (using tracing GC). We could make each request an
actor, but it would have to communicate asynchronously with all
surrounding state. Rust does not support regions, but could \eg{}
build a unique object holding unique values and thread this object
through computation. While more complicated, Rust's values would
have individual lifetimes.

\subsubsection*{Key--Value Store}

The key--value store is implemented as a single region containing
a skip list (Line 5, could as well be a hash table, B-tree, etc.).
As it is a large long-living data structure of objects with
different lifetimes, arena allocation is not a suitable strategy.
Furthermore, if the resources (values) stored in the skip list are
costly, \emph{reference counting} is a good choice as it allows
the resources in the list to be recycled immediately when they
become garbage. Alternatively (the path we chose) is to make each
element a region of its own, with independently managed memory (Line 1).

If the store is large enough to warrant parallel
accesses, it can be divided into several smaller regions with one
list each. For a compile-time guarantee that no reference count
manipulations in the skip list are needed during lookups, lookups
can be implemented using \c{explore} rather than \c{enter} as the
former opens the skip list region directly as \Paused{} (Line 23,
and its dynamic extent \c{foo()}). Note that since the elements in
the skip list are regions of their own, these can be entered
separately and thus mutated, even if the list structure
is immutable (Line 35).

\paragraph{Comparison}

Cyclone's dynamic regions are a good fit for the key--value store.
The objects that make up the skip list would require manual
reference counting, which can be laborious. Failure to properly
manage reference counts in Cyclone will also lead to ``memory
leaks'' which will not be reclaimed until the region holding the
store is (manually) destroyed. Pony can wrap the skip list inside
an actor with an asynchronous interface, and manage its memory
using tracing of the entire list leading to more time spent
tracing memory and more floating garbage. Skip lists (hash tables,
B-trees, etc.) cannot be constructed in Rust without judicious use
of \c{unsafe}. Safe Rust's reference counting does not relax its
ownership rules, so mutation of aliased values is not permitted.

\begin{figure}[p]
\begin{code}[style=tinyverona,numbers=left,caption={Skeleton code for \emph{Po}. To save space, we permit constructor arguments to \cc{iso} objects to take \cc{mut} arguments (Lines 38, 45, 52, and 54). This is safe and can desugar to an extra \cc{enter}. These lines all create an object that escapes the active region and is merged into the enclosing region, making them \cc{mut} in \cc{r} (\cc{append} expects a \cc{mut} argument). The \cc{ServerSocket} is created elsewhere. It likely does not use arena allocation since it allocates on each turn of the loop (Line 13). If it did, those allocations would only be free'd on Line 31. The default annotation on \cc{new} \cc{iso} is \cc{<Arena>} (\cf{}~\cref{sec:supp-diff-memory}). Notice how code is agnostic to how memory is managed.},label={lst:skeleton}]
type_alias KV = Skiplist[imm Id, iso Value] // To shorten code horisontally for this presentation
type_alias Response = UpdateOK | InsertOK | DeleteOK | Failure

def start_po(fn : imm String, server_socket : iso ServerSocket) : Unit {
  let kv : iso KV = new iso<RC> KV // create empty key-value store; (*@<RC>@*)=reference counting, see (*@\it\cref{sec:supp-diff-memory}@*)
  enter kv { list => // Populate key-value store from persistent storage
    let data : tmp File = open(fn, "r")
    ... list.insert(...) ... // read contents from data and add to list
  } // data goes out of scope, so get's free'd and closed

  enter server_socket { ss =>
    while (ss.is_open()) {
      let socket : mut Socket = ss.accept() // new connection
      let raw_request : imm String = socket.read_request() // get incoming request

      let r : iso Response = enter new iso<Arena> Unit { _ => // arena-managed region for tmp allocations
        let work : mut List[mut Tasks] = RequestParser::parse(raw_request) // parse request
        let response = new mut List[mut Response] // holds all responses to tasks in request

        while (!work.empty()) {
          response.append(merge match work.pop() { // merges iso result from match into r since append expects mut
            case mut StopTask => return // stop service, no response to client
            case mut UpdateTask(id) =>          explore kv { kv' => update(kv', id) }
            case mut InsertTask(id, payload) => enter kv { kv' => insert(kv', id, payload) }
            case mut DeleteTask(id) =>          enter kv { kv' => delete(kv', id) }
          })
        }
        response.accumulate(new iso Message) // chained through accumuator, moves out of arena
      } // arena effectively goes out of scope, allocs on line 17, 18 + any tmp objects are freed
      socket.respond(r); // render response object
}}}

def update(kv : paused KV, id : imm Id) : iso Response // process UpdateTasks, inside suspended kv
  let value : paused Store[iso Value] = kv.get(id) // reference to a link's reference to a Value
  enter *value { v => // Requires a dynamic check - because of aliasing cannot rule out v is already open
    v.add_log_entry() // adds surviving object to v's region
    v.remove_some_token() // makes object in v's region garbage
    return new iso UpdateOK(v) // moves out of *value and kv regions and merged into r on line 21
}}

def insert(kv : mut KV, id : imm Id, payload : imm Payload) : iso Response // process InsertTasks, inside kv
  let new_value = Factory::create(id, payload) // decides memory management for new_value dynamically
  enter new_value { v => v.tokens = new mut List; v.log = new mut List }
  kv.insert(id, new_value) // buries new_value
  return new iso InsertOK() // moves out of kv region and merged into r on line 21
}

def delete(kv : mut KV, id : imm Id) : iso Response // process DeleteTasks, inside kv
  enter new iso<Arena> Unit { _ => // create new tmp region, the one created on Line 16 is not accessible here
    let q : mut SQL_Query = Factory::make_can_delete_query(id) // q cannot refer to kv because it is mut
    let r : tmp SQL_Result = Backend::execute_query(q) // because it is tmp, r could refer to kv if it needed to
    if (!r.OK) return new iso Failure(...) // moves out of tmp and kv regions and merged into r on line 21
  } // arena goes out of scope, allocs on lines 50, 51 are free'd
  return enter kv.remove(id) { v => new iso DeleteOK(v) } // Creates garbage in kv, drops a Value region
}
\end{code}
\end{figure}

\subsubsection*{Values in the Store}

Finally, the elements in the store are suitable for either
reference counting or tracing GC because of their graph-like
structure (Line 42 delegates this decision to a factory method).
GC is especially favoured in the (possible) presence of cycles
which are expensive to detect with reference
counting~\cite{jones_garbage_2016}.

\paragraph{Comparison}

Later versions of Cyclone and MLKit support a global arena where
memory is managed using tracing GC\@. Thus, all elements in the
store contribute to pressure on the same GC, and GC requires
tracing through all elements to free garbage objects in one
element. MLKit's region propagation requires all elements to be
put in a single region. Pony can handle this pattern by making
each element an actor, which makes each element aliasable, and use
an asynchronous interface. Finally, Rust will not be able to
express and manage lifetimes of these structures automatically. A
combination of \c{unsafe} and manual memory management is needed.

\subsubsection*{Navigating Regions}

\Cref{lst:zip} shows a zip computation involving three unrelated regions.

\newcommand{\TP}{[\,]}
\begin{wrapfigure}{r}{.5\textwidth}
  \vspace*{-2ex}
\begin{code}[numbers=right,firstnumber=1,caption={Opening two non-nested regions, computing a result in an active region. \TP{} is type parameters.},label=lst:zip]
let staff : iso List[imm Employee] = ...
let reviews : iso List[imm Review] = ...
var zip = new iso List[imm (String, (*@Int@*))]

explore staff { s => // open as immutable
	explore reviews { r => // open as immutable
		enter zip { z => // open as mutable
			let si = s.iterator() // si is tmp (*@\label{si}@*)
			let ri = r.iterator() // ri is tmp (*@\label{ri}@*)
      while (si.has_more() && ri.has_more()) {
			  z.append(new imm Pair(si.next().name(), (*@\label{append}\label{next1}@*)
          ri.next().calculate_salary())) (*@\label{next2}@*)
      }}}} (*@\label{zip:last}@*)
\end{code}
  \vspace*{-4ex}
\end{wrapfigure}

Using \c{explore}, we open the staff and reviews regions to make
them temporarily immutable and their contents accessible. Finally,
we open the zip region using \textsf{enter}. This makes the region
active which allows allocation of the two iterators on
Lines~\ref{si} and \ref{ri} and any allocation needed by the call
to append on Line~\ref{append} to extend the list. It also allows
the mutation in the \c{next()} calls to advance the iterators, and
the mutation necessary to add the new pair to the zip list.
Allocating the iterators inside the same region as their
corresponding list is not useful as it would make the
iterators immutable on Lines~\ref{next1} and \ref{next2}. This
would cause the program to not typecheck---as the \c{next()} method
needs to update the iterator, it needs to be called on a \Mut{} or
\Tmp{} receiver. Since the iterators need to store \Paused{} references,
they must be \Tmp{} (\cf{} \cref{tab:vpa}). This can be handled in
\c{iterator()} by overloading on the self type, letting the
implementation with \Paused{} self-type return a \Tmp{} reference.
Elements are immutable so \eg{} \c{next()} returns \Imm{}.

\subsubsection*{Design Thoughts on Explore vs. Enter---And Invariants}
\label{sec:explore}

The \c{explore} construct is essentially syntactic sugar for two
nested \c{enter} blocks, the outermost entering the region to be
explored and the innermost entering a fresh region:
\begin{center}
  \c{explore x \{ y => $\ \ldots\ $ \} }
  \quad\emph{desugars to}\quad
  \c{enter x \{ y => \{ enter (new iso Unit) \{ _ => $\ \ldots\ $ \} \} }
\end{center}
The first \c{enter} activates the region, and the second suspends
it. The new region (\c{new iso Unit}) is independent from the rest
of the program. As it is active, it serves all allocations that
appear inside the \c{explore} block (as suspended regions do not
permit allocation or deallocation, \cref{tab:overview}). What
\c{explore} guarantees that unprincipled nesting of \c{enter}s
does not, is that the explored region was not mutated before
suspended. Conceptually, this is a big difference as we will
explain next.

Similar to object invariants, we expect invariants of a closed
region to hold at the time of opening. While active, invariants
may temporarily be broken and then restablished before the region
is closed. As a nested \c{enter} can reference any
enclosing region, it will be able to observe any invariants broken
by mutation following the opening of the enclosing regions. By opening
regions directly in a suspended state, \c{explore} ensures that
region invariants continue to hold. We are considering using a
separate capability to capture this statically. 
We are also considering an
``eager'' explore construct that opens a region along
with all its subregions as suspended in one fell swoop. This
would avoid the need for explicit opening of subregions thus
further simplifying working with immutable objects. The
cost is more complexity in the type system. Time will tell
whether this complexity is warranted or not.

With respect to memory management, \c{explore} allows opening a
region for reading, and navigating through it, without any memory
management overhead as the region's object structure is invariant
and there are neither allocations nor deallocations in the region.

\subsubsection*{\Venice{}'s Borrowing Capabilities}

\begin{wrapfigure}{r}{.45\linewidth}
\vspace*{-1em}
\begin{code}[style=tinyverona,numbers=none]
var x = new iso Cell // x : var Store[iso Cell]
enter y { => _ // now x : paused Store[iso Cell]
  enter *x { => z
    ... // Can still mutate z!
  }
}
\end{code}
\vspace*{-1em}
\end{wrapfigure}

Traditional borrowing as originally introduced
by~\citet{wadler_linear_1990}, explored deeply by
\eg{}~\citet{boyland_alias_2001} and
\citet{boyland_capabilities_2001}, and popularised by Rust relaxes
uniqueness of a value in a well-defined lexical scope. We can
express a similar form of borrowing through the type \c{paused
  Store[iso T]}, \ie{} a reference to a storage location in a
suspended region storing a reference to a closed region. Such a
reference (\eg{} \c{x}) can be shared freely inside a single
thread, allowing it to flow to a place where it can be opened
(\c{enter *x}) \emph{with} mutation rights, including swapping the
bridge object (as long as the new bridge object is a subtype of \c{T}).

\section{Formalising \Venice{}}%
\label{sec:formalism}

We formalise \Venice{} through two interacting languages:
\emph{region} and \emph{command}, and their respective semantics. The former
controls all accesses to memory (loads and stores), allocation of objects in
regions, creating, merging, freezing---and importantly entering and
exiting---regions. The most important properties of the region language is
expressed as a \emph{topology invariant} (\cf{} \cref{sec:meta-theory}).
The command language is essentially ``what the programmer wrote''.
This separation makes it possible to specify \eg{} under what
conditions a store is valid, irrespective of what caused the
store.

During execution, the command language emits effects which the region language
performs. The static semantics of the command language ensures,
modulo one dynamic check, that the topology invariant is preserved.
We present most of the rules of the region language and key type system rules.
Additional details are available in the appendix.

\subsection{Dynamic Semantics of the Region Language}%
\label{sec:outer-semantics}

\begin{wrapfigure}{r}{.25\textwidth}
  \vspace*{-3em}\small
  $\begin{array}{@{~}rcl}
     \RCfg & ::= & \langle \RS; H; H; H\rangle        \\
     \RS   & ::= & \RF \mathop{::} \RS
                   ~|~ \epsilon                             \\
     \RF   & ::= & (r, S, F)                          \\
     H     & ::= & R ~|~ H * H ~|~ \epsilon
   \end{array}$
   \vspace*{-1em}
   \caption{Configuration in region language (1/2)}
   \vspace*{-1em}
 \end{wrapfigure}
A configuration in the region language has four components: a LIFO region stack
$\RS$ and the sub-heaps of open regions \OpenRegs, closed regions \ClosedRegs,
and frozen regions \FrozenRegs. The latter is an unimportant simplification
(conceptually, only mutable objects live in regions).
A heap $H$ is a collection of disjoint regions $R$. Opening a
closed region moves it from \ClosedRegs{} to \OpenRegs{} and pushes a new stack
frame on top of $\RS$. Closing the top-most region in $\RS$ returns it back to
\ClosedRegs{}. Freezing a region moves it, and all regions reachable from it,
permanently to \FrozenRegs{}.
As an example, consider the region topology depicted in
\cref{fig:explanation}. We can write down the corresponding configuration as $\langle \RS,
\OpenRegs, \ClosedRegs, \FrozenRegs \rangle$ where
\[
  \begin{array}{lcl}
    \RS &=& \RF_4 \mathop{::} \RF_3 \mathop{::} \RF_2 \mathop{::} \RF_1
    \mathop{::} \epsilon \\
    \OpenRegs &=& R_1 * R_2 * R_3 * R_4 \\
    \ClosedRegs &=& R_5 * R_6 \\
    \FrozenRegs &=& \epsilon
  \end{array}
\]
For the region sub-heap $R_i$, if $R_i$ is open (\ie{} part of $\OpenRegs$),
$\RF_i$ is its region frame (depicted in~\cref{fig:explanation} as a white box) that holds the stack variables created in the scope of the corresponding \c{enter} block. Opening the closed region
$R_6$ would push a new frame $\RF_6$ above $\RF_4$ in $\RS$, and move $R_6$ from
$\ClosedRegs$ to $\OpenRegs$. Similarly we could freeze (merge) $R_6$, which
would move it from $\ClosedRegs$ to $\FrozenRegs$ (remove it from $\ClosedRegs$
and merge it into $R_4$).

\begin{wrapfigure}{r}{.25\textwidth}
  \vspace*{-1em}\small
  $\begin{array}{@{~}rcl}
     S     & ::= & \iota \mapsto o, S ~|~ \epsilon    \\
     F     & ::= & f\mapsto v?, F  ~|~ \epsilon       \\
     v?    & ::= & v ~|~ \mathbf{undef}               \\
     v     & ::= & (k, \iota)                         \\
     R     & ::= & (r, S)                             \\
     o     & ::= & (\#CL, F)                          \\
   \end{array}$
   \vspace*{-1em}
   \caption{Configuration in region language (2/2)}
   \vspace*{0pt}
 \end{wrapfigure}
%
In this model, an inter-region reference into an open region is permissible iff
it points downwards in the region stack (from left to right according to
$\RS$), or it is the unique (\Iso) reference through which the region was
opened. The LIFO region stack constitutes a ``path'' through the region forest
that corresponds to the opening order of \c{enter} blocks (\cf{} the region LIFO
order in \cref{fig:explanation}). Thus, in \cref{fig:explanation}, any reference
from $R_4$ to $R_3$ is permissible (as long as we do not close $R_4$), while a
reference from $R_2$ to $R_3$ must necessarily be the reference from object $e$
to object $f$. We model \c{explore} as nested \c{enter}s.
%
%

%
A region stack frame \RF{} contains a region identifier $r$, a temporary store
$S$ for objects whose lifetimes are bounded by the scope of the region's \c{enter}
block (values with capability \Tmp{}), and a map $F$ from variable names $f$ to
values $v$, representing the local variables in that \c{enter} block (we model
destructive reads of a variable $x$ by remapping it to $\mathbf{undef}$, at
which point reading $x$ again will lead to the program getting stuck).
A region $R$ is a tuple of a (unique) region identifier $r$ and a store $S$
containing the objects in that region.

Objects are identified by $\iota$.
Values $v$ are object identifiers $\iota$ tagged with a capability $k$.
Stores $S$ map object ids $\iota$ to objects $o$ which store their
class tag $\#CL$ and fields (for simplicity we reuse the same $F$
as for local variables, although a field will never contain $\mathbf{undef}$).

The command language communicates with the region language via
effects. The relation
$\RCfg \yrightarrow{\Eff}[-1pt] \RCfg'$ should
be understood as performing the effect $\Eff$ in $\RCfg$,
resulting in $\RCfg'$.
%
Effects include entering and exiting a region, loading a value from an object
store, writing (swapping) a value for another in an object store, merging or
freezing a region, etc.
%
We now describe a selection of rules for these effects.

\newcommand{\RN}[1]{\textsc{#1}}
\newcommand{\RNP}[1]{(\RN{#1})}


{\footnotesize
  \drulesnohead{
    regionXXload
  }
}\noindent
In every effect, the first parameter should be understood as the name of the
variable where the results should be stored. For example, the effect
$\mathit{load}(x, y.f)$ is handled in rule \RN{region-load} by binding the value
of field $y.f$ to variable $x$.
First, the value of $y$ is looked up in the top stack frame as
$(k, \iota)$. The object id $\iota$ is used to find the
corresponding object $o$ in the configuration---it may be stored
in the subheap of an open or frozen region, or in one of the
temporary stores on the region stack. The capability $k$ is used
for viewpoint adaptation of the (capability of the) value $v$ of
field $f$ in $o$ before it is inserted into the top frame.


\begin{wrapfigure}{r}{.3\textwidth}
\vspace*{-1ex}\small
$\begin{array}{l}
   \mathbf{get}(F[x \mapsto v], \text{\c{drop}}\ x) =\\
   \quad (v, F[x \mapsto \mathbf{undef}])\\
   \mathbf{get}(F[x \mapsto (k, \iota)], x) =\\
   \quad ((k, \iota), F[x \mapsto (k, \iota)])\\
   \quad \text{ if } k \not\eq \Var \land k \not\eq \Iso
 \end{array}$
\vspace*{-1ex}
\caption{(Non-)destructive reads}
\label{fig:gethelper}
\vspace*{-1ex}
\end{wrapfigure}
{\footnotesize
\drulesnohead{
  regionXXswapXXtemp,
  regionXXswapXXheap
}
}\noindent
Field assignments are caused by the effect $\mathit{swap}(x, y.f, \Use)$, which
writes the value of $\Use$ to $y.f$ and binds the
\emph{old} value of $y.f$ to $x$. A $\Use$ is a potentially destructive variable
access ($z$ or $\kw{drop}~z$, see \cref{fig:syntax}). Rules \RN{region-swap-temp} and
\RN{region-swap-heap} handle the cases where the object being assigned to is in
the temporary store or on the heap. In both cases, we perform the $\Use$ (which
may make a variable invalid) with the helper function $\mathbf{get}$ (see
\cref{fig:gethelper}). We then proceed just as when loading a field, but finish
by updating the object being assigned to and update its containing store $S$.
Note that assigning and loading mutable variables are special
cases of the $\mathit{swap}$ and $\mathit{load}$ effects since we model mutable
variables as single-field objects.

{\footnotesize
\drulesnohead{
  regionXXallocXXheapXXmut,
  regionXXallocXXheapXXiso
}
}\noindent
Allocation on the heap is caused by the effect
$\mathit{halloc}(x, k, C, \Use_1 ... \Use_n)$, which instructs the
region language to \underline{h}eap \underline{alloc}ate a new
$C$ object with fields initialized according to $\Use_1 ... \Use_n$ and bind it
to the name $x$. The capability $k$ denotes whether to allocate in the current
region \RNP{region-alloc-heap-mut} or in a new region
\RNP{region-alloc-heap-iso}.
Since each $\Use$ is a possibly destructive variable access the ordering
matters. We begin by performing these one by one with the local variables $F_1$.
Each value $v_i$ is paired up with the corresponding field $f_i$ of the class
and put into an object $o$.
We then add $o$ at location $\iota$ to the subheap of the
currently active region, or add a new region $r'$ in the closed
regions containing only the object $o$ at location $\iota$.
Finally we bind the object to $x$ in the top frame with capability
\Mut{} or \Iso{}.
We omit the rule \RN{region-alloc-temp} which allocates objects
with capabilities \Tmp{} or \Var{} in the temporary store $S$ of
the currently active region frame.

The key rules of the region language govern entering and exiting
a region.
{\footnotesize%
\drulesnohead{
  regionXXenterXXok
}%
\drulesnohead{
  regionXXenterXXfail
}%
}\noindent
\RN{region-enter-ok} shows successfully entering a region $r'$
through its bridge object $\iota'$ stored in the field $f$ of the
variable $y$.
(This operation can fail if $R'$ is already opened. This will not change the
state in the region language, as seen in \RN{region-enter-fail}, and it is up
to the command language to choose how to handle this: by exception,
having a construct like \c{if-enter-else}, etc. For simplicity, the command
language steps to a failure state.)
The $\mathit{enter}$ effect supplies four things: the name $w$ and capability
$k$ of the parameter of the \c{enter} block, the field $y.f$
through which we are entering, and a list of bindings
$\overline{z = \Use}$ denoting the block's captured variables.
Going back to \cref{lst:isolation}, the \kw{enter} block captures \c{i}, so the
corresponding effect would include $z = \text{\c{i}}$. Note that $z$ is chosen by the
command language and due to variable renaming is not necessarily \c{i}.
Considering the rule again, we first use the $\mathbf{get}$ helper function to
perform the $\Use$s.
For each resulting value $(k_i, \iota_i)$, we apply the \Paused{}
viewpoint adaptation when $k_i$ is not \Iso{} and create a new
mapping $F$ of the captured variables.
We then get the value $\iota$ of $y$, load its corresponding
object $o$ and extract the value $\iota'$ of field $f$ (our bridge
object).
On the last line of the premises we check that $\iota'$ is an
object in a closed region $r'$; this region will be moved into
the collection of open regions. We extend $F$ with a mapping from
$w$ to a fresh location $\iota''$, and finally install this
extended $F$ into a region frame $\RF$ where $\iota''$ is the
identifier of a ref cell object pointing to our bridge object
$\iota'$. We push this region frame onto the region stack.

{\footnotesize%
  \drulesnohead{
    regionXXexitXXheap,
    regionXXexitXXtemp
  }%
}\noindent
\RN{region-exit-heap} and \RN{region-exit-temp} describe exiting the region
$r'$, popping its region frame from the top of the region frame stack. After
exiting, the region frame corresponding to $r$ will be on the top of the stack
and thus active. The $\mathit{exit}$ effect provides $\Use$ and $x$ which
correspond to the return value and the variable to which this will be bound (in
the stack frame of $r$). $z.f'$ specifies a location in $r'$ where a reference
to the new bridge object can be found. Finally, $y.f$ specifies a location where
this reference will be written.
The only difference between \RN{region-exit-heap} and \RN{region-exit-temp} is
where the object pointed to by $y$ is located. In the former it is on the heap
of some open region, while for the latter it is in a temporary store in
the region stack.

For simplicity, we do not implicitly reinstate \Iso{} variables
captured from the previous region even if they are still valid
upon exit from a region (this would be sound, \cf{} \cref{lst:zip}
where variables \c{reviews} and \c{zip} are reinstated in the
top-level scope after line \ref{zip:last}). This is without loss
of generality as we can return them in an object together with the result $v$
and reinstate them manually.

{\footnotesize%
  \drulesnohead{
    regionXXfreeze,
    regionXXmerge
  }
}\noindent
The rules for merging \RNP{region-merge} and freezing
\RNP{region-freeze} are similar. Both perform a $\Use$ to get an
object identifier $\iota$ and find its containing region $R$ among
the closed regions.
For merges, the subheap of $R$ is merged with the subheap of the
currently active region, and $\iota$ is bound to $x$ as \Mut{}.
For freezes, all the reachable regions of $R$ are moved from the
closed to the frozen regions together with $R$, and $\iota$ is
bound to $x$ as \Imm{}.

In addition to allocation in the temporary store, we have omitted
the rules for type casts and rebinding of variables.

\subsection{Static Semantics of the Command Language}
\label{sec:static-semantics}

\begin{wrapfigure}{r}{.5\textwidth}\small
\vspace*{-1em}
$\begin{array}{@{~}rcl}
     e & ::= & \Use ~|~ \text{\c{let}}~x = b~\text{\c{in}}~e\\
       & ~|~ & \text{\c{if}~\c{typetest}}(\Use, t) \SB y => e \FB \SB y => e \FB\\
  \Use & ::= & x ~|~ \text{\c{drop}}~x\\
     b & ::= & \c{*}\lval
         ~|~ \lval~\c{:=}~\Use
         ~|~ \fnc(\many{use})
         ~|~ \text{\c{var}}~\Use\\
       & ~|~ & \text{\c{new}}~k~C(\many{use})
         ~|~ \text{\c{freeze}}~\Use
         ~|~ \text{\c{merge}}~\Use \\
       & ~|~ & \text{\c{enter}}~\lval~[\many{y=\Use}] \SB z~\c{=>}~e \FB
         ~|~ e\\
 \lval & ::= & x ~|~ x.f\\
     t & ::= & k~CL ~|~\, t\,\c{|}\,t
\end{array}$
\vspace*{-1em}
\caption{Syntax of the command language}
\vspace*{-1ex}
  \label{fig:syntax}
\end{wrapfigure}

The command language is an imperative language in A-normal form.
The syntax is shown in \cref{fig:syntax}.
We encode mutable variables of type $t$ as ref cells. For
uniformity we model these as objects of
type \c{Cell}\c{[}$t$\c{]} with a single field \c{val}.
The \c{if typetest} expression is a dynamic type test similar to
Java~16 style pattern matching, \c{drop}~$x$ denotes a
destructive read, $\texttt{*}\lval$ dereferences a field or ref
cell, and \c{var} allocates a new ref cell with the capability
\Var{} and initializes its value from $\Use$. For simplicity we provide \c{enter}
blocks with an explicit capture list, but this could also be inferred from
variable use.
Types $t$ are unions $t_1\,\texttt{|}\,t_2$ or $k~CL$, where $k$ is a
capability and $CL$ is \c{Cell} or a class name $C$.

The static semantics is a flow-sensitive type system producing judgements of the form
\\ $\Gamma_1 \vdash r : t \dashv \Gamma_2$ ($r \in e \cup b \cup \Use$). Thus it
statically tracks destructive reads and strong updates of unique variables.
We discuss a few of the rules below.


{\footnotesize
  \drulesnohead{
    cmdXXtyXXuseXXkeep,
    cmdXXtyXXuseXXdrop,
    cmdXXtyXXderefXXfield
  }
}\noindent
{\footnotesize
  \drulesnohead{
    cmdXXtyXXassign,
    cmdXXtyXXassignXXvar
  }
}\noindent
Reading a variable $x$ that is not a \Var{} or \Iso{} is
straightforward and introduces an alias \RNP{cmd-ty-use-keep}.
When $x$ is \Var{} or \Iso{}, \RN{cmd-ty-use-drop} allows reading
the variable but undefines it in the environment as a side-effect
to ensure its single use.
When accessing a field $x.f$ \RNP{cmd-ty-deref-field}, its type is
subject to viewpoint adaptation $\VPA{k}{t}$ where $k$ is the
capability of $x$ and $t$ the type of $f$. Note that viewpoint
adaptation disallows reading an \Iso{} field unless $k$ is \Imm{}, expressing
the fact that freezing a region is deep (all nested regions will be frozen as
well).
A field $x.f$ can be updated through assignment \RNP{cmd-ty-assign}
when the capability of $x$ is \Mut{} or \Tmp{},
\ie{} internal references in the currently active region (note that assignment
returns the \emph{old} value of the field). Viewpoint adaptation
is not needed as we are moving values rather than copying them, allowing
swapping of \Iso{} references.
Local variables allow strong updates \RNP{cmd-ty-assign-var}. As we model them
as ref cells we update the type parameter for $x$ after assignment.

{\footnotesize
  \drulesnohead{
    cmdXXtyXXenter,
    cmdXXtyXXenterXXvar
  }
}\noindent
The predicate {$\Hcap{k, t}$} asserts that the type $t$ has capability $k$;
the predicate {$\Hopen{k}$} is true if the capability denotes an
open region, \ie{} $k$ is \Mut{}, \Tmp{}, \Var{} or \Paused{}.
Finally, {$\Hmakemut{t}$} and {$\Hmakeiso{t}$} return a $t$ whose
\Iso{} capabilities have been replaced by \Mut{} and vice versa.

Opening a region through a field $x.f$ \RNP{cmd-ty-enter} requires that
$x$'s capability is $\mathsf{open}$, and $f$'s capability is
\Iso{}.
We create a new environment $\Gamma'$ with the captured variables
$y_1,...,y_n$, using viewpoint adaptation to suspend the types of
all non-\Iso{} variables, as well as a \Tmp{} ref cell holding
the bridge object.
We use {$\Hmakemut{t}$} to change the type of the bridge
from \Iso{} to \Mut{} as control is moving \emph{inside} the
opened region. Finally, the \c{enter} block may only return
\Iso{}'s and \Imm{}'s.
(Note that entering a region through a field incurs a dynamic
check to see if the region is already open. )

Opening a region through a \Var{} ref cell \RNP{cmd-ty-enter-var}
is similar to a field \RNP{cmd-ty-enter}, but allows strong updates of
the ref cell holding the bridge object by retaining its \Var{}
capability. This allows changing the bridge object's type from
within the \c{enter} block.

{\footnotesize
  \drulesnohead{
    cmdXXtyXXmerge,
    cmdXXtyXXfreeze
  }
}\noindent
The rules for merging and freezing a region (\RN{cmd-ty-merge} and
\RN{cmd-ty-freeze}) are straightforward. Both demand that the
value that we operate on is an \Iso{} reference (\ie{} bridge object to a closed
region), and produce either a \Mut{} or \Imm{} depending on the operation.




\subsection{Dynamic Semantics of the Command Language}

A configuration $\SB de \FB$ in the command language is a dynamic
expression $de$, which is an extension of $e$ by ``entered blocks'' that propagates syntactically the nesting structure of enters and exits,
and thus dynamically tracks the nesting of open regions, and
$\mathbf{Failure}$, used to report failed dynamic checks when
entering an already open region.
The dynamic semantics steps a configuration and produces an effect
of the same kind consumed by the region language.
For example, the expression \c{let}~$x = \texttt{*}y.f~$\c{in}$~e$
produces the effect $\textit{load}(x, y.f)$, which tells the
region language to load the field $f$ from the object stored in
$y$ and store it in $x$.

\subsection{Interaction Between the Region and Command Languages}

A complete configuration is a product of the configurations of
the region and command languages. It steps if there is an effect
that steps both of them in tandem:

{\footnotesize
  \drulesnohead{
    tandemXXstep
  }
}
\noindent
A dynamic expression is typed under a stack of typing contexts
$\many{\Gamma}$, corresponding to the nesting of
entered blocks. We lift the static semantics of the
command language to define well-formed effects:
the relation $\many{\Gamma} \vdash \Eff \dashv \many{\Gamma}'$
statically describes the effect $\Eff§$ and how it
changes the typing context.
For example, the static description of the effect
$\mathit{load}(x, y.f)$ states that the type of $y$ in the
top-most entered block is $k~CL$, that $CL$ has a field
$f$ of type $t$, and that the viewpoint adapted type $\VPA{k}{t}$
is well-formed (\cf{} \RN{cmd-ty-deref-field}).
%

In order to reason about soundness, we define well-formedness of a
configuration in the region language as the relation
$\many{\Gamma} \vdash \langle \RS; \OpenRegs; \ClosedRegs;
\FrozenRegs\rangle$. A well-formed configuration ensures four
things.
First, the stack of environments $\many{\Gamma}$ mirrors the
region stack $\RS$ so that each environment describes the local
variables of a region frame in $\RS$.
Second, each field of every object in the configuration contains a
value that corresponds to its static type.
Third, we have invariants about the reference
capabilities: \Var{} references are unique, objects in frozen
regions only refer to other references in frozen regions, \Mut{}
references point within the same region and to the heap,
\Tmp{} and \Var{} point within the same region and to the
temporary store, \Paused{} references point downwards in the
region stack, etc.
Finally, we have the invariant that the object graph and its
regions have the expected topology. We describe this invariant in
detail in the following section.





\subsection{The Topology Invariant}%
\label{sec:meta-theory}

The most important properties of the object (and region) graph
are captured in a single invariant that we call the
topology invariant (\cref{fig:topology_invariant}).
We express this as a property that holds for any pair of
references $\REF{1}$ and $\REF{2}$ in a well-formed configuration.
The helper functions \src{} and \dst{} denote the storage location
and referee of a reference respectively; \reg{} denotes the region
of an object (or variable); \regions{} projects the region
identifiers out of a set of regions $H$.

For all references $\REF{1}$ and $\REF{2}$, either:
they are the same reference, \eg{} both are stored in the same $\iota.f$ or variable $x$ (1);
both refer to objects in different regions (2); or at least one of them
is an intra-region reference (3);
refers to a permanently immutable object (4); or
is a reference outwards in the nesting hierarchy, downwards in the region stack (5).
The relation $\RS\vdash \dst{\REF{}}\preceq\src{\REF{}}$
holds if $\dst{\REF{}}$ is higher up in $\RS$ than $\src{\REF{}}$
and $\REF{}$ originates from the temporary store. In other words, we
allow temporary references into suspended regions from open
regions.

\begin{wrapfigure}{r}{.48\textwidth}\small
\vspace*{-1em}
$\begin{array}{l}
   \forall \REF{1}, \REF{2} \in \references{\langle \RS; \OpenRegs; \ClosedRegs; \FrozenRegs\rangle}.\\[1ex]
   \quad\bigvee\left\{\begin{array}{l@{\quad}l}
                        \REF{1} = \REF{2} & (1) \\[1ex]
                        \reg{\dst{\REF{1}}} \neq \reg{\dst{\REF{2}}} & (2) \\[1ex]
                        \reg{\src{\REF{1}}} = \reg{\dst{\REF{1}}}\ \lor \\ \qquad \reg{\src{\REF{2}}} = \reg{\dst{\REF{2}}}  & (3) \\[1ex]
                        \reg{\dst{\REF{1}}} \in \regions{\FrozenRegs}\ \lor \\ \qquad \reg{\dst{\REF{2}}} \in \regions{\FrozenRegs} & (4) \\[1ex]
                        \RS\vdash\dst{\REF{1}} \preceq \src{\REF{1}}\ \lor \\ \quad \RS\vdash\dst{\REF{2}} \preceq\src{\REF{2}} & (5) \\
                      \end{array}\right.
 \end{array}$
\vspace*{-0.8em}
\caption{The topology invariant}
\label{fig:topology_invariant}
\vspace*{-1.7em}
\end{wrapfigure}

The topology invariant has several important implications:
The \emph{object graph} inside a region is unconstrained (3). The object graph
of the permanently immutable objects is unconstrained (4). Temporary objects in
an open region $R$ are allowed to refer to objects in an open region $R'$ as
long as $R'$ was opened before $R$ (5).
Finally, considering the whole invariant, if we have two external references
((3) does not hold) pointing into the same non-frozen region ((2) and (4)
do not hold), and neither of them points downwards in the region stack ((5)
does not hold), then they must be the same reference ((1) holds).
In particular, this means that there is at most one external reference into any
closed region, implying that the region graph of closed regions forms a forest.


\subsubsection*{The Topology Invariant and \cref{fig:isolation}}

Applying the topology invariant to all pairs of references in
\cref{fig:isolation}, assuming $R$ was opened after $R'$, the
reference \Ptr{e}{n} is allowed to co-exist with any other alias
of $n$ since $\RS \vdash n \preceq e$ (5).
The reference \Ptr{m}{e} is \emph{not} allowed to co-exist with
\Ptr{m}{a} since there would be two external references into the
same region (1--5). However, \Ptr{e}{a} can co-exist with \Ptr{m}{a}
since the former stays within its region (3).
The references \Ptr{a}{i} and \Ptr{o}{i} are allowed to co-exist
because i is in a frozen region (4).
Finally, \Ptr{o}{a} is illegal both because it cannot co-exist
with \Ptr{m}{a}, and since references in frozen regions cannot
point to non-frozen regions.
%

\subsection{\Venice{} is Sound}

%
We prove soundness of our system by proving variants of progress
and preservation for the respective language. (The full proofs
are available in the appendix.)
\begin{lemma}
  \textbf{Command Language Progress} A well-formed command
  configuration is done, has failed or can step:
  $\many{\Gamma} \vdash \SB de \FB \implies de = \Use \lor de =
  \mathbf{Failed} \lor \exists \Eff, de'.~ \SB de \FB
  \yrightarrow{\Eff}[-1pt] \SB de' \FB$.
\end{lemma}

\begin{lemma}
  \textbf{Command Language Preservation} The command language
  preserves well-formedness and produces well-formed effects:
  $\many{\Gamma} \vdash \SB de \FB ~\land~ \SB de \FB \yrightarrow{\Eff}[-1pt] \SB de' \FB \implies
  \exists \many{\Gamma}'.~ \many{\Gamma}'\vdash \SB de' \FB ~\land~
  \many{\Gamma} \vdash \Eff \dashv \many{\Gamma}'$.
\end{lemma}

The command language is more permissive than the region
language, since it has no way of inspecting the state of the
global configuration. For example, an enter can \emph{always} both
fail and succeed in the command language, whereas the region
language always permits exactly one of the behaviours. This
affects the formulation of progress:

\begin{lemma}
  \textbf{Region Language Progress} In a well-formed
  configuration where the command configuration can step, there is
  some effect which steps both configurations:
  $\vdash \langle\SB de \FB~\RCfg\rangle \land \SB de \FB \yrightarrow{\Eff}[-1pt] \SB de' \FB \implies
  \exists \Eff', de'', \RCfg'. \SB de \FB \yrightarrow{\Eff'}[-1pt] \SB de'' \FB ~\land~ \RCfg \yrightarrow{\Eff'}[-1pt] \RCfg'$.
\end{lemma}

\begin{lemma}\label{lem:region_preservation}
  \textbf{Region Language Preservation} The region language
  preserves well-formedness for well-formed effects:
  $\many{\Gamma} \vdash \RCfg ~\land~ \RCfg \yrightarrow{\Eff}[-1pt] \RCfg'
  ~\land~ \many{\Gamma} \vdash \Eff \dashv \many{\Gamma}' \implies \many{\Gamma}'
  \vdash \RCfg'$.
\end{lemma}

Note that Lemma~\ref{lem:region_preservation} includes
preservation of the topology invariant.
Together, these lemmas prove the final soundness theorem:

\begin{theorem}
  \textbf{Soundness} A program never gets stuck and
  it preserves well-formedness: \\
  $\vdash \langle\SB de \FB~\RCfg\rangle \implies de = \Use \lor
  de = \mathbf{Failed} \lor \exists de', \RCfg'. \langle\SB de \FB~\RCfg\rangle
  \rightarrow \langle\SB de' \FB~\RCfg'\rangle \land \\ \vdash
  \langle\SB de' \FB~\RCfg'\rangle$.
\end{theorem}

\section{\Venice{} in \Verona{}}

While \Venice{} regions are a stand-alone language design
component, they were developed specifically for the \Verona{}
programming language, from where the overarching
goal~\ref{goal:local} stems. In this section, we describe
\Verona{}-specific aspects and revisit concurrency-related goals
\ref{goal:concurrent} and \ref{goal:drf}.

\subsection{Safe Concurrency}
\label{sec:concurrency}

While regions and isolation can form the backbone of a ``safe
concurrency'' story for a language, concurrency is an orthogonal
aspect to our region design. \Venice{} regions can be integrated
with different concurrency models. The necessary feature missing
from this paper is a way to share regions across threads of
control.

\Verona{} uses a concurrency model based on \emph{behaviours}
(tasks that do not join or have a return value) that operate on
\emph{cowns}, short for concurrent owners. A cown is a wrapper
around an \Iso{} that permits regions to be indirectly shared
across multiple threads of control, but importantly does not
permit direct access to its contents. 
Cowns and \Iso{}'s are similar in that an explicit operation is
needed to access their contents. In the case of \Iso{}'s, access
is immediate and synchronous as exclusivity is already
established. In the case of cowns, access is asynchronous and will
only commence after exclusive access has been established
dynamically. This check requires region isolation for
soundness~\cite{BoC} and as a result, any mutable reference
accessible to a thread of control is safe to access
synchronously~\ref{goal:drf}.

For a complete introduction to \Verona{}'s concurrency model, see
work by~\citet{BoC}.

\subsection{Concurrent Memory Management}

As memory management must only consider objects inside the active
region~\ref{goal:incremental} when determining liveness (\cf{}
\cref{sec:gc-example}), and regions are always exclusive to one
thread, reference count manipulations do not need atomic
instructions, tracing GC does not need barriers, and there is no
need to momentarily stop all threads as in concurrent
GC's~\cite{click_pauseless_2005, flood_shenandoah_2016,
  liden_z_2018}.

By extension, a thread in \Verona{} is free to mediate between
program work or memory management work without informing or
synchronising with other threads. Thus, we achieve concurrent
memory management~\ref{goal:concurrent}.


\subsection{Support for Different Memory Management Strategies}%
\label{sec:supp-diff-memory}

\Verona{} currently supports three different memory management
strategies for regions:
\emph{arena allocation},
\emph{reference counting}, and
\emph{tracing GC}\@.

How a region manages its memory is decided at use-site at
creation time using a qualifier on the \c{new} keyword: \c{new
  iso<Arena>}, \c{new iso<RC>} and \c{new iso<GC>}. As liveness is
a local property, different regions' memory management does not
interact, so we do not need to \eg{} propagate this information
further in the program.

Selecting memory management at use-site is desirable since it lets
a programmer implement a data structure or library without having
to commit to decisions that could limit its future use. Such a
design also allows straightforward support for libraries that
consist of multiple nested regions whose memory management can be
controlled when the library is instantiated by programmatic means,
\eg{} through a strategy pattern or equivalent.

\subsection{Memory Management of Immutable Objects}

Note that the memory management offered by regions does not extend
to immutable objects, at least not conceptually. One option for
implementation is going the way of Erlang and let a region have
a copy of each immutable object it references. This may facilitate
fast reclamation, but increases memory pressure. Furthermore, it
introduces a $O(n)$ copying overhead for transferring immutable
objects across region boundaries, in sharp contrast
with~\ref{goal:ownership-transfer}.

In the \Verona{} run-time, we permit immutable objects to be
shared between regions. Thus, when a region is collected, we must
detect its implications for liveness of immutable objects.
Immutable objects must also consider roots across multiple
threads. On the other hand, tracing immutable objects is easy and
efficient as the structures are guaranteed not to change
underfoot~\cite{clebsch_orca_2017}. How \Verona{} manages
immutable objects is out of scope of this paper.

\subsection{Propagating Capabilities Through Self Typing}

\Verona{} is a class-based programming language. As an instance's
capability is determined at use-site, methods declare an
explicit self-capability, \eg{} \texttt{self\,:}\,\Mut{} to
propagate the external view of the instance into the instance. A
class may provide several different implementations of a method
overloaded on the self-capability.

\begin{wrapfigure}{r}{0.42\textwidth}
\vspace*{-3ex}
\begin{code}[numbers=right,firstnumber=1,caption={Self typing in \c{Cell}.},label=lst:self-typing]
class Cell {
  var value : I64 // (*@\BLUE{mut Store[mut I64]}@*)
  def set_value(self:mut, v:I64) {
    // value : (*@\BLUE{($\VPA{\Mut}{\Mut}$)\,Store[mut I64]}@*)
    value := v
  }
  def set_value(self:paused, v:I64) {
    // value : (*@\BLUE{($\VPA{\Paused}{\Mut}$)\,Store[mut I64]}@*)
    value := v // (*@\RED{\it does not typecheck!}@*)
  }
  def get_value(self:mut) = value
  def get_value(self:paused) = value
}
\end{code}
\vspace*{-2em}
\end{wrapfigure}

A method can only be called on a receiver if its self-capability
matches the receiver's static type, which means that the object's
treatment of itself internally will match the external view, both
in terms of restrictions and abilities. For example, a method
whose self-capability is \Paused{} can only be called when the
receiver's region is \Paused{}. 
Notably, it is
\emph{not} permitted to call a \Paused{} method on a \Mut{} receiver because
this would lead to aliasing between \Mut{} and \Paused{} (which
would weaken \Paused{} from temporarily immutable to read-only).

Methods that are polymorphic in their self capability can be used
to avoid multiple near-identical versions of a single method like
in~\cref{lst:self-typing}. For brevity, we refrain from discussing
this further.

\section{Related Work}
\label{sec:related}

We started out by describing related work leading up to Rust. We
now extend this picture by going beyond Rust and also relating
our work to garbage collection work before revisiting novelty.

\subsubsection*{Beyond Rust}

In addition to what we have already covered, there is continuing
research into Rust to alleviate its restrictions, including
incorporating a garbage collector~\cite{rustGC}, careful library
design~\cite{RustMSc}, phantom types~\cite{GhostCell}, or proving
unsafe Rust code correct~\cite{RustBelt18, StackedBorrows,
  rusty-ftfjp2022}.

Recent research has focused on techniques for ``post-Rust
languages'', building on Rust's use of ownership types, but supporting
more flexible program topologies (and hopefully more efficient
execution), typically by increasing the complexity of the type
system. This remains an active research area: the tradeoffs between
regions, ownership, types, capabilities, effects, topologies,
restrictions etc are complex and multifaceted
\cite{gordon-ecoop2020,systemc-oopsla2022}.

Pony~\cite{franco_correctness_2018, clebsch_orca_2017} employs
  implicit regions, external uniqueness and ownership to offer high
  performance for actor programs by concurrent execution on multicore
  CPUs, while maintaining data-race and memory safety.  Building on
  capabilities used to describe what programs can do with particular
  references \cite{boyland_capabilities_2001} Pony offers at least six
  ``reference capabilities'': unique, thread-local, read-only,
  write-only, and identity-only, (globally) immutable, plus type
  modifiers for ephemeral (Hogg's ``free'') and aliased references.
  Reading or writing a field depends on the capabilities of both the
  reference to the object, and of the field within the object: there
  are 43 valid cases from 72 possible combinations of capabilities.

\citet{dala_onward2021} design Dala
as a simplified alternative to Pony, based on three
different kinds of objects---immutable, unique (aka isolated), and
thread-local---rather than six different kinds of references. Dala
programs are also data-race free, however this guarantee may be
provided by a race detector at runtime, or by an optional/gradual
type system.

Milano et al.'s \citeyearpar{gallifrey-pldi2022} Gallifrey aims to be more flexible than
Rust, by relying at least as much on MLKit style region inference as
on ownership annotations.  Rather than an explicit global ownership
model, Gallifrey programmers have to annotate unique (aka isolated)
object fields, and identify parameters that will be consumed by a
method invocation or that should be in the same region as other
parameters or the method result. A dynamic ``if disconnected''
predicate searches the program's heap at runtime to determine of two
references are mutually disjoint.

Cogent~\cite{cogent2021} is a derivative of Haskell for systems
programming. It adopts a Rust-like discipline, permitting either
multiple read-only references to objects, or a single read-write
reference. Cogent uses annotations to support both a formally
defined operational semantics, generation of executable C source
code, and a proof certificate proving that the generated code
accurately implements the semantics.

\subsubsection*{Garbage Collection}

Of the GC design goals, \ref{goal:incremental} and
\ref{goal:concurrent} can be met to some extent \emph{without}
region isolation. Here, \Venice{}'s contribution is trading
additional work to manage regions (at development time) for
reduced overheads of managing memory (at run-time) due to avoiding
GC techniques like remembered sets, barrier synchronisation, and
stop-the-world pauses. Region isolation guarantees that a
region's remembered set will always be empty. Thus, there is no
cost associated with additional region partitioning due to
tracking of inter-region references.


With respect to \ref{goal:incremental}, generational GC's (\eg{}
G1~\cite{G1}) and thread-local GC's (\eg{}~\cite{Domani02})
support collecting only a portion of the heap (\eg{} just the
young generation or one particular thread), but the shape and size
of this heap is beyond programmer control (\eg{} all young or
thread-local objects will take part of GC, not just particular
data structures). Furthermore, in the absence of (something like)
region isolation, inter-region aliases must be tracked dynamically
to be able to correctly compute liveness. Actor GC's that rely on
actor isolation using types (\eg{}
Pony~\cite{franco_correctness_2018, clebsch_orca_2017}) or copying
(\eg{} Erlang~\cite{armstrong_history_2007}) are close as they
allow individual actor-local heaps to be collected. This is
similar to \Venice{}'s regions, but \Venice{} supports additional
partitioning of the heap without an imposed asynchronous
indirection.

With respect to \ref{goal:mix-match}, while it may be
possible to run different GC's in different generations (or
threads, actors, etc.), GC's typically use the same algorithm for
the entire heap, with minor tweaks (\eg{} to account for different
object characteristics due to age) as do actor GC's.
HRTGC is a real-time GC for mixed-criticality
work-loads~\cite{HRTGC} that hierarchically decomposes the heap
into regions that each run a different tracing GC, tuned
differently and with different collection frequency. HRTGC permits
inter-region references and tracks them dynamically. In the actor
world, Isolde~\cite{yang_type-assisted_2017} permits actors
implemented using type-enforced actor
isolation~\cite{castegren_relaxed_2017, EncoreTS}
to manage their memory concurrent with their execution, using a
reference-counting based scheme.

With respect to \ref{goal:concurrent}, garbage collectors like
C4~\cite{tene_c4_2011},
She\-nan\-doah~\cite{flood_shenandoah_2016} and
ZGC~\cite{liden_z_2018} provide concurrent collection with brief
stop-the-world pauses to coordinate phase changes, with pause times invariant
of heap sizes. Their heaps have unrestricted references, and
instead rely on dynamic checks in read and write barriers. The
aforementioned actor GC's~\cite{franco_correctness_2018,
  clebsch_orca_2017, armstrong_history_2007} support ``fully
concurrent'' collection: an actor can choose to collect its local
garbage without synchronising with any other concurrent activity.
\Venice{}'s regions additionally allow us to statically detect
when an entire region is invalidated, without the need for a
specific actor collector to eventually detect the floating garbage
(\eg{} \cite{clebsch_fully_2013}). Explicitly killing an actor to
instantly free its heap is a common programming pattern in Erlang
where it is made possible by copying objects on transfer, \ie{}
giving up \ref{goal:ownership-transfer}.

\section{Discussion}

First, we place \Verona's design concepts
%
%
into the context of all related work. Almost any ownership
system paired with external uniqueness will support region isolation and
dynamic reconfiguration. \Verona's key
contribution here is a negative contribution, but important
nonetheless. Almost all other systems provide one or more top,
global, or shared heap region, and in various ways permit
references from inner/encapsulated/shorter-lived regions back to
outer/enclosing/longer-lived regions. (Generational garbage
collection works on a very similar principle
\cite{jones_garbage_2016}). \Verona{} does not, and this enhanced
decoupling of regions is critical to achieving many of our goals,
especially about concurrency, and independent GC.

\Verona's dynamic mutability and relaxed isolation is
novel and differs from other ownership and region systems.
Inasmuch as region systems like the MLKit are based on inference, and
are sound for all legal programs, questions of mutability and
isolation don't apply---if the program is type-correct, the
inference system can always place objects into regions such that no
region errors will arise at runtime.  Programming with more explicit
regions, or with ownership and capability annotations, either lack
polymorphism (\eg{} Dala) or require complex resolution or viewpoint
adaptation rules, or asynchronous indirections (\eg{} Pony).
In a way, \Verona{}'s region system trades precision for
simplicity. It cannot construct data structures that consist of
several morally overlapping regions, as is possible in \eg{} C++
or Rust. We believe all programs written in \eg{} C++ or Rust pay
the price for that precision---even though most programs do not
need it.

\Verona's ``single window of mutability'' is
probably its most novel concept. In pretty much every language, from
FORTRAN to LISP to ML to Haskell to C$++$ to Pony to Gallifrey, if
some code can finagle a mutable reference to an object, the program
can always update the object through that reference.  In Rust for
example, programs can collect up ``mutable borrows'' (\verb+&mut+) of
any number of objects, pass them around as method arguments to
anywhere in the program, and then mutate all the borrowed
objects. Rust's ``interior mutability'' (aka C$++$'s const-cast) just
increases the scope for potential mutation.  This kind of
indiscriminate mutation is exactly what the ``single window of
mutability'' in \Verona{} prevents.  Once a program departs---even
temporarily---from the scope of an opened region, (\eg{} by opening
some \textit{other} closed region) the program can no longer modify
anything in that first opened region, no matter what kind of objects
are in the region, nor what kind of capability or reference the
program has to those objects.  Perhaps a single window of mutability
will prove too restrictive in practice, which may be why no other
system has yet adopted it.  \Verona{} demonstrates that it is possible
to build a system with a single mutability window as a core design
concept; exploring that concept further must necessarily be further work.

The single window of mutability is key to simplicity, both with
respect to the type system that enforces region isolation and our
invariants for memory management. We need only distinguish objects
in the active region, from objects in suspended regions and
objects in closed regions. As suspended regions are immutable and
closed regions inaccessible, we do not need to distinguish objects
belonging to different regions as nothing can be done to them that
affects or is affected by their region membership. By having only
a single mutable region at a time, non-local operations cannot
effect object liveness in the active region, or the liveness of an
entire active region. This permits optimisations at the
implementation level and simplifies the task of the programmer
wishing to reason about---and control---memory management
performance.

In contrast to other works which use types to enforce (or impose)
a structure on the heap, we let the path of the program through
the heap dictate the permissible pointer structure---not the other
way around. For example, \Verona{} allows a single point in the
program to access the contents of two mutually isolated regions
$A$ and $B$, simply by virtue of opening them in a nested fashion.
The key insight is the decoupling of accessing from mutating,
implemented through the movable window of mutability. Rather than
alternating between accessing $A$ and $B$, we can gain access to
first $A$ and then $B$ without giving up access to $A$, just the
rights to mutate it. Opening $B$ after $A$ allows references to
objects in $A$ from $B$ to be created freely, but these references
may only persist as long as $B$ remains open. When $B$ is closed,
the references to $A$ are invalidated. This retains the
flexibility of navigating heap structures in any order, \eg{} we
could close $B$ then $A$ and open them immediately in the opposite
order, allowing pointers from $A$ to $B$. It also ensures that an
object is either mutable or immutable at any moment in time.

\section{Conclusion}

We have presented \Venice{}, a region system enforced by reference capabilities
that partitions a program's heap into a forest
of isolated regions. Memory in different regions
can be managed differently~\ref{goal:mix-match},
incrementally~\ref{goal:incremental} and
concurrently~\ref{goal:concurrent}. The single external reference
to each region plus their full encapsulation enable cheap
ownership transfer~\ref{goal:ownership-transfer} and guarantees
freedom from data races~\ref{goal:drf}. The ability to temporarily
trade mutability for access on the region stack allows any region
to (temporarily) reference any other region, and
also allows ``external code'' to operate inside a region, which is
crucial for libraries and reuse---not all uses of a region can be
predicted and supplied in its interface. In combination with the
region isolation and the single window of
mutability this allows the formulation of
topological invariants which are useful for a programmer to
control and reason about object liveness and implications of
memory management~\ref{goal:local} and can be leveraged for
efficient implementation of memory management. Memory management
costs are only incurred by the active region (one per thread), and
data accesses within that region---whether reading, writing,
tracing, or reference counting---never need atomic operations to
coordinate with other threads.

\begin{acks}                            
  This work was partially supported by a grant from the
  \grantsponsor{VR}{Swedish Research Council}{https://www.vr.se/}
  (\grantnum{VR}{2020-05346}), and partially by the Royal Society of
  New Zealand Te Ap\={a}rangi  Marsden Fund Te P\={u}tea Rangahau a
  Marsden grants CRP1801 and CRP2101, and Agoric.
  We thank the anonymous reviewers at OOPSLA'23 for their input that
  greatly improved the presentation of this paper.
\end{acks}

\bibliography{main}

\appendix

\section{Introduction to Appendix}
This appendix contains the static and dynamic semantic of
\Venice{}, and the full proof of soundness. The dynamic semantics
is defined as two systems executing in tandem: the region
semantics and the command semantics.
The region semantics is the most important dynamic semantics in
terms of understanding the system. It is responsible for entering
and exiting regions, handling allocations and variable bindings.
(We do not model deallocation.)
The command semantics drives execution by emitting effects which
are then performed by the region semantics.

Any differences between the main paper and the appendix is purely
for presentational reasons.

\section{Tandem Semantics}

\[
  \begin{array}{rcll}
    \Cfg & ::= & \langle \SB de \FB~\RCfg\rangle & \text{Tandem configuration}
  \end{array}
\]

A \Verona{} configuration is a tuple of a command configuration
and a region configuration. It steps if the two configurations can
both step with the same effect, and is well-formed if both
configurations are well-formed with the same $\Gammas$.

\begin{DisplayRule}
  \drules{$\Cfg \rightarrow \Cfg'$}{Tandem Semantics}{
    tandemXXstep
}
\end{DisplayRule}

\begin{DisplayRule}
  \drules{$\vdash \Cfg$}{Tandem Semantics}{
    tandemXXwf
  }
\end{DisplayRule}

\section{Region Semantics}

\[
  \begin{array}{ll}
    r \in \Reg                                       & \text{Region ids}                                     \\
    \iota \in \Oid                                   & \text{Object ids}                                     \\
    x,y,z,w,f,g \in \X                               & \text{Field and variable names}\\
    C \in \Cls                                       & \text{Class names}                                    \\
  \end{array}
\]

\[
  \begin{array}{rcll}
    \CL &  ::= & C ~|~ \Cell[t] & \text{Static types} \\
    \CLTag & ::= & \#C ~|~ \#\Cell    & \text{Dynamic type tags} \\
    k & ::= & \Iso ~|~ \Var ~|~ \Mut ~|~ \Tmp ~|~ \Paused ~|~ \Imm & \text{Capabilities}                        \\
    t & ::= & k~\CL ~|~ t | t & \text{Types} \\
    v & ::= & (k, \iota)                                & \text{Values (references tagged with capabilities)}   \\
    o & ::= & (\CLTag, F)                                  & \text{Objects (class tag and field)}                 \\
    F & ::= & x \mapsto v, F ~|~ x \mapsto \Undef, F ~|~ \varepsilon & \text{Field and variable bindings}                    \\
    R & ::= & (r, S)                                    & \text{Region (region id and sub-heap)}                \\
    S & ::= & \iota \mapsto o, S ~|~ \varepsilon        & \text{Store}                                          \\
    H & ::= & H * H ~|~ R ~|~ \varepsilon               & \text{Heap (collection of Regions)}                   \\
    \RF & ::= & (r, S, F)                               & \text{Region frame}                                   \\
    \RS & ::= & \RF \mathop{::} \RS
              ~|~ \varepsilon                           & \text{Region stack}                                  \\
    \RCfg & ::= & \langle \RS; H; H; H \rangle & \text{Configuration}

  \end{array}
\]

The heap $H$ is a collection of disjoint regions $R$, each
containing a region identifiers $r$ and a store $S$ mapping object
identifiers $\iota$ to objects $o$. An object consists of the
object's class tag (without the type parameter of \Cell) and a
mapping $F$ from field names $f$ to values $v$ (consisting of a
capability $k$ and an object identifier $\iota$).
The region configuration \RCfg{} is a region stack \RS{} and three
heaps $H$, for the open, closed and frozen regions respectively.
The region stack is a stack of frames \RF{} holding a region
identifier, a (local) store $S$ and a mapping $F$ from variables
to values. Note that for simplicity we reuse $F$ for storing field
mappings and variable mappings, meaning fields and variables are
in the same syntactic category $\X$.

We assume the existence of a table of classes and their fields
such that $\mathbf{ftypes}(CL) = f_1 : t_1, ..., f_n : t_n$ and
$\mathbf{ftype}(CL, f_i) = t_i$, meaning that the class $CL$ has a
field $f_i$ with type $t_i$. We also assume that no field has
capability $\Var{}$. We use $\mathbf{fields}(CL)$ to get just the
names of the fields.
We further assume that we can always create \emph{fresh}
identifiers that do not appear elsewhere in a configuration.

\subsection{Dynamic Semantics}

In order to avoid duplicating rules depending on whether reads are
destructive or not, we define the following helper function:
\[
  \begin{array}{lrcll}
    \Get(F[x \mapsto v],             & \mathbf{drop}\ x) & = & (v, F[x \mapsto \mathbf{undef}]) \\
    \Get(F[x \mapsto (k, \iota)],    & x)                & = & ((k, \iota), F[x \mapsto (k, \iota)]) & \text{ if } k \not\eq \Var \land k \not\eq \Iso \\
    \Get(F,                          & use)              & = & \FUndef & \text{ otherwise } \\
  \end{array}
\]

where $\FUndef$ denotes the \emph{undefined} value.

\begin{DisplayRule}\small
  \drules{\RCfg $\xrightarrow{\Eff}$ \RCfg'}{Region semantics}{
    regionXXload,
    regionXXswapXXtemp,
    regionXXswapXXheap,
    regionXXallocXXheapXXmut,
    regionXXallocXXheapXXiso,
    regionXXallocXXtemp,
    regionXXfreeze,
    regionXXmerge,
    regionXXcast,
    regionXXnocast,
    regionXXenterXXok,
    regionXXenterXXfail,
    regionXXexitXXtemp,
    regionXXexitXXheap,
    regionXXbind,
    regionXXeps
}
\end{DisplayRule}

\subsubsection{Helper Functions and Predicates}

We use the following helper definitions:



Viewpoint adaptation of a value amounts to viewpoint adaptation on
its capability (defined in Section~\ref{sec:cmd:static}).

\[
  \VPA{k}{(k'~\iota)} = (\VPA{k}{k'})~\iota
\]

We can look up objects in a store $S$ given their identifier.

\[
  \begin{array}{rcl}
    \mathsf{load}((\iota \mapsto o, S), \iota') & = &
    \begin{cases}
      o & \text{if } \iota = \iota'\\
      \mathsf{load}(S, \iota') & \text{otherwise}
    \end{cases}\\
    \mathsf{load}(\varepsilon, \iota') & = & \FUndef
  \end{array}
\]

Looking up an object on the region stack goes through the local
stores of each frame from the top down.

\[
  \begin{array}{rcl}
    \mathsf{stack\_load}((r, S, F) :: RS, \iota) & = &
    \begin{cases}
      o & \text{if } \mathsf{load}(S, \iota) = o\\
      \mathsf{stack\_load}(S, \iota) & \text{otherwise}
    \end{cases}\\
    \mathsf{stack\_load}(\varepsilon, \iota) & = & \FUndef
  \end{array}
\]

Looking up an object on the heap goes through the regions left to
right.

\[
  \begin{array}{rcl}
    \mathsf{heap\_load}(H_1 * H_2, \iota) & = &
    \begin{cases}
      o & \text{if } \mathsf{heap\_load}(H_1, \iota) = o\\
      \mathsf{heap\_load}(H_2, \iota) & \text{otherwise}
    \end{cases}\\
    \mathsf{heap\_load}((r, S), \iota) & = & \mathsf{load}(S, \iota)
  p\end{array}
\]

The top-level lookup function $\mathsf{cfg\_load}$ performs a
lookup on a given stack and heap using the helper functions above.
Note that an object identifier $\iota$ can only appear in the
domain of a single store $S$ in a well-formed configuration.

\[
  \mathsf{cfg\_load}(RS, H, \iota) =
  \begin{cases}
    o & \text{if } \mathsf{stack\_load}(RS, \iota) = o\\
    \mathsf{heap\_load}(H, \iota) & \text{otherwise}
  \end{cases}
\]

Similarly, we can update a store $S$ with a new object for a given
object identifier.

\[
  \begin{array}{rcl}
    \mathsf{store}((\iota \mapsto o, S), \iota', o') & = &
    \begin{cases}
      \iota \mapsto o', S & \text{if } \iota = \iota'\\
      \iota \mapsto o, \mathsf{store}(S, \iota', o') & \text{otherwise}
    \end{cases}\\
    \mathsf{store}(\varepsilon, \iota', o') & = & \FUndef
  \end{array}
\]

When storing an object on the stack, we always store in the
top-most frame, so we don't need a function for traversing the
stack. When storing to the heap, we go from left to right.

\[
  \begin{array}{rcl}
    \mathsf{heap\_store}(H_1 * H_2, \iota, o) & = &
    \begin{cases}
      H_1' * H_2 & \text{if } \mathsf{heap\_store}(H_1, \iota, o) = H_1'\\
      H_1 * \mathsf{heap\_store}(H_2, \iota, o) & \text{otherwise}
    \end{cases}\\
    \mathsf{heap\_store}((r, S), \iota, o) & = &
    \begin{cases}
      (r, S') & \text{if } \mathsf{store}(S, \iota, o) = S'\\
      \FUndef    & \text{otherwise}
    \end{cases}
  \end{array}
\]

When freezing a region, we need to calculate the reachable regions
from the region being frozen. We define a relation
$\mathsf{reachable\_step}$ which holds when a region $R$ has an
object with a reference to an object in another region $R'$.
Calculating the reachable regions amounts to calculating the
transitive closure of $\mathsf{reachable\_step}$.

\[
  \begin{array}{l}
    \mathsf{reachable\_regions}(R, H) =
    \{R' ~|~ \mathsf{reachable\_step}^+(R, R')\}\\
    \quad\textit{where}\\
    \quad\mathsf{reachable\_step}(R, R') \equiv
         R' \in H
         \land o[f\mapsto \iota] \in \mathbf{rng}(R.S)
         \land \iota \in \mathbf{dom}(R'.S)

  \end{array}
\]

\subsection{Well-Formedness Rules}

\begin{DisplayRule}
  \drules{$\Gammas \vdash \RCfg$}{Region configuration}{
    wfXXrcfg
  }
\end{DisplayRule}

\begin{DisplayRule}
  \drules{$\Gammas;\Delta;\Psi \vdash \RCfg$}{Region configuration types}{
    wfXXtyXXrcfg
  }
\end{DisplayRule}

\begin{DisplayRule}
  \drules{$\Gammas;\Delta;\Delta;\Psi \vdash \RS$}{Region stack types}{
    wfXXrsXXcons,
    wfXXrsXXnil
  }
\end{DisplayRule}

\begin{DisplayRule}
  \drules{$\Delta;\Delta;\Psi \vdash H$}{Region heap types}{
    wfXXheapXXprod,
    wfXXheapXXsing
  }
\end{DisplayRule}

\begin{DisplayRule}
  \drules{$\Delta;\Delta \vdash S$}{Region subheap types}{
    wfXXsubheapXXcons,
    wfXXsubheapXXnil
  }
\end{DisplayRule}

\begin{DisplayRule}
  \drules{$\Delta;\Gamma \vdash F$}{Region variable types}{
    wfXXvarsXXcons,
    wfXXvarsXXnil
  }
\end{DisplayRule}

\begin{DisplayRule}
  \drules{$\CLTag \subtag \CL$}{Subtagging}{
    subtagXXclass,
    subtagXXcell
  }
\end{DisplayRule}

\subsection{Predicates on Graphs}

We will now define some predicates used to define the invariants of our system.
These are defined in terms of the object graph of the configuration. Therefore
we first define the set of graphs $\GraphSet$, and a mapping $G: \RCfgSet \to
\GraphSet$, where $\RCfgSet$ is the set of region configurations.

\subsubsection{Region Configuration Graph}

The the elements of $\GraphSet$ are tuples $(\LocSet, \RefSet)$ of vertices $\LocSet$
(locations) and edges $\RefSet$ (refereces). We represent the vertices
and edges as

\[
  \begin{array}{rcll}
    \Loc  & ::= & \Heap(r, \iota) ~|~ \Temp(r, \iota) ~|~ \Root(r) & \text{Location} \\
    \REF & ::= & \Loc \RefTo{f, k} \Loc' & \text{Reference, marked with field and
                                               capability}
  \end{array}
\]

We call the set of all locations and references $\AllLocs$ and
$\AllRefs$ respectively.

\[
  \begin{aligned}
    &\Rid(\Heap(r, \iota)) &=\quad r \\
    &\Rid(\Temp(r, \iota)) &=\quad r \\
    &\Rid(\Root(r)) &=\quad r
  \end{aligned}
\]

Since all $\Loc$ contains a region id $r$, we can define a shorthand $\Loc[\cdot]$ with a hole for $r$ as

\[
  \Loc[\cdot] ::= \Heap(\cdot, \iota) ~|~ \Temp(\cdot, \iota) ~|~ \Root(\cdot)
\]

We define $\cupdisj$ as the union of two disjoint sets.
\[
  A \cupdisj B = \begin{cases}
                   A \cup B  & \text{ if } A \cap B = \emptyset \\
                   \FUndef      & \text{ otherwise}
                 \end{cases}
\]
where $\FUndef$ denotes the \emph{undefined} value. In particular,
if any of $A$ or $B$ is $\FUndef$ we let $A \cap B = \FUndef$, and thus $A
\cupdisj B = \FUndef$. For sets $S, S'$, and $A, B \subseteq S$, $f: S \to S'$,
we define $\cupdisj_f$ by
\[
  A \cupdisj_f B = \begin{cases}
                     A \cup B  & \text{ if } I_f(A) \cap I_f(B) = \emptyset \\
                     \FUndef      & \text{ otherwise}
                   \end{cases}
\]
where $I_f(A)$ is the image of $f$ restricted to the set $A$. Furthermore
$I_f(\FUndef) = \FUndef$. We call $f$ above a \emph{separating function}.


For the definition of $G$ we define the separating function $\SepLoc$ as
\[
  \begin{array}{lcl}
    \SepLoc(\Heap(r, \iota)) & = & \ObjSep(\iota) \\
    \SepLoc(\Temp(r, \iota)) & = & \ObjSep(\iota) \\
    \SepLoc(\Root(r)) & = & \Root(r)
  \end{array}
\]
Assuming $\LocSet^* = \LocSet \cupdisj_\SepLoc \LocSet' \neq \FUndef$ we see that
all object ids in $\LocSet^*$ are unique, and furthermore that there is at most
one $\Root(r)$ for each region id $r$.

We are now ready to start defining the graph of a region configuration $\RCfg$.
We write
\[
  G(\RCfglong{RS; \Hop; \Hcl; \Hfr}) = (\LocSet, \RefSet)
\]
where
\[
  \LocSet = \LocRS(RS) \cupdisj_{\SepLoc} \LocH(\Hop) \cupdisj_{\SepLoc} \LocH(\Hcl)
  \cupdisj_{\SepLoc} \LocH(\Hfr)
\]

\[
  \begin{array}{lcl}
    \LocRS((r, S, F)::RS) & = & \LocS^\Temp(S) \cupdisj_\SepLoc \{ \Root(r) \}
                                \cupdisj_\SepLoc \LocRS(RS) \\
    \LocRS(\NilRS) & = & \emptyset \vspace{1em} \\
    \LocH(H_1 * H_2) & = & \LocH(H_1) \cupdisj_\SepLoc \LocH(H_2) \\
    \LocH((r, S))    & = & \LocS^\Heap(r; S) \vspace{1em} \\
    \LocS^\Heap(r; S, \iota \mapsto \_) & = & \LocS^\Heap(r; S) \cupdisj_\SepLoc \{
                                              \Heap(r, \iota) \} \\
    \LocS^\Temp(r; S, \iota \mapsto \_) & = & \LocS^\Temp(r; S) \cupdisj_\SepLoc \{
                                              \Temp(r, \iota) \} \\
    \LocS^*(\NilS)    & = & \emptyset
  \end{array}
\]

Because of the properties of $\SepLoc$ mentioned above (in particular the
uniqueness of object ids $\iota$), given that $\LocSet \neq \FUndef$ we can
define a partial function $\Loc_\LocSet: \Oid \partialto \LocSet$ as
\[
  \Loc_\LocSet(\iota) = \begin{cases}
                          \Heap(r, \iota) & \text{ if } \Heap(r, \iota) \in
                                            \LocSet \\
                          \Temp(r, \iota) & \text{ if } \Heap(r, \iota) \in
                                            \LocSet
                        \end{cases}
\]

We move on to $\RefSet$. We define the separating function $\SepRef$ as
\[
  \SepRef\left(\Loc \RefTo{f, k} \Loc'\right) = (loc, f)
\]
We can use this as a separating function to ensure that uniqueness constraints
on field names in a well-formed configuration will be mirrored in the
construction of $\RefSet$. I.e. only configurations where each object or frame
has at most one field with a certain name will have a well-defined graph.

Given $\Loc_\LocSet$ as above we have

\[
  \RefSet = \RefRS(RS) \cupdisj^\SepRef \RefH(\Hop) \cupdisj^\SepRef \RefH(\Hcl)
  \cupdisj^\SepRef \RefH(\Hfr)
\]

\[
  \begin{array}{lcl}
    \RefRS((r, S, F)::RS) & = & \RefS(S) \cupdisj^\SepRef \RefF(\Root(r); F)
                                \cupdisj^\SepRef \RefRS(RS) \\
    \RefRS(\NilRS) & = & \emptyset \vspace{1em} \\
    \RefH(H_1 * H_2) & = & \RefH(H_1) \cupdisj^\SepRef \RefH(H_2) \\
    \RefH((r, S)) & = & \RefS(S) \vspace{1em} \\
    \RefS(S, \iota \mapsto (CL, F)) & = & \RefS(S) \cupdisj^\SepRef
                                          \RefF(\Loc_\LocSet(\iota); F) \\
    \RefS(\NilS) & = & \emptyset \vspace{1em} \\
    \RefF(\Loc; F, f \mapsto (k, \iota)) & = & \RefF(F) \cupdisj^\SepRef \{ \Loc
                                               \RefTo{f, k} \Loc_\LocSet(\iota)
                                               \} \\
    \RefF(\Loc; F, f \mapsto \Undef) & = & \RefF(F) \\
    \RefF(\NilF) & = & \emptyset
  \end{array}
\]

We want to have some kind of sanity check on a graph. Therefore we define
well-formedness of a graph $\Graph$, $\WFG(\Graph)$
as a predicate on a graph, as

\[
  \begin{array}{lcl}
    \WFG((\LocSet, \RefSet)) & \iff & \WFSrcDst(\LocSet, \RefSet) \AN
                                      \WFRefs(\RefSet) \\
    \WFSrcDst(\LocSet, \RefSet) & \iff & \begin{aligned}[t]
                                           \forall & \Loc \RefTo{f, k} \Loc' \in
                                         \RefSet. \\
                                                   &\Loc \in \LocSet \AN \Loc'
                                                    \in \LocSet
                                         \end{aligned} \\
    \WFRefs(\RefSet) & \iff & \begin{aligned}[t]
                                \forall & \REF_1 = (\Loc \RefTo{f, \_} \_) ~, \REF_2
                                          = (\Loc \RefTo{f', \_} \_) \in \RefSet. \\
                                        & \REF_1 = \REF_2 \OR f \neq f'
                              \end{aligned}
  \end{array}
\]
For a well-formed graph $\Graph = (\LocSet, \RefSet)$ we can define the partial
function
\[
  \begin{array}{lcl}
    \REF_\RefSet(\Loc, f) & = & \begin{cases}
                                  \Loc \RefTo{f, k} \Loc' & \text{ if } \Loc
                                                            \RefTo{f, k} \Loc'
                                                            \in \RefSet \\
                                  \FUndef & \text{ otherwise}
                                \end{cases} \\

  \end{array}
\]
Furthermore we define shorthands
\[
  \begin{array}{lcl}
    \Loc_\Graph(\iota) & = & \Loc_\LocSet(\iota) \\
    \REF_\Graph(\Loc, f) & = & \REF_\RefSet(\Loc, f) \\
    \REF_\Graph(\iota, f) & = & \REF_\Graph(\Loc_\Graph(\iota), f)
  \end{array}
\]

\subsubsection{Region Orders and Sets}

We define $\rho$ to talk about ordering of regions.

\[
  \begin{array}{rccl}
    \rho & ::= & r \rhosep{x.f} \rho ~|~ \Nilrho & \text{Region ordering}
  \end{array}
\]

If we do not care about $x.f$ we write a shorthand $r \rhosepe \rho$.

From a well formed region stack $\Gammas \vdash \RS$,  we can generate a $\rho$:

\[
  \begin{array}{lcl}
    \rho(\Gamma \rhosep{x.f} \Gammas, (r, S, F)::RS) & = & r \rhosep{x.f} \rho(RS) \\
    \rho(\NilGammas, \NilRS) & = & \Nilrho
  \end{array}
\]

We extend this definition to region configurations, $\rho(\Gammas, \RCfg)$, in the
obvious way.

We define ordering of region ids under $\rho$ as
\begin{DisplayRule}
  \drules{$\rho \vdash r < r' \quad \rho \vdash r \leq r'$}{Region ordering}{
    regionXXorderXXltXXcons,
    regionXXorderXXleqXXstrict,
    regionXXorderXXleqXXeq
  }
\end{DisplayRule}

We define
\[
  \begin{array}{lcl}
    \Regions(H_1 * H_2) &=& \Regions(H_1) \cupdisj \Regions(H_2) \\
    \Regions((r, S)) & = & \{ r \}
  \end{array}
\]

\[
  \begin{array}{lcl}
    \Open(\RCfglong{RS; \Hop; \Hcl; \Hfr}) &= & \Regions(\Hop) \\
    \Closed(\RCfglong{RS; \Hop; \Hcl; \Hfr}) &= & \Regions(\Hcl) \\
    \Frozen(\RCfglong{RS; \Hop; \Hcl; \Hfr}) &= & \Regions(\Hfr)
  \end{array}
\]

We define intersection of $\rho$ and a set of region ids in the obvious way. The
relation $\vdash \rho, \Cl, \Fr$ is simply disjointness of $\rho$ and the region id
sets $\Cl, \Fr$:

\[
  \vdash \rho, \Cl, \Fr  \iff \rho \cap \Cl = \rho \cap \Fr = \Cl \cap \Fr = \emptyset
\]

\subsubsection{Capability Semantics}
The predicate \CapOK(\RCfg) is defined using three auxilliary
predicates as follows.
\[
  \begin{aligned}
    \CapOK(\Gammas, \RCfg) \iff & \VarUnique(G(\RCfg)) \AN \\
                       &\RegionOrder(\rho(\Gammas, \RCfg), \Closed(\RCfg), \Frozen(\RCfg), G(\RCfg)) \AN \\
                       &\LocationOK(G(\RCfg)) \\
                       &\DeepFreeze(\Frozen(\RCfg), G(\RCfg))
  \end{aligned}
\]

The three auxilliary predicates are defined using the object graph $G(\RCfg)$ as
follows.

\[
  \RegionOrder(\rho, \Cl, \Fr, \Graph) \iff \forall \REF \in \Graph.
  \regionorder1(\rho, \Cl, \Fr, \REF)
\]

\[
  \begin{aligned}
    \regionorder1(\rho, \Cl, \Fr,
    &(\Loc[r] \RefTo{f, k} \Loc'[r']))\\
    &\iff \\
    &\hspace*{-1cm}\begin{aligned}
       & (k = \Mut &\implies& r = r') &\AN \\
       & (k = \Tmp &\implies& r = r') &\AN \\
       & (k = \Var &\implies& r = r') &\AN \\
       & (k = \Paused &\implies& \rho \vdash r' < r) &\AN \\
       & (k = \Iso &\implies& r \neq r' \AN (r' \in \Cl \OR ((\rho \vdash r < r') \OR r' \in
                              \Fr \AN r \in \Fr ))) &\AN \\
       & (k = \Imm &\implies& r' \in \Fr)) &
     \end{aligned}
  \end{aligned}
\]

\vspace{1em}

\[
  \locationok(\Graph) \iff \forall \REF \in \Graph. \locationok1(\REF)
\]
\vspace{1ex}

\[
  \begin{aligned}
    \locationok1(&\Loc \RefTo{f, k} \Loc') \\
    &\iff \\
    &\begin{aligned}
       & (k = \Mut &\implies& \Loc' = \Heap(r', \iota')) &\AN \\
       & (k = \Tmp &\implies& ((\Loc = \Root(r) \OR \Loc = \Temp(r, \iota))
                              \AN \Loc' = \Temp(r', \iota'))) &\AN \\
       & (k = \Var &\implies& (\Loc = \Root(r) \AN \Loc' = \Temp(r', \iota')))
                                                         & \AN \\
       & (k = \Paused &\implies& ((\Loc = \Root(r) \OR \Loc =
                                 \Temp(r, \iota)) \AN \Loc' \neq \Root(r')))
                                                         &\AN \\
       & (k = \Iso &\implies& \Loc' = \Heap(r', \iota')) &\AN \\
       & (k = \Imm &\implies& \Loc' = \Heap(r', \iota')) &
     \end{aligned}
  \end{aligned}
\]

\vspace{1em}

\[
  \deepfreeze(\Fr, \Graph) \iff \forall \REF \in \Graph. \deepfreeze1(\Fr, \REF)
\]

\vspace{1em}

\[
  \deepfreeze1(\Fr, \Loc[r] \refto{\_, \_} \loc'[r']) \iff r \in \Fr \implies r'
  \in \Fr
\]

\vspace{1em}
\[
  \VarUnique(\Graph) \iff \begin{aligned}[t]
                            \forall &\REF_1, Ref_2\in \Graph. \\
                                    &\VPU(\REF_1, \REF_2)
                          \end{aligned}
\]

\[
  \VPU(\REF_1, \REF_2) \iff
  \begin{aligned}[t]
    &\REF_1 = \_ \RefTo{\_, \Var} \Loc \AN &\\
    &\REF_2 = \_ \RefTo{\_, \_} \Loc &\implies \REF_1 = \REF_2
  \end{aligned}
\]

\subsubsection{Topology Invariant}
Finally we define $\TopOK(\RCfg)$. We again use the graph representation to
formulate it.

\[
  \begin{aligned}
    \TopOK(&\Gammas, \RCfg)\iff  \\
           &\topokgraph(\rho(\Gammas, \RCfg), \Frozen(\RCfg), G(\RCfg)) \AN \\
           &\entrypointsok(\rho(\Gammas, \RCfg), G(\RCfg))
  \end{aligned}
\]

\[
  \begin{aligned}
    \topokgraph(\rho, \Fr, \Graph) &\iff \\
                  &\begin{aligned}
                     \forall &\REF_1,\REF_2 \in \Graph. \\
                             &\ToppOK(\rho, \Fr, \REF_1, \REF_2)
                   \end{aligned}
  \end{aligned}
\]

\[
  \begin{aligned}
    \ToppOK(&\rho, \Fr, \REF_1, \REF_2) \iff \\
    &\begin{aligned}[t]
       \REF_1 = &~\Loc_1[r_1] \RefTo{\_, \_} \Loc_1'[r_1'] \AN \\
       \REF_2 = &~\Loc_2[r_2] \RefTo{\_, \_} \Loc_2'[r_2'] \\
       \implies & \begin{aligned}[t]
                    &\REF_1 = \REF_2 &\OR \\
                    &r_1' \neq r_2' &\OR \\
                    &r_1' \in \Fr \OR r_2' \in \Fr &\OR \\
                    &\rho \vdash r_1' \leq r_1 \OR \rho \vdash r_2' \leq r_2 &
                  \end{aligned}
     \end{aligned}
  \end{aligned}
\]

Note that the topology invariant in the paper also relies on the source of
backward pointing references being $\temp$ locations. This is not included in
this version, but still holds because of $\capok$.

\[
  \begin{aligned}
    &\entrypointsok(r' \rhosep{x.f} r :: \rho, \Graph) \iff \\
    &\ruut(r) \refto{x, k_x} \loc_x[r^*] \in \Graph \AN \\
    &\loc_x[r^*] \refto{f, k_f} \Heap(r', \_) \in \Graph \AN \\
    &\entrypointsok(r :: \rho, \Graph)
  \end{aligned}
\]
\[
  \begin{aligned}
    &\entrypointsok(r :: \Nilrho, \Graph) \iff \text{true}
  \end{aligned}
\]

\section{Effects}

\[
  \begin{array}{rcl}
    \Use & ::= & x ~|~ \kw{drop}~x\\
    \Eff & ::= & \load{x, y.f}\\
    & ~|~ & \swap{x, y.f, \Use}\\
    & ~|~ & \bind{\many{x = \Use}}\\
    & ~|~ & \halloc{x, k, \#C, \many{\Use}}\\
    & ~|~ & \salloc{x, k, \CLTag, \many{\Use}}\\
    & ~|~ & \enter{w, k, y.f, CL, \many{x = \Use}}\\
    & ~|~ & \badenter{x.f}\\
    & ~|~ & \exit{x, \Use, y.f, w.g}\\
    & ~|~ & \cast{x, \Use, k~C}\\
    & ~|~ & \nocast{x, \Use, k~C}\\
    & ~|~ & \bind{\many{x = \Use}}\\
    & ~|~ & \epsilon\\
  \end{array}
\]

\subsection{Well-Formed Effects}

We use the following helper predicate for a capability pointing
into an open region:

\[
  \mathbf{open}(k) \iff k \in \{\Mut, \Tmp, \Var, \Paused\}
\]

Effects are typed under a stack of contexts $\Gammas$, defined in
Section~\ref{sec:cmd:static}.

\begin{DisplayRule}
  \drules{$\Gammas \vdash \Eff \dashv \Gammas$}{Well-formed Effects}{
    wfXXeffXXenter,
    wfXXeffXXbadenter,
    wfXXeffXXexit,
    wfXXeffXXload,
    wfXXeffXXswapXXclass,
    wfXXeffXXswapXXvar,
    wfXXeffXXhallocXXmut,
    wfXXeffXXhallocXXiso,
    wfXXeffXXsalloc,
    wfXXeffXXfreeze,
    wfXXeffXXmerge,
    wfXXeffXXcast,
    wfXXeffXXnocast,
    wfXXeffXXbind,
    wfXXeffXXsplit,
    wfXXeffXXeps
  }
\end{DisplayRule}

\section{Command Semantics}
\label{sec:cmd}

\[
  \begin{array}{rcll}
     e & ::= & \Use ~|~ \kw{let}~x = b~\kw{in}~e
         ~|~ \kw{if}~\kw{typetest}(\Use, t) \SB y => e \FB \SB y => e \FB & \text{Expressions} \\
  \Use & ::= & x ~|~ \kw{drop}~x & \text{Variable usage}\\
     b & ::= & \c{*}\lval
         ~|~ \lval~\c{:=}~\Use
         ~|~ \fnc(\many{use})
         ~|~ \kw{var}~\Use
         ~|~ \kw{new}~k~C(\many{use})& \text{Bound expressions}\\
       & ~|~ & \kw{enter}~\lval~[\many{y=\Use}] \SB z~\c{=>}~e \FB
         ~|~ \kw{freeze}~\Use
         ~|~ \kw{merge}~\Use
         ~|~ e\\
 \lval & ::= & x ~|~ x.f & \text{L-values}\\
    de & ::= & e ~|~ \kw{let}~x = db~\kw{in}~e ~|~ \textbf{Failure} & \text{Dynamic expressions}\\
    db & ::= & b ~|~ de ~|~ \kw{entered}~x.f~y.f' \SB de \FB & \text{Dynamic bound expressions}\\
    de[\bullet] & ::= & \bullet ~|~ \kw{let}~x = \bullet~\kw{in}~e
         ~|~ \kw{let}~x = \kw{entered}~y.f~w.f' \SB \bullet \FB~\kw{in}~e & \text{Execution context}\\
    \CCfg & ::= & \SB de \FB & \text{Dynamic configuration}
  \end{array}
\]

The syntax of the command language is in A-normal form (ANF).
An expression is a \Use{} of a name (a potentially destructive
read), the binding of an expression $b$ to a name, or a dynamic
type test, where the value is rebound in the chosen branch in
order to track the type.
Note that we model mutable variables as objects, created with the
syntax $\kw{var}~\Use$. A bound expression is a field access, a
field update, a function call, the creation of a variable or an
object, an \kw{enter} block, or a \kw{freeze} or \kw{merge}.
An enter block enters through a variable $x$ or a field $x.f$ and
explicitly captures variables $\many{\Use}$, potentially
destructively, as $\many{y}$. The bridge object is bound to $z$
and the body $e$ of the block is executed. The capture list is for
convenience only, and could be inferred based on variable usage in
$e$.

During execution, an \kw{enter} block reduces to a \kw{entered}
block, which stores the field entered through and the field in
which the new bridge object can be found when exiting the region.
The only difference between dynamic expressions $de$ and static
expressions $e$ is that dynamic expressions can contain
\kw{entered} blocks in the currently bound expression being
evaluated, or denote failure.
We use an execution context $de[\bullet]$ to find the inner-most
dynamic expression to evaluate.
A dynamic configuration is just a dynamic expression.

We assume the existence of a table of well-formed functions such
that if
$\mathbf{fnctype}(\fnc) = (t_1, ..., t_n) \rightarrow t$,
then
$\mathbf{fnclookup}(\fnc) = (x_1, ..., x_n) \rightarrow e$
and $x_1 : t_1, ..., x_n : t_n \vdash e : t \dashv \Gamma'$ for
some $\Gamma'$. We further assume that function lookup renames all
let-bound variables to globally fresh names to avoid accidental
shadowing.

\subsection{Dynamic Semantics}

Because the command language is in ANF, many rules differ only in
which effect they emit. In order to avoid duplicating rules, for
expressions with the shape $\kw{let}~x~=~e_1~\kw{in}~e_2$ we define a
helper function to calculate the effect of $b$:
\[
\begin{array}{lcl}
  \mathbf{effect}(x, \c{*y.f})                     & = & \load{x, y.f}\\
  \mathbf{effect}(x, \c{*y})                       & = & \load{x, y.val}\\
  \mathbf{effect}(x, \c{y.f := }\Use)              & = & \swap{x, y.f, \Use}\\
  \mathbf{effect}(x, \c{y   := }\Use)              & = & \swap{x, y.val, \Use}\\
  \mathbf{effect}(x, \kw{new}~\Mut~C(\many{\Use})) & = & \halloc{x, \Mut, \#C, \many{\Use}}\\
  \mathbf{effect}(x, \kw{new}~\Iso~C(\many{\Use})) & = & \halloc{x, \Iso, \#C, \many{\Use}}\\
  \mathbf{effect}(x, \kw{new}~\Tmp~C(\many{\Use})) & = & \salloc{x, \Tmp, \#C, \many{\Use}}\\
  \mathbf{effect}(x, \kw{var}~\Use)                & = & \salloc{x, \Var, \#\Cell, \Use}\\
  \mathbf{effect}(x, \kw{freeze}~\Use)             & = & \freeze{x, \Use}\\
  \mathbf{effect}(x, \kw{merge}~\Use)              & = & \merge{x, \Use}\\
  \mathbf{effect}(x, \Use)                         & = & \bind{x = \Use}\\
\end{array}
\]

\begin{DisplayRule}
  \drules{$de \ComStep{\Eff} de$}{Command semantics}{
    cmdXXletXXeff,
    cmdXXenterXXfieldXXfail,
    cmdXXenterXXvarXXfail,
    cmdXXenterXXfield,
    cmdXXenterXXvar,
    cmdXXexit,
    cmdXXifXXtypetestXXtrue,
    cmdXXifXXtypetestXXfalse,
    cmdXXcall,
    cmdXXec,
    cmdXXecXXfail
  }
\end{DisplayRule}

\subsection{Static Semantics}
\label{sec:cmd:static}

\[
  \begin{array}{rcll}
    t & ::= & k~C ~|~ k~\c{Cell}[t] ~|~ t\c{|}t & \text{Types}\\
    \Gamma & ::= & \Gamma, x : t ~|~ \Gamma, x : \Undef ~|~ \varepsilon & \text{Typing context}\\
    \Gammas & ::= & \Gammas\underset{x.f}{::}\Gamma ~|~ \varepsilon & \text{Dynamic typing context}
  \end{array}
\]

A type is a class $C$ with a capability $k$, a \c{Cell} with a
given type (\cf{} ML-style ref cells), or a union $t_1\c{|}t_2$ of
two types.
A typing context maps variables to types, or remembers that they
have been consumed.
When typing a dynamic expression, we use a stack of typing
contexts mirroring the stack of open regions. Each $::$ between
two contexts in the stack is marked with $x.f$ denoting the field
through which we entered the next region. Whenever we don't care
about this field, we omit it for readability.
When dealing with union types, we sometimes need to merge two
different contexts with the same domain but (possibly) different
codomains. This is done with the following function:

\[
  (\Gamma_1\c{|}\Gamma_2) =
  \left\{
  \begin{array}{llcl}
    (\Gamma_1'\c{|}\Gamma_2'), x : t_1\c{|}t_2 &
      \text{if } \Gamma_1 = \Gamma_1', x : t_1 & \text{ and }& \Gamma_2 = \Gamma_2', x : t_2\\
    (\Gamma_1'\c{|}\Gamma_2'), x : \Undef &
      \text{if } \Gamma_1 = \Gamma_1', x : \Undef & \text{ and }& \Gamma_2 = \Gamma_2', x : t_2\\
    (\Gamma_1'\c{|}\Gamma_2'), x : \Undef &
      \text{if } \Gamma_1 = \Gamma_1', x : t_1 & \text{ and }& \Gamma_2 = \Gamma_2', x : \Undef\\
    (\Gamma_1'\c{|}\Gamma_2'), x : \Undef &
      \text{if } \Gamma_1 = \Gamma_1', x : \Undef & \text{ and }& \Gamma_2 = \Gamma_2', x : \Undef\\
    \varepsilon & \text{if } \Gamma_1 = \varepsilon & \text{ and } & \Gamma_2 = \varepsilon
  \end{array}
  \right.
\]

\[
  (\many{\Gamma}_1\c{|}\many{\Gamma}_2) =
  \left\{
  \begin{array}{llcl}
    (\Gamma_1'\c{|}\Gamma_2') :: (\many{\Gamma}_1'\c{|}\many{\Gamma}_2') &
      \text{if } \many{\Gamma}_1 = \Gamma_1' :: \many{\Gamma_1}' & \text{ and }& \many{\Gamma}_2 = \Gamma_2' :: \many{\Gamma}_2'\\
    \varepsilon & \text{if } \many{\Gamma}_1 = \varepsilon & \text{ and } & \many{\Gamma}_2 = \varepsilon
  \end{array}
  \right.
\]

We sometimes need to compare what variables are defined for a certain $\Gamma$
or $\Gammas$. We use define $\defined$ as follows

\[
  \begin{array}{lcl}
    \defined(\Gamma, x: t) & = & \defined(\Gamma) \cup \{ x \} \\
    \defined(\Gamma, x : \Undef) & = & \defined(\Gamma) \\
    \defined(\NilGamma) & = & \emptyset \\
  \end{array}
\]

To compare $\Gammas_1$ and $\Gammas_2$, we check pairwise set inclusion of
defined sets:
\[
  \begin{aligned}
    \Gamma :: \Gammas &\gammadefincl \Gamma' :: \Gammas' \iff \\
                      &\defined(\Gamma) \subseteq \defined(\Gamma') \AN \\
                      &\Gammas \gammadefincl \Gammas'
  \end{aligned}
  \]

\noindent
We further use the following helper functions and predicates
(viewpoint adaptation is in Table~\ref{tab:vpa}):

We use $\mathit{make\_iso}$ and $\mathit{make\_mut}$ to change all
occurrences of \Mut{} to \Iso{} and vice versa for a type (all
other capabilities remain unchanged).
\[
  \begin{array}{rcl}
    \mathit{make\_iso}(k~CL) & = &
    \begin{cases}
      \Iso~CL & \text{if } k = \Mut\\
      k~CL    & \text{otherwise}
    \end{cases}\\
    \mathit{make\_iso}(t_1~|~t_2) & = & \mathit{make\_iso}(t_1) ~|~ \mathit{make\_iso}(t_2)
  \end{array}
\]

\[
  \begin{array}{rcl}
    \mathit{make\_mut}(k~CL) & = &
    \begin{cases}
      \Mut~CL & \text{if } k = \Iso\\
      k~CL    & \text{otherwise}
    \end{cases}\\
    \mathit{make\_mut}(t_1~|~t_2) & = & \mathit{make\_mut}(t_1) ~|~ \mathit{make\_mut}(t_2)
  \end{array}
\]



We use the predicate $\Kap(\many{k}, t)$ to check if all
capabilities in a type are in $\many{k}$. We omit brackets for
singleton sets (the first case below), and use the complement
notation $\Kap(\{k\}^c, t)$ for checking that capability $k$ does
\emph{not} occur anywhere in $t$.

\[
  \begin{array}{rcl}
    \Kap(k, t) & \equiv & \Kap(\{k\}, t)\\
    \Kap(\many{k}, k' CL)     & \iff & k' \in \many{k}\\
    \Kap(\many{k}, t_1~|~t_2) & \iff & \Kap(\many{k}, t_1) \land \Kap(\many{k}, t_2)
  \end{array}
\]

We use $\mathbf{classtype}(t)$ to check if a type is definitely a
class type:

\[
  \mathbf{classtype}(t) \iff \forall k~CL \in t.~ \exists C. CL = C
\]

\begin{table}
  \vspace*{-3ex}
  \caption{Viewpoint adaptation.
    If the capabilities of \c{x} and \c{f} are $\alpha$ and $\beta$,
    then the capability of \c{x.f} is $\VPA{\alpha}{\beta} = \gamma$,
    which we read as ``$\alpha$ sees $\beta$ as $\gamma$.''
    Note that this table is slightly different from the one in the
    paper since we only use it for non-destructive reads. }
  \label{tab:vpa}
  \centering
  \begin{tabular}{r|ccccc}
    \toprule
       Capability & \multicolumn{5}{c}{Capability on \c{f}}             \\ %
       on \c{x}   & \Mut{}     & \Tmp{}     & \Imm{}    & \Iso{}      & \Paused{} \\ 
    \midrule                                                              %
    \Mut{}        & \Mut{}     & $\FUndef$  & \Imm{}    & $\FUndef{}$ & $\FUndef$ \\ 
    \Tmp{}        & \Mut{}     & \Tmp{}     & \Imm{}    & $\FUndef{}$ & \Paused{} \\ 
    \Var{}        & \Mut{}     & \Tmp{}     & \Imm{}    & $\FUndef{}$ & \Paused{} \\ 
    \Imm{}        & \Imm{}     & \Imm{}     & \Imm{}    & \Imm{}      & \Imm{}    \\ 
    \Iso{}        & $\FUndef$  & $\FUndef$  & $\FUndef$ & $\FUndef{}$   & $\FUndef$     \\ 
    \Paused{}     & \Paused{}  & \Paused{}  & \Imm{}    & $\FUndef{}$ & \Paused{} \\ 
    \bottomrule
  \end{tabular}
  \[
    \begin{array}{rcl}
      \VPA{k}{(k'~CL)} & = &
      \begin{cases}
        k''~CL & \text{if } \VPA{k}{k'} = k''\\
        \FUndef{} & \text{otherwise}
      \end{cases}\\
      \VPA{k}{(t_1~|~t_2)} & = &
      \begin{cases}
        t_1'~|~t_2' & \text{if } \VPA{k}{t_1} = t_1' \text{ and } \VPA{k}{t_2} = t_2'\\
        \FUndef{} & \text{otherwise}
      \end{cases}
    \end{array}
  \]
\end{table}

We extend viewpoint adaptation to $k \sees \Gamma$ and $k \sees F$ by mapping
over the elements of $\Gamma$ and $F$ respectively.

\begin{DisplayRule}
  \drules{$\Gamma \vdash e \dashv \Gamma\quad\Gamma \vdash b \dashv \Gamma$}{Command Language Typing}{
    cmdXXtyXXuseXXkeep,
    cmdXXtyXXuseXXdrop,
    cmdXXtyXXlet,
    cmdXXtyXXtypetest,
    cmdXXtyXXderefXXfield,
    cmdXXtyXXderefXXvar,
    cmdXXtyXXassign,
    cmdXXtyXXassignXXvar,
    cmdXXtyXXcreateXXvar,
    cmdXXtyXXcall,
    cmdXXtyXXnew,
    cmdXXtyXXenter,
    cmdXXtyXXenterXXvar,
    cmdXXtyXXfreeze,
    cmdXXtyXXmerge,
    cmdXXtyXXsplit,
    cmdXXtyXXsub
  }
\end{DisplayRule}

\begin{DisplayRule}
  \drules{$\Gammas \vdash de \dashv \Gamma\quad\Gammas \vdash db \dashv \Gamma$}{Command Language Dynamic Typing}{
    cmdXXdynXXtyXXexpr,
    cmdXXdynXXtyXXlet,
    cmdXXdynXXtyXXentered,
    cmdXXdynXXtyXXsub,
    cmdXXdynXXtyXXsplit,
    cmdXXdynXXtyXXfailure
  }
\end{DisplayRule}

\begin{DisplayRule}
  \drules{$\vdash t$}{Well-formedness of types}{
    tyXXclass,
    tyXXcell,
    tyXXdisj
  }
\end{DisplayRule}

\begin{DisplayRule}
  \drules{$t <: t$}{Subtyping}{
    subXXrefl,
    subXXdisjXXrightXXOne,
    subXXdisjXXrightXXTwo,
    subXXdisjXXleft,
    subXXcell
  }
\end{DisplayRule}

\begin{DisplayRule}
  \drules{$\Gammas \vdash \SB de \FB$}{Well-formedness of dynamic configuration}{
    wfXXccfg
  }
\end{DisplayRule}

\begin{DisplayRule}
  \drules{$\vdash \Gamma$}{Well-formedness of typing contexts}{
    wfXXctxXXcons,
    wfXXctxXXconsXXundef,
    wfXXctxXXempty
  }
\end{DisplayRule}

\begin{DisplayRule}
  \drules{$\vdash \Gammas$}{Well-formedness of dynamic typing contexts}{
    wfXXdynXXctxXXcons,
    wfXXdynXXctxXXempty
  }
\end{DisplayRule}

\section{Proofs}

\subsection{Theorem: Progress of Command Language}
\label{theorm:cmd_progress}

\[
  \begin{array}{l}
    \many{\Gamma} \vdash \SB de \FB \implies\\
    \quad de = \Use \OR de = \mathbf{Failure} \OR
          \exists \Eff, de'.~ de \ComStep{\Eff} de'\\
  \end{array}
\]

\subsubsection{Proof Sketch}
The proof of progress is very simple since there are very few
rules that have premises that can be false: function calls must
refer to existing functions and objects can only be constructed
with capability \Mut{}, \Tmp{} or \Iso{}. All other rules are
always applicable when the expression has the right shape.

\subsubsection{Proof of Theorem}
\pf

We begin by inversion of $\many{\Gamma} \vdash \SB de \FB$ and get
$\many{\Gamma} \vdash de : t \dashv \Gamma'$. We proceed by
induction over the dynamic typing judgment.

\begin{pfcases}
  \pfcase{\rn{cmd-dyn-ty-failure}}, $de = \mathbf{Failure}$\\
  Proof holds trivially

  \pfcase{\rn{cmd-dyn-ty-expr}}, $de = e$
  \begin{enumerate}[label=A\arabic*]
  \item $\Gammas = \Gamma$
  \end{enumerate}
  We proceed by induction over the static typing judgment
  $\Gamma \vdash de : t \dashv \Gamma'$:
  \begin{pfsubcases}
    \pfsubcase{\rn{cmd-ty-use-keep}}, $de = x$\\
    Proof holds trivially.

    \pfsubcase{\rn{cmd-ty-use-drop}}, $de = \kw{drop}~x$\\
    Proof holds trivially.

    \pfsubcase{\rn{cmd-ty-typetest}}, $de = \kw{if}~\kw{typetest}(\Use, t_1)\{y \c{=>} e_1\}\{y \c{=>} e_2\}$\\
    We pick any $k~CL$.
    \begin{itemize}
    \item If $k~CL<:t$, the configuration steps by
      \rn{cmd-if-typetest-true}.
    \item If $k~CL\not<:t$, the
      configuration steps by \rn{cmd-if-typetest-false}.
    \end{itemize}

    \pfsubcase{\rn{cmd-ty-let}}\llabel[pf]{case:let}, $de = \kw{let}~x=b~\kw{in}~e$\\
    By \rn{cmd-ty-let} we have
    \begin{enumerate}[label=A\arabic*]
    \item $\Gamma \vdash b : t' \dashv \Gamma''$ \llabel{ass:wft}
    \end{enumerate}
    We proceed by cases on the shape of $b$.
    \begin{itemize}
    \item If $b = \fnc(\Use_1, ..., \Use_n)$, then by the
      inversion lemma with \lref{ass:wft} we get
      $\mathbf{fnctype}(\fnc) = (t_1, ..., t_n) \rightarrow t''$,
      and thus by assumptions about the function table
      $\mathbf{fnclookup}(\fnc) = (x_1, ..., x_n) \rightarrow
      e_{body}$. The configuration steps by \rn{cmd-call}.
    \item If $b = \kw{new}~k~C(\Use_1, ..., \Use_n)$, then by the
      inversion lemma with \lref{ass:wft} we get
      $k \in \{\Mut, \Tmp, \Iso\}$. The configuration steps by
      \rn{cmd-let-eff}
    \item If $b = \kw{enter}...$, the configuration steps by
      \rn{cmd-enter-field}, \rn{cmd-enter-var} or
      \rn{cmd-enter-fail}
    \item For all remaining cases, the configuration steps by
      \rn{cmd-let-eff}.
    \end{itemize}
  \end{pfsubcases}

  \pfcase{\rn{cmd-dyn-ty-let}, $de = \kw{let}~x=db~\kw{in}~e$}
  \begin{enumerate}[label=A\arabic*]
  \item $\Gammas = \many{\Gamma}' :: \Gamma$
  \end{enumerate}
  We proceed by cases on whether $db$ is in the syntactic category
  of $b$ or not.
  \begin{itemize}
  \item If it is, the proof collapses to case \lref[pf]{case:let},
    since every non-dynamic typing derivation needs to use
    \rn{cmd-dyn-ty-expr}, which requires an empty $\many{\Gamma}$.
  \item If it is not, we have either $db = de'$ or
    $db = \kw{entered}~y.f~w.\c{val} \SB de' \FB$. We proceed by
    cases.
    \begin{pfsubcases}
      \pfsubcase{$db = de'$}
      \begin{itemize}
      \item By \rn{cmd-ty-let} we have
        $\Gammas \vdash de' : t' \dashv \Gamma''$.
      \item By the induction hypothesis we have $de' = \Use$, or
        $de' = \mathbf{Failure}$ or
        $\exists \Eff, de''.~ de' \ComStep{\Eff} de''$. We
        proceed by cases.
        \begin{itemize}
        \item If $de' = \Use$, configuration steps by
          \rn{cmd-eff}.
        \item If $de' = \mathbf{Failure}$, configuration steps by
          \rn{cmd-ec-fail}.
        \item If $de'\ComStep{\Eff}de''$, configuration steps
          by \rn{cmd-ec}.
        \end{itemize}
      \end{itemize}

      \pfsubcase{$db = \kw{entered}~x.f~y.\c{val} \SB de' \FB$}
      \begin{enumerate}[label=B\arabic*]
      \item $\Gammas = \many{\Gamma}'' :: \Gamma' :: \Gamma$
      \end{enumerate}
      \begin{itemize}
      \item By \rn{cmd-ty-let} we have $\many{\Gamma} \vdash \kw{entered}~x.f~y.\c{val} \SB de' \FB : t' \dashv \Gamma''$.
      \item By the inversion lemma, we have $\many{\Gamma}'' :: \Gamma' \vdash de' : t' \dashv \Gamma_1'$
      \item By the induction hypothesis we have $de' = \Use$, or
        $de' = \mathbf{Failure}$ or
        $\exists \Eff, de''.~ de' \ComStep{\Eff} de''$. We
        proceed by cases.
        \begin{itemize}
        \item If $de' = \Use$, configuration steps by \rn{cmd-ec}
          with \rn{cmd-exit}.
        \item If $de' = \mathbf{Failure}$, configuration steps by
          \rn{cmd-ec} with \rn{cmd-ec-fail}.
        \item If $de'\ComStep{\Eff}de''$, configuration steps
          by \rn{cmd-ec} with \rn{cmd-ec}
        \end{itemize}
      \end{itemize}
    \end{pfsubcases}
  \end{itemize}
\end{pfcases}

\subsection{Theorem: Preservation of Command Language}
\label{theorm:cmd_preservation}

\[
  \begin{array}{l}
    \many{\Gamma} \vdash \SB de \FB \AN \SB de \FB \ComStep{\Eff} \SB de' \FB \implies\\
    \quad \exists \many{\Gamma}'.~ \many{\Gamma}'\vdash \SB de' \FB \AN
          \many{\Gamma} \vdash \Eff \dashv \many{\Gamma}'
  \end{array}
\]

\subsubsection{Proof Sketch}
The proof of preservation ensures that stepping preserves
well-formedness and that it produces a well-formed effect. The
former is straightforward since the only changes we make to the
configuration is alpha conversion of subexpressions, introduction
of (well-formed) function bodies, and introduction of \kw{entered}
blocks from (well-formed) \kw{enter} blocks.
The latter comes from lifting the rules for expression typing to
the rules for well-formed effects, via the inversion lemma.

\subsubsection{Proof of Theorem}
\pf

We state an alpha conversion property, the observation that we can
change the name of a variable without affecting well-formedness of
an expression: \TODO{Prove as lemma?}

\[
  \begin{array}{l}
    \Gamma[x : t_1] \vdash e : t \dashv \Gamma'[x : t_2] \AN y~\mathit{fresh} \implies\\
    \quad \Gamma[y : t_1] \vdash e[x \mapsto y] : t \dashv \Gamma'[y : t_2]\\\\
    \Gamma[x : t_1] \vdash e : t \dashv \Gamma'[x : \Undef] \AN y~\mathit{fresh} \implies\\
    \quad \Gamma[y : t_1] \vdash e[x \mapsto y] : t \dashv \Gamma'[y : \Undef]\\
  \end{array}
\]

We also note that the type system supports weakening of the typing
context as long as the bound variables in the expression are not
in the domain of the added context.

\[
  \begin{array}{l}
  \Gamma_1 \vdash e : t \dashv \Gamma_2 \AN \vdash \Gamma, \Gamma_1 \AN \mathbf{BV}(e) \cap \mathbf{dom}(\Gamma) = \emptyset\implies\\
    \quad \Gamma, \Gamma_1 \vdash e : t \dashv \Gamma, \Gamma_2
  \end{array}
\]

We start the proof of preservation by inversion of well-formedness
of the configuration and get
$\many{\Gamma} \vdash de : t \dashv \Gamma'$.
We proceed by induction over the dynamic typing judgment.

\begin{pfcases}
  \pfcase{\rn{cmd-dyn-ty-failure}}, $de = \mathbf{Failure}$\\
  Proof holds vacuously as the configuration does not step.

  \pfcase{\rn{cmd-dyn-ty-expr}}, $de = e$
  \begin{enumerate}[label=G\arabic*]
  \item $\many{\Gamma} = \Gamma$
  \end{enumerate}
  We proceed by induction over the static typing judgment
  $\Gamma \vdash de : t \dashv \Gamma'$:
  \begin{pfsubcases}
    \pfsubcase{\rn{cmd-ty-use-keep}}, $de = x$\\
    Proof holds vacuously as the configuration does not step.

    \pfsubcase{\rn{cmd-ty-use-drop}}, $de = \kw{drop}~x$\\
    Proof holds vacuously as the configuration does not step.

    \pfsubcase{\rn{cmd-ty-typetest}}, $de = \kw{if}~\kw{typetest}(\Use, t')\{y \c{=>} e_1\}\{y \c{=>} e_2\}$\\
    By \rn{cmd-ty-typetest} we have
    \begin{enumerate}[label=A\arabic*]
    \item $\Gamma \vdash \Use : t'' \dashv \Gamma''$ \llabel{ass:use}
    \item $\Gamma'', y : t' \vdash e_1 : t_1 \dashv \Gamma_1, y : \_$ \llabel{ass:then}
    \item $\Gamma'', y : t''  \vdash e_2 : t_2 \dashv \Gamma_2, y : \_$ \llabel{ass:else}
    \end{enumerate}

    There are two rules that step the configuration.
    \begin{itemize}
    \item \rn{cmd-if-typetest-true} \llabel[pf]{case:true}
      \begin{enumerate}[label=B\arabic*]
      \item $\Eff = \cast{y', \Use, k~CL}$
      \item $k~CL <: t'$ \llabel{ass:sub}
      \item $y'~\mathit{fresh}$
      \end{enumerate}
      \begin{itemize}
      \item By the alpha conversion property with \lref{ass:then},
        we have $\Gamma'', y' : t' \vdash e_1[y \mapsto y'] : t_1 \dashv \Gamma_1, y' : \_$.
      \item The resulting configuration is well-formed by \rn{cmd-dyn-ty-expr}.
      \item The effect is well-formed under $\Gamma$ By \rn{wf-eff-cast} with \lref{ass:use} and \lref{ass:sub}.
      \end{itemize}

    \item \rn{cmd-if-typetest-false}\\
      Similar to the above.
    \end{itemize}

    \pfsubcase{\rn{cmd-ty-let}}\llabel[pf]{case:let}, $de = \kw{let}~x=b~\kw{in}~e$\\
    By \rn{cmd-ty-let} we have
    \begin{enumerate}[label=A\arabic*]
    \item $\Gamma \vdash b : t' \dashv \Gamma''$ \llabel{ass:wfb}
    \item $\Gamma'', x : t' \vdash e : t \dashv \Gamma', x : \_$ \llabel{ass:wfe}
    \end{enumerate}
    There are seven rules that step the configuration.
    \begin{itemize}
    \item \rn{cmd-let-eff}
      \begin{enumerate}[label=B\arabic*]
      \item $x'~\mathit{fresh}$
      \item $\mathbf{effect}(x', b) = \Eff$
      \end{enumerate}
      \begin{itemize}
      \item By the alpha conversion property with \lref{ass:wfe},
        we have
        $\Gamma'', x' : t' \vdash e[x \mapsto x'] : t \dashv
        \Gamma'', x' : \_$, and the resulting configuration is
        well-formed by \rn{cmd-dyn-ty-expr}.
      \item Depending on the shape of $b$ there are eleven cases
        of $\mathbf{effect}(x', b) = \Eff$:
        \begin{itemize}
        \item $\mathbf{effect}(x', \c{*y.f}) = \load{x, y.f}$\\
          The effect is well-formed by \rn{wf-eff-load}, by the
          inversion lemma with \lref{ass:wfb}, followed by
          induction over the shape of $t'$.

        \item $\mathbf{effect}(x', \c{*y}) = \load{x, y.val}$\\
          Similar to the previous case.

        \item $\mathbf{effect}(x', \c{y.f \c{:=} }\Use) = \swap{x, y.f, \Use}$\\
          The effect is well-formed by \rn{wf-eff-swap-class}, by
          the inversion lemma with \lref{ass:wfb}, followed by
          induction over the shape of $t'$.

        \item $\mathbf{effect}(x', \c{y \c{:=} }\Use) = \swap{x, y.val, \Use}$\\
          Similar to the previous case, but using \rn{wf-eff-swap-var}.

        \item $\mathbf{effect}(x', \kw{new}~\Mut~C(\many{\Use})) = \halloc{x, \Mut{}, \#C, \many{\Use}}$\\
          The effect is well-formed by \rn{wf-eff-halloc-mut}, by
          the inversion lemma with \lref{ass:wfb}. For the typing
          of each $\Use$, we use \rn{cmd-ty-sub} since the
          inversion lemma gives us that the arguments are subtypes
          of the field types.

        \item $\mathbf{effect}(x', \kw{new}~\Iso~C(\many{\Use})) = \halloc{x, \Iso{}, \#C, \many{\Use}}$\\
          Similar to the previous case, but using
          \rn{wf-eff-halloc-iso} and without the need for
          \rn{cmd-ty-sub} (note that all the arguments are \Iso{}
          or \Imm{}).

        \item $\mathbf{effect}(x', \kw{new}~\Tmp~C(\many{\Use})) = \salloc{x, \Tmp{}, \#C, \many{\Use}}$\\
          Similar to the case for $\Mut{}$ allocation, but using
          \rn{wf-eff-salloc}.

        \item $\mathbf{effect}(x', \kw{var}~\Use) = \salloc{x, \Var{}, \#\Cell, \Use}$\\
          Similar to the previous case.

        \item $\mathbf{effect}(x', \kw{freeze}~\Use) = \freeze{x, \Use}$\\
          The effect is well-formed by \rn{wf-eff-freeze}, by the
          inversion lemma with \lref{ass:wfb}.

        \item $\mathbf{effect}(x', \kw{merge}~\Use) = \merge{x, \Use}$\\
          The effect is well-formed by \rn{wf-eff-merge}, by the
          inversion lemma with \lref{ass:wfb}.

        \item $\mathbf{effect}(x', \Use) = \bind{x = \Use}$\\
          The effect is well-formed by \rn{wf-eff-bind} with \lref{ass:wfb}.
        \end{itemize}
      \end{itemize}

    \item \rn{cmd-enter-field}, $b = \kw{enter}~y.f~[\many{z=\Use}] \SB w~\c{=>}~e' \FB$
      \begin{enumerate}[label=B\arabic*]
      \item $w'~\mathit{fresh}$
      \item $z'_1,...,z'_n~\mathit{fresh}$
      \item $\Eff = \enter{w', \Tmp{}, y.f, z'_1 = \Use_1, ..., z'_n = \Use_n}$
      \item $de' = \kw{let}~x = \kw{entered}~y.f~w'.\c{val} \SB e'[w\mapsto w', z_1 \mapsto z'_1, ..., z_n \mapsto z'_1]\FB$
      \end{enumerate}
      By the inversion lemma with \lref{ass:wfb} we get
      \begin{enumerate}[resume*]
      \item $\forall i \in [1,n]. \Gamma_i \vdash \Use_i : t_i \dashv \Gamma_{i+1}$ \llabel{ass:uses}
      \item $\Gamma_{n+1}(y) = t_y$ \llabel{ass:yty}
      \item $\Kap(\{\Mut, \Tmp, \Var, \Paused\}, t_y)$ \llabel{ass:open}
      \item $\mathbf{fresult}(t_y, f)) = t_f$ \llabel{ass:ftype}
      \item $\Kap(\Iso, t_f)$ \llabel{ass:iso}
      \item $\Gamma' = y_1 : t'_1, ..., y_n : t'_n \text{ where } t'_i =
               \begin{cases}
                 t_i & \text{if } \Kap(\Iso, t_i)\\
                 \VPA{\Paused}{t_i} & \text{otherwise}
               \end{cases}$ \llabel{ass:gamma}
      \item $t_f' = \mathit{make\_mut}(t_f)$
      \item $\Kap(\{\Iso, \Imm\}, t')$ \llabel{ass:ret}
      \item $ t' <: t$ \llabel{ass:sub}
      \item $\Gamma', z : \Tmp~\Cell[t_f'] \vdash e : t' \dashv \Gamma', z : \Tmp~\Cell[t_f']$ \llabel{ass:inner}
      \end{enumerate}
      \begin{itemize}
      \item $\kw{entered}~y.f~w'.\c{val} \SB e'[w\mapsto w', z_1 \mapsto z'_1, ..., z_n \mapsto z'_1]$
        is well-formed by induction over the shape of $t_y$. In
        the inductive case we use \rn{cmd-dyn-ty-split}. In the
        base case we use the alpha conversion property and
        \rn{cmd-dyn-ty-entered} with \lref{ass:open},
        \lref{ass:ftype}, \lref{ass:iso}, \lref{ass:inner} and
        \lref{ass:ret}.
      \item $de'$ is well-formed by \rn{cmd-dyn-ty-sub} with
        \lref{ass:sub} and \rn{cmd-dyn-ty-let} with the above and
        \lref{ass:wfe}.
      \item The effect is well-formed by induction over the shape
        of $t_y$. In the inductive case we use \rn{wf-eff-split}.
        In the base case we use \rn{wf-eff-enter} with
        \lref{ass:uses}, \lref{ass:yty}~, \lref{ass:open}
        and \lref{ass:iso}, with resulting dynamic context
        $(\Gamma', z : \Tmp~\Cell[t_f']) :: \Gamma_{n+1}$.
      \end{itemize}

    \item \rn{cmd-enter-var}, $b = \kw{enter}~y~[\many{z=\Use}] \SB w~\c{=>}~e' \FB$\\
      Similar to the previous case.

    \item \rn{cmd-enter-field-fail}
      \begin{enumerate}[label=B\arabic*]
      \item $\Eff = \badenter{y.f}$
      \item $de' = \mathbf{Failure}$
      \end{enumerate}
      By the inversion lemma with \lref{ass:wfb} we get
      \begin{enumerate}[resume*]
      \item $\forall i \in [1,n]. \Gamma_i \vdash \Use_i : t_i \dashv \Gamma_{i+1}$
      \item $\Gamma_{n+1}(y) = t_y$ \llabel{fail:ass:yty}
      \item $\Kap(\{\Mut, \Tmp, \Var, \Paused\}, t_y)$ \llabel{fail:ass:open}
      \item $\mathbf{fresult}(t_y, f)) = t_f$ \llabel{fail:ass:ftype}
      \item $\Kap(\Iso, t_f)$ \llabel{fail:ass:iso}
      \end{enumerate}
      \begin{itemize}
      \item The resulting configuration $\SB \mathbf{Failure} \FB$
        is always well-formed.
      \item The effect is well-formed by induction over the shape
        of $t_y$. In the inductive case we use \rn{wf-eff-split}.
        In the base case we use \rn{wf-eff-badenter} with
        \lref{fail:ass:yty}, \lref{fail:ass:open}, \lref{fail:ass:ftype} and
        \lref{fail:ass:iso}.
      \end{itemize}

    \item \rn{cmd-enter-var-fail}\\
      Similar to the previous case.

    \item \rn{cmd-call}, $b = \fnc(\Use_1, ..., \Use_n)$
      \begin{enumerate}[label=B\arabic*]
      \item
        $\mathbf{fnclookup}(\fnc) = (x_1, ..., x_n) \rightarrow
        e_{body}$
      \item $x'_1, ..., x'_n~\mathit{fresh}$
      \item $\many{x} = x_1, ..., x_n$
      \item $\many{x}' = x'_1, ..., x'_n$
      \item $\Eff = \bind{x'_1 = \Use_1, ..., x'_n = \Use_n}$
      \end{enumerate}
      By the inversion lemma with \lref{ass:wfb} we get
      \begin{enumerate}[resume*]
      \item $t'' <: t'$ \llabel{call:ass:sub}
      \item $\mathbf{fnctype}(\fnc) = (t_1, ..., t_n) \rightarrow t''$
      \item $\Gamma = \Gamma_1$
      \item $\forall i \in [1,n]. \Gamma_i \vdash \Use_i : t_i \dashv \Gamma_{i+1}$ \llabel{ass:wfargs}
      \item $\Gamma'' = \Gamma_{n+1}$
      \end{enumerate}
      \begin{itemize}
      \item By assumptions about the function table we have
        $x_1 : t_1, ..., x_n : n_n \vdash e : t'' \dashv \Gamma_{res}$.
      \item By the alpha conversion property we have
        $x'_1 : t_1, ..., x'_n : n_n \vdash e[\many{x} \mapsto \many{x'}] : t'' \dashv \Gamma_{res}$.
      \item By the weakening property (we assume function lookup
        dynamically renames all variables to avoid clashes) we
        have
        $\Gamma'', x'_1 : t_1, ..., x'_n : n_n \vdash e[\many{x}
        \mapsto \many{x'}] : t'' \dashv \Gamma'', \Gamma_{res}$,
        and the resulting configuration is well-formed by
        \rn{cmd-dyn-ty-expr}, with \rn{cmd-ty-sub} with
        \lref{call:ass:sub} and \rn{cmd-ty-let}.
      \item The effect is well-formed by \rn{wf-eff-bind} with \lref{ass:wfargs}.
      \end{itemize}

    \item \rn{cmd-ec}\\
      Preservation holds by \rn{cmd-dyn-ty-let} with the induction
      hypothesis and \lref{ass:wfe}.
    \end{itemize}
  \end{pfsubcases}

  \pfcase{\rn{cmd-dyn-ty-let}, $de = \kw{let}~x=db~\kw{in}~e$}
  \begin{enumerate}[label=A\arabic*]
  \item $\Gammas = \many{\Gamma}' :: \Gamma$
  \item $\many{\Gamma}' :: \Gamma \vdash db : t' \dashv \Gamma''$ \llabel[pf]{ass:wfb}
  \item $\Gamma'', x : t' \vdash e : t \dashv \Gamma', x : \_$ \llabel[pf]{ass:wfe}
  \end{enumerate}
  We proceed by cases on whether $db$ is in the syntactic category
  of $b$ or not.
  \begin{itemize}
  \item If it is, the proof collapses to case \lref[pf]{case:let},
    since every non-dynamic typing derivation needs to use
    \rn{cmd-dyn-ty-expr}, which requires a singleton $\many{\Gamma}$.
  \item If it is not, we have either $db = de''$, where $de''$ is
    \emph{not} in $b$, or
    $db = \kw{entered}~y.f~w.\c{val} \SB de'' \FB$. We proceed by
    cases.
    \begin{pfsubcases}
      \pfsubcase{$db = de''$}\\
      Since $de''$ is not in $b$, there are only two dynamic rules
      that apply:
      \begin{itemize}
      \item \rn{cmd-ec}\\
        Preservation holds by \rn{cmd-dyn-ty-let} with the
        induction hypothesis and \lref[pf]{ass:wfe}.

      \item \rn{cmd-ec-fail}\\
        Preservation holds since the resulting expression
        $\SB \mathbf{Failure} \FB$ and the empty effect are
        trivially well-formed.
      \end{itemize}

      \pfsubcase{$db = \kw{entered}~y.f~z.\c{val} \SB de'' \FB$}\\
      By the inversion lemma with \lref[pf]{ass:wfb} we get
      \begin{enumerate}[label=B\arabic*]
      \item $\Gamma = \Gamma_0[y : t_y]$
      \item $\many{\Gamma} = \many{\Gamma}''::\Gamma_1\underset{y.f}{::}\Gamma$
      \item $\many{\Gamma}' \vdash de'' : t' \dashv \Gamma_1'$ \llabel{ass:inner}
      \item $\Kap(\{\Mut, \Tmp, \Var, \Paused\}, t_y)$
      \item $\mathbf{fresult}(t_y, f) = t_f$
      \item $\Kap(\Iso{}, t)$ \llabel{ass:iso}
      \item $\Kap(\{\Iso{}, \Imm{}\}, t')$ \llabel{ass:isoimm}
      \item $t' <: t$
      \item $\Gamma_1'(z) = t_z$
      \item $\mathbf{fresult}(t_z, \c{val}) = t''$
      \item $\Kap(\Mut{}, t'')$ \llabel{ass:mut}
      \item $\Kap(\{\Tmp{}, \Var{}\}, t_z)$\llabel{ass:tmpvar}
      \item $\Gamma' = \Gamma_0[y : t_y']$
      \item$t_y' =
               \begin{cases}
                 \mathit{make\_cell}(\mathit{make\_iso}(t'')) & \text{if } t_y = \mathit{make\_cell}(\_)\\
                 t_y & \text{otherwise}
               \end{cases}$ \llabel{ass:cases}

      \end{enumerate}
      There are two rules that step the configuration:
      \begin{itemize}
      \item \rn{cmd-exit}
        \begin{enumerate}[resume*]
        \item $de'' = \Use$
        \item $x~\textit{fresh}$
        \item $de' = e[x\mapsto x']$
        \end{enumerate}
        \begin{itemize}
        \item The resulting configuration is well-formed by the
          alpha conversion property with \lref[pf]{ass:wfe}.
        \item Since the $de''$ is non-dynamic, we have
          $\many{\Gamma}'' = \varepsilon$.
        \item The effect is well-formed by induction over the
          shape of $t_z$. In the inductive cases we use
          \rn{wf-eff-split}. In the base case we use
          \rn{wf-eff-exit} with \lref{ass:inner},
          \lref{ass:isoimm}, \lref{ass:mut}, \lref{ass:iso},
          \lref{ass:tmpvar} and \lref{ass:cases}.
        \end{itemize}

      \item \rn{cmd-ec}\\
        Preservation holds by induction on the shape of $t_z$. In
        the inductive cases we use \rn{wf-eff-split}. In the base
        case we use \rn{cmd-dyn-ty-let} with
        \rn{cmd-dyn-ty-entered} with the assumptions given by the
        inversion lemma.
        Note that the induction hypothesis only gives
        well-formedness for $\Eff$ for $\many{\Gamma}'$, but
        adding $\Gamma'_1$ to the \emph{right} of $\many{\Gamma}'$
        does not affect well-formedness of the effect since the
        well-formedness rules for effects looks at the dynamic
        typing context from the left.
      \end{itemize}
    \end{pfsubcases}
  \end{itemize}
\end{pfcases}

\subsection{Theorem: Progress of Region Semantics}
\label{theorm:region_progress}
\[
  \begin{array}{l}
    \vdash \langle\SB de \FB~\langle\RS; H_{op}; H_{cl}; H_{fr}\rangle \land \SB de \FB \xrightarrow{\Eff} \SB de' \FB \implies\\
    \quad\exists \Eff', \SB de \FB \xrightarrow{\Eff'} \SB de' \FB ~\land~ \langle\RS; H_{op}; H_{cl}; H_{fr}\rangle \xrightarrow{\Eff'} \langle\RS'; H'_{op}; H'_{cl}; H'_{fr}\rangle
  \end{array}
\]

\subsubsection{Proof Sketch}
The proof of progress ensures that if the command semantics
produces a well-formed effect, there is an effect that steps both
the command semantics and the region semantics. In practice the
two effects will only be different for dynamic type tests, where
the command semantics always allows both effects $\mathit{cast}$
and $\mathit{nocast}$ with any type argument, and when
\kw{enter}ing a region, where the command semantics always allows
both the effects $\mathit{enter}$ and $\mathit{badenter}$. The
region semantics is always able to ``select'' an effect that works
for both semantics.
The rest of the proof is straightforward and comes down to the
fact that looking up a well-defined name in a well-formed state
always succeeds.

\subsubsection{Proof of Theorem}
\pf

Whenever we have a well-formed $\Use$ of a variable, the
corresponding $\Get$ in the region semantics produces a value of
the corresponding type, and looking up that value in the region
stack or heap produces an object of that type. We call this the
well-formed lookup property and it follows from lemmas
\ref{lemma:get_total}, \ref{lemma:wf_get} and
\ref{lemma:wf_cfg_load}.
We will sometimes use the $\CapOK$ property to narrow where we are
looking for an object. For example, we know that a \Mut{}
reference points to an open heap region.
Similarly, an update of an existing object via a $\mathsf{store}$,
$\mathsf{stack\_store}$ or $\mathsf{heap\_store}$ always succeeds.
We call this property well-formed stores.

We start the proof by inversion of \rn{tandem-wf} and get
$\many{\Gamma} \vdash \SB de \FB$. By preservation of the command
language we get $\many{\Gamma} \vdash \Eff \dashv \many{\Gamma}'$.
We proceed by induction over the well-formed effect judgment.

\begin{pfcases}
  \pfcase{\rn{wf-eff-eps}}, $\Eff = \epsilon$\\
  The configuration steps trivially by \rn{region-eps}.

  \pfcase{\rn{wf-eff-load}}, $\Eff = \load{x, y.f}$
  \begin{enumerate}[label=A\arabic*]
  \item $\many{\Gamma} = \Gamma :: \many{\Gamma}_2$\llabel{ass:gammas}
  \item $x~\mathit{fresh}$\llabel{ass:fresh}
  \item $\Gamma(y) = k~CL$\llabel{ass:y}
  \item $\mathbf{ftype}(CL, f) = t$\llabel{ass:ftype}
  \item $k \neq \Iso{}$\llabel{ass:niso}
  \end{enumerate}
  \begin{itemize}
  \item By \rn{wf-rs-cons} with \lref{ass:gammas} we have
    $\RS = (r, S, F) :: \RS'$.
  \item By well-formed lookup with \lref{ass:y} we get
    $F(y) = (k, \iota)$.
  \item Since only \Iso{} references can point to closed regions
    by \CapOK{}, by \lref{ass:niso} and well-formed lookup we have
    $\CfgLoad((r, S, F) :: \RS', H_{op} \c{*} H_{fr}, \iota) = o[f
    \mapsto v]$.
  \item The configuration steps with $\Eff$ by \rn{region-load} with $\Eff$.
  \end{itemize}

  \pfcase{\rn{wf-eff-swap-class}}, $\Eff = \swap{x, y.f, \Use}$\llabel[pf]{case:swap}
  \begin{enumerate}[label=A\arabic*]
  \item $\many{\Gamma} = \Gamma :: \many{\Gamma}_2$\llabel{ass:gammas}
  \item $x~\mathit{fresh}$\llabel{ass:fresh}
  \item $\Gamma \vdash \Use : t \dashv \Gamma'$\llabel{ass:use}
  \item $\Gamma'(y) = k~CL$\llabel{ass:y}
  \item $\mathbf{ftype}(CL, f) = t$\llabel{ass:ftype}
  \item $k \in \{\Mut{}, \Tmp{}\}$\llabel{ass:k}
  \end{enumerate}
  \begin{itemize}
  \item By \rn{wf-rs-cons} with \lref{ass:gammas} we have
    $\RS = (r, S, F) :: \RS'$.
  \item By well-formed lookup with \lref{ass:use} we get
    $\Get(F, \Use) = (v, F')$ with a well-formed $F'$ under
    $\Gamma'$.
  \item By well-formed lookup with \lref{ass:y} we get
    $F'(y) = (k, \iota)$.
  \item We proceed by cases on \lref{ass:k}
    \begin{pfsubcases}
    \pfsubcase{$k = \Mut{}$}
      \begin{itemize}
      \item By \CapOK{} $\iota$ points into region $r$ on the heap
        and we have $H_{op} = (r, S')$.
      \item By well-formed lookup we have
        $\mathsf{load}(S', \iota) = o[f \mapsto v']$.
      \item By well-formed stores we have
        $\mathsf{store}(S', \iota, o[f \mapsto v]) = S''$.
      \item The configuration steps by \rn{region-swap-heap} with $\Eff$.
      \end{itemize}

    \pfsubcase{$k = \Tmp{}$}\llabel[pf]{case:swap_tmp}
      \begin{itemize}
      \item By \CapOK{} $\iota$ points into $S$ on the region
        stack.
      \item By well-formed lookup we have
        $\mathsf{load}(S, \iota) = o[f \mapsto v']$.
      \item By well-formed stores we have
        $\mathsf{store}(S, \iota, o[f \mapsto v]) = S'$.
      \item The configuration steps by \rn{region-swap-temp} with $\Eff$.
      \end{itemize}
    \end{pfsubcases}
  \end{itemize}

  \pfcase{\rn{wf-eff-swap-var}}, $\Eff = \swap{x, y.\c{val}, \Use}$\\
  Similar to \lref[pf]{case:swap} going into \lref[pf]{case:swap_tmp}.

  \pfcase{\rn{wf-eff-halloc-mut}}, $\Eff = \halloc{x, \Mut{}, \#C, \Use_1, ..., \Use_n}$\llabel[pf]{case:halloc_mut}
  \begin{enumerate}[label=A\arabic*]
  \item $\many{\Gamma} = \Gamma_1 :: \many{\Gamma}_2$\llabel{ass:gammas}
  \item $x~\mathit{fresh}$\llabel{ass:fresh}
  \item $\mathbf{ftypes}(C) = f_1 : t_1, ..., f_n : t_n$\llabel{ass:ftypes}
  \item $\forall i \in [1,n]. \Gamma_i \vdash \Use_i : t_i \dashv \Gamma_{i+1}$ \llabel{ass:args}
  \end{enumerate}
  \begin{itemize}
  \item By \rn{wf-rs-cons} with \lref{ass:gammas} we have
    $\RS = (r, S, F_1) :: \RS'$.
  \item By induction over $n$, and by well-formed lookup we have
    $\forall i \in [1,n]. \Get(F_i, \Use_i) = (v_i, F_{i+1})$.
  \item By \lref{ass:ftypes} and the assumptions of the class
    table we have $\mathbf{fields}(C) = f_1, ..., f_n$.
  \item The configuration steps by \rn{region-alloc-heap-mut} with $\Eff$.
  \end{itemize}

  \pfcase{\rn{wf-eff-halloc-iso}}, $\Eff = \halloc{x, \Iso{}, \#C, \Use_1, ..., \Use_n}$\\
  Similar to \lref[pf]{case:halloc_mut}.

  \pfcase{\rn{wf-eff-salloc}}, $\Eff = \salloc{x, k, \#CL, \Use_1, ..., \Use_n}$\\
  Similar to \lref[pf]{case:halloc_mut}, but the allocation happens on
  the region stack instead.

  \pfcase{\rn{wf-eff-freeze}}, $\Eff = \freeze{x,\Use}$\llabel[pf]{case:freeze}
  \begin{enumerate}[label=A\arabic*]
  \item $\many{\Gamma} = \Gamma :: \many{\Gamma}_2$\llabel{ass:gammas}
  \item $x~\mathit{fresh}$\llabel{ass:fresh}
  \item $\Gamma \vdash \Use : t \dashv \Gamma'$\llabel{ass:use}
  \item $\Kap(\Iso{}, t)$\llabel{ass:iso}
  \end{enumerate}
  \begin{itemize}
  \item By \rn{wf-rs-cons} with \lref{ass:gammas} we have
    $\RS = (r, S, F) :: \RS'$.
  \item By well-formed lookup with \lref{ass:use} we get
    $\Get(F, \Use) = ((k, \iota), F')$.
  \item By \CapOK{} an \Iso{} reference in a variable, like
    $\iota$, must point into a closed region $R$, so that
    $H_{cl} = R \c{*} H'_{cl}$.
  \item By \CapOK{}, outgoing references from a closed region
    cannot go into an open region. This means that any regions
    reachable from $R$ must also be closed, allowing us to write
    $H_{cl}$ as $R \c{*} (H \c{*} H''_{cl})$ where
    $H = \mathsf{reachable\_regions}(R, (H \c{*} H''_{cl})\c{*}
    H_{op})$.
  \item The configuration steps by \rn{region-freeze} with $\Eff$.
  \end{itemize}

  \pfcase{\rn{wf-eff-merge}}, $\Eff = \merge{x,\Use}$\\
  Similar to \lref[pf]{case:freeze}.

  \pfcase{\rn{wf-eff-bind}}, $\Eff = \bind{x_1 = \Use_1, ..., x_n = \Use_n}$
  \begin{enumerate}[label=A\arabic*]
  \item $\many{\Gamma} = \Gamma_1 :: \many{\Gamma}_2$\llabel{ass:gammas}
  \item $\forall i \in [1,n]. x_i~\mathit{fresh}$\llabel{ass:fresh}
  \item $\forall i \in [1,n]. \Gamma_i \vdash \Use_i : t_i \dashv \Gamma_{i+1}$ \llabel{ass:uses}
  \end{enumerate}
  \begin{itemize}
  \item By \rn{wf-rs-cons} with \lref{ass:gammas} we have
    $\RS = (r, S, F) :: \RS'$.
  \item By induction over $n$, and by well-formed lookup we have
    $\forall i \in [1,n]. \Get(F_i, \Use_i) = (v_i, F_{i+1})$.
  \item The configuration steps by \rn{region-bind} with $\Eff$.
  \end{itemize}

  \pfcase{\rn{wf-eff-cast}}, $\Eff = \cast{x, \Use, k~C}$
  \begin{enumerate}[label=A\arabic*]
  \item $\many{\Gamma} = \Gamma :: \many{\Gamma}_2$\llabel{ass:gammas}
  \item $x~\mathit{fresh}$\llabel{ass:fresh}
  \item $\Gamma \vdash \Use : t \dashv \Gamma'$\llabel{ass:use}
  \end{enumerate}
  \begin{itemize}
  \item By \rn{wf-rs-cons} with \lref{ass:gammas} we have
    $\RS = (r, S, F) :: \RS'$.
  \item By well-formed lookup with \lref{ass:use} we get
    $\Get(F, \Use) = ((k, \iota), F')$ with a well-formed $F'$
    under $\Gamma'$.
  \item By well-formed lookup we get
    $\CfgLoad((r, S, F') :: \RS', H_{op} \c{*} H_{cl} \c{*}
    H_{fr}, \iota) = (\#C', \_)$.
  \item We proceed by cases on whether $k C' <: t$ or not
    \begin{itemize}
    \item If $k C' <: t$, the command configuration steps by
      \rn{cmd-if-typetest-true} with effect $\cast{x, \Use, k~C}'$,
      and the region configuration steps by \rn{region-cast} with
      the same effect.
    \item If $k C' \not<: t$, the command configuration steps by
      \rn{cmd-if-typetest-false} with effect
      $\nocast{x, \Use, k~C'}$, and the region configuration steps
      by \rn{region-nocast} with the same effect.
    \end{itemize}
  \end{itemize}

  \pfcase{\rn{wf-eff-nocast}}, $\Eff = \nocast{x, \Use, k~C}$\\
  Similar to the previous case.

  \pfcase{\rn{wf-eff-enter}}, $\Eff = \enter{w, k, y.f, x_1 = \Use_1, ..., x_n = \Use_n}$
  \begin{enumerate}[label=A\arabic*]
  \item $\many{\Gamma} = \Gamma :: \many{\Gamma}_2$\llabel{ass:gammas}
  \item $w~\mathit{fresh}$\llabel{ass:fresh}
  \item $\forall i \in [1,n]. x_i~\mathit{fresh}$\llabel{ass:freshs}
  \item $\forall i \in [1,n]. \Gamma_i \vdash \Use_i : t_i \dashv \Gamma_{i+1}$ \llabel{ass:uses}
  \item $\Gamma_{n+1}(y) = k'~CL$ \llabel{ass:yty}
  \item $k' \in \{\Mut, \Tmp, \Var, \Paused\}$ \llabel{ass:open}
  \item $\mathbf{ftype}(CL, f)) = t$ \llabel{ass:ftype}
  \item $\Kap(\Iso, t)$ \llabel{ass:iso}
  \item $k \in \{\Tmp, \Var\}$\llabel{ass:k}
  \end{enumerate}
  \begin{itemize}
  \item By \rn{wf-rs-cons} with \lref{ass:gammas} we have
    $\RS = (r, S, F) :: \RS'$.
  \item By induction over $n$, and by well-formed lookup we have
    $\forall i \in [1,n]. \Get(F_i, \Use_i) = (v_i, F_{i+1})$.
  \item By well-formed lookup we have $F_{n+1}(y) = (k', \iota)$.
  \item By \CapOK{} references which are $\mathbf{open}$ point to an
    open region on the stack or heap. By well-formed lookup and by
    \lref{ass:open} we have
    $\CfgLoad((r, S, F') :: \RS', H_{op}, \iota) = o[f \mapsto (\_, \iota')]$.
  \item We proceed by cases on whether $\iota'$ is in the domain
    of any closed region.
    \begin{itemize}
    \item If it is, the configuration steps by \rn{region-enter-ok}.
    \item If it is not, the configuration steps by \rn{region-enter-fail}.
    \end{itemize}
  \end{itemize}

  \pfcase{\rn{wf-eff-badenter}}\\
  Similar to the previous case, noting that any destructive reads
  will not have made $y$ undefined, as this would not have been
  well-typed in the command semantics.

  \pfcase{\rn{wf-eff-exit}}
  \begin{enumerate}[label=A\arabic*]
  \item $\many{\Gamma} = \Gamma' :: \Gamma[y : k~CL]:: \many{\Gamma}_2$\llabel{ass:gammas}
  \item $x~\mathit{fresh}$\llabel{ass:fresh}
  \item $k \in \{\Mut, \Tmp, \Var, \Paused\}$ \llabel{ass:open}
  \item $\Gamma' \vdash \Use : t \dashv \Gamma''$ \llabel{ass:use}
  \item $\Gamma''(w) = k'~CL'$ \llabel{ass:wty}
  \item $\mathbf{ftype}(CL', g)) = t'$ \llabel{ass:gtype}
  \end{enumerate}
  \begin{itemize}
  \item By \rn{wf-rs-cons} with \lref{ass:gammas} we have
    $\RS = (r', S', F') :: (r, S, F) :: \RS'$.
  \item By well-formed lookup with \lref{ass:use} we have
    $\Get(F', \Use) = (v, F'')$.
  \item By well-formed lookup with \lref{ass:wty} we have
    $F''(z) = (\_, \iota')$ and
    $\mathsf{load}(S', \iota') = o'[f' \mapsto(\_, \iota'')]$.
  \item By well-formed lookup with \lref{ass:gammas} we have
    $F(y) = (\_, \iota)$.
  \item Since $y$ is $\mathbf{open}$ by \lref{ass:open}, by
    \CapOK{} $\iota$ points to the open heap or the region stack.
    We proceed by cases.
    \begin{itemize}
    \item If $\iota$ points to the stack we have
      $\mathsf{stack\_load}(r, S, F) :: \RS, \iota = o[f \mapsto
      (k, \_)]$, and by well-formed stores
      $\mathsf{stack\_store}((r, S, F) :: \RS, \iota, o[f \mapsto
      (k, \iota'')]) = (r, S'', F)::\RS'$. The configuration steps
      by \rn{region-exit-temp}.
    \item If $\iota$ points to the heap we have
      $\mathsf{heap\_load}(H_{op}, \iota) = o[f \mapsto
      (k, \_)]$, and by well-formed stores that
      $\mathsf{heap\_store}(H_{op}, \iota, o[f \mapsto
      (k, \iota'')]) = H'_{op}$. The configuration steps by
      \rn{region-exit-heap}.
    \end{itemize}
  \end{itemize}

  \pfcase{\rn{wf-eff-split}}
  \begin{enumerate}[label=A\arabic*]
  \item $\many{\Gamma} = \many{\Gamma}_0[x : t|t']$
  \item $\many{\Gamma}_0[x : t] \vdash \Eff \dashv \many{\Gamma}_1$
  \item $\many{\Gamma}_0[x : t'] \vdash \Eff \dashv \many{\Gamma}_1$
  \end{enumerate}
  \begin{itemize}
  \item By \rn{wf-vars-cons}, the type $t_x$ of the value of $x$
    must be a subtype of $t|t'$.
  \item By the subtyping rules, we must have $t_x <: t$ \emph{or}
    $t_x <: t'$. We proceed by cases.
  \item Without loss of generality, we assume $t_x <: t$ (the other case is symmetric).
  \item We have
    $\many{\Gamma}_0[x : t] \vdash \langle\RS; H_{op}; H_{cl}; H_{fr}\rangle$
  \item Progress holds by the induction hypothesis.
  \end{itemize}

\end{pfcases}

\subsection{Theorem: Preservation of Region Semantics}
\label{theorm:region_preservation}
The region semantics has a property similar to preservation that we express like
this.

\[
  \begin{array}{l}
    \many{\Gamma}; \Delta; \Psi \vdash \RCfg \AN
    \CapOK(\RCfg) \AN
    \TopOK(\RCfg) \AN
    \RCfg \RegStep{\Eff} \RCfg' \AN
    \many{\Gamma} \vdash \Eff \dashv \many{\Gamma}'
    \implies \\
    \quad\exists \Delta', \Psi'.\\
    \qquad\many{\Gamma}'; \Delta'; \Psi' \vdash \RCfg' \AN
          \CapOK(\RCfg') \AN
          \TopOK(\RCfg')
  \end{array}
\]

\subsubsection{Proof Sketch}

This is by far the most complex proof for this system. We proceed by induction
on the well-formedness relation for $\Eff$. The main problem here is the base
cases, for which we first prove well-typedness and then the invariants $\capok$ and
$\topok$. The proof of well-typedness is done by a standard induction
argument on the well-typedness relations for each part of the configuration. For
the invariants we argue about the restrictions of each reference in the object
graph, with regards to its capability.

Each step of the region configuration will have a corresponding action
(removal/addition of locations and references) on the region graph. Furthemore,
because of well-typedness we can express this action using only the object graph
$G(\RCfg)$, the region ordering $\rho(\Gammas, \RCfg)$  and $\Eff$. We use this
fact and the constraints imposed by the reference capabilities to argue that the
invariants hold in the resulting configuration.

\subsubsection{Proof of Theorem}
\pf

\begin{enumerate}[label=\textbf{Main \arabic*}]
\item $\RCfg \RegStep{\Eff} \RCfg'$
\item $\many{\Gamma} \vdash \Eff \dashv \many{\Gamma}'$
\item \llabel[pf]{wf_rcfg}$\many{\Gamma}; \Delta; \Psi \vdash \RCfg$
\item \llabel[pf]{capok} $\CapOK(\RCfg)$
\item \llabel[pf]{topok} $\TopOK(\RCfg)$
\end{enumerate}

We proceed by structural induction on the derivation of $\Gammas \vdash \Eff
\dashv \Gammas'$

\begin{pfcases}
\pfcase{\rn{wf-eff-cast}}
\llabel[pf]{case_eff_cast}

  From \rn{wf-eff-cast} we get
  \begin{enumerate}[label=A\arabic*, align=left]
  \item $\Eff = \cast{x, use, k~CL}$
  \item $\many{\Gamma} = \Gamma :: \many{\Gamma}_0$
  \item \llabel{gammausegamma} $\Gamma \vdash use : t \dashv \Gamma'$
  \item $k~C <: t$
  \item $\many{\Gamma}' = \Gamma', x : t :: \many{\Gamma}_0$
  \end{enumerate}
  From \rn{region-step-cast}
  \begin{enumerate}[resume*]
  \item $\RCfg = \RCfglong{(r, S, F)::RS; H_{op}; H_{cl}; H_{fr}}$
  \item $\Get( \Use, F ) = ((k, \iota), F')$
  \item $\CfgLoad((r, S, F)::RS, \Hop * \Hcl * \Hfr, \iota) = (C, \_)$
  \item $F'' = F', x \mapsto (k, \iota)$
  \item \llabel{newrcfg} $\RCfg' = \RCfglong{(r, S, F'')::RS; \Hop; \Hcl; \Hfr}$
  \end{enumerate}
  We apply lemma \ref{lemma:wf_get}, and get
  \begin{enumerate}[resume*]
  \item \llabel{a1} $\Delta; \Gamma' \vdash F'$
  \item \llabel{a2} $\iota : \CL \in \Delta$
  \item \llabel{a3} $k \CL <: t$
  \end{enumerate}
  We apply \rn{wf-F-cons} on \lref{a1}, \lref{a2}, \lref{a3} to
  get
  \begin{enumerate}[resume*]
  \item $\Delta; \Gamma', x : t \vdash F''$
  \end{enumerate}
  The rest of the configuraton remains unchanged. Therefore we can conclude that
  \begin{enumerate}[resume*]
  \item $\Gamma' :: \many{\Gamma}_0; \Delta; \Psi \vdash \RCfg'$
  \end{enumerate}
  We now move to prove $\CapOK(\RCfg')$ and $\TopOK(\RCfg')$. Given that
  $G(\RCfg) = \Graph = (\LocSet, \RefSet)$, $G(\RCfg') = \Graph' = (\LocSet', \RefSet')$ it follows
  from \ref{lemma:wd_graph_act} that
  \begin{enumerate}[resume*]
  \item $\LocSet' = \LocSet$
  \item
      $\Graph' = \begin{cases}
                 \Graph + (\Root(r) \RefTo{x, k} \Loc_z) & \text{ if } \Use = z \\
                 \Graph - (\Root(r) \RefTo{z, k} \Loc_z)  +  (\Root(r) \RefTo{x,
                   k} \Loc_z) & \If \Use = \Drop z
               \end{cases}$
  \item \llabel{refgraph} $\REF_\Graph(\Root(r), z) = (\Root(r) \RefTo{z, k} \Loc_z)$
  \item $\Loc_\Graph(\iota) = \Loc_z$
  \end{enumerate}
  By \lref[case]{newrcfg}, \lref[pf]{capok}, \lref{refgraph}
  \begin{enumerate}[resume*]
  \item $\rho = \rho(\Gammas, \RCfg) = \rho(\Gammas, \RCfg')$
  \item $\Cl = \Closed(\RCfg) = \Closed(\RCfg')$
  \item $\Fr = \Frozen(\RCfg) = \Frozen(\RCfg')$
  \item \llabel{rordold} $\RegionOrder1(\rho, \Cl, \Fr, \Root(r) \RefTo{z, k} \Loc_z)$
  \item \llabel{locokold} $\LocationOK1(\Root(r) \RefTo{z, k} \Loc_z)$
  \item \llabel{varuniqold} $\VarUnique(G(\RCfg))$
  \end{enumerate}
  We proceed by cases of the value of $k$
  \begin{pfsubcases}
  \pfsubcase{$k = \Mut$}
    \llabel[case]{scasemut}

    \begin{itemize}
    \item $\varunique(\Graph')$ because of \lref[pf]{obs:varuniq_neq_var}.
    \item $\regionorder(\rho, \Cl, \Fr, \Graph')$ By \lref[case]{rordold},
      $\RegionOrder1(\rho, \Cl, \Fr, \Root(r)) \RefTo{x,
        k} \Loc_z)$. It follows.
    \item $\locationok(\Graph')$ By \lref[case]{locokold},
      $\LocationOK1(\Root(r) \RefTo{x, k} \Loc_z)$. It follows.
    \item $\deepfreeze(\Fr, \Graph')$ since no frozen regions have changed.
    \item $\topokgraph(\rho, \Fr, \Graph')$ since $k = \mut$ implies
      $\rho \vdash \Rid(loc_z) \leq r$.
    \item $\entrypoints(\rho, \Graph')$ since we are not affecting entry
      points.
    \end{itemize}
  \pfsubcase{$k \in \{\Paused, \Tmp\}$} Similar to \lref[case]{scasemut}.
  \pfsubcase{$k = \Imm$}
    Similar to \lref[case]{scasemut}, observing that $\Frozen(\RCfg) =
    \Frozen(\RCfg')$ and that $\CapOK(\RCfg)$ implies $\Rid(\Loc_z) \in \Frozen(\RCfg)$.
  \pfsubcase{$k = \Var$}
    Similar to \lref[case]{scasemut}, with \lref[pf]{obs:varuniq_swap} to get
    $\varunique(\Graph')$
  \pfsubcase{$k = \Iso$}
    Similar to previous cases, using \lref[pf]{obs:topok_swap} to get
    $\topokgraph(\rho, \Graph')$
  \end{pfsubcases}
\pfcase{\rn{wf-eff-nocast}} Similar to \lref[pf]{case_eff_cast}.
\pfcase{\rn{wf-eff-split}}
\llabel[pf]{case_eff_split}
  By rule \rn{wf-eff-split}
  \begin{enumerate}[label=A\arabic*]
  \item $\Gammas = \Gammas[x : t | t']$
  \item $\Gammas[x : t] \vdash \Eff \dashv \Gammas_1$
  \item $\Gammas[x : t'] \vdash \Eff \dashv \Gammas_2$
  \item $\Gammas' = \Gammas_1 | \Gammas_2$
  \end{enumerate}
  By \lref[pf]{obs:wf_rcfg_split}, and WLOG
  \begin{enumerate}[resume*]
    \item $\Gammas[x : t] \vdash \RCfg$
  \end{enumerate}
  By induction hypothesis
  \begin{enumerate}[resume*]
  \item \llabel{gammas1rcfg} $\Gammas_1 \vdash \RCfg'$
  \item $\CapOK(\RCfg')$
  \item $\TopOK(\RCfg')$
  \end{enumerate}
  By $\Gammas[x : t] \gammadefincl \Gammas[x : t']$,
  \lref[pf]{obs:wf_eff_result_subgamma}
  \begin{enumerate}[resume*]
  \item \llabel{domgammas} $\Gammas_1 \gammadefincl \Gammas_2$
  \end{enumerate}
  By \lref{domgammas}, \lref{gammas1rcfg}, \lref[pf]{obs:wf_rcfg_gammas_extend}
  \begin{enumerate}[resume*]
  \item $\Gammas_1 | \Gammas_2 \vdash \RCfg'$.
  \end{enumerate}

\pfcase{\rn{wf-eff-enter}}
\llabel[pf]{case_eff_enter}
  By \rn{wf-eff-enter}
  \begin{enumerate}[label=A\arabic*]
  \item $\Eff = \enter{w, k, y.f, x_1 = \Use_1, \dots, x_n = \Use_n}$
  \item $\Gammas = \Gammas_1 :: \Gammas^*$
  \item $\forall i \in [1, n]. \Gamma_i \vdash \Use_i \dashv \Gamma_{i+1}$
  \item $\Gamma_{n+1}(y) = k' \CL$
  \item $\open(k')$
  \item $\FType(CL, f) = t$
  \item \llabel{kapt} $\Kap(iso, t)$
  \item $k \in \{\Tmp, \Var\}$
  \item $k = \Var \implies k' = \Var$
  \item $t' = \Hmakemut(t)$
  \item \llabel{typevpa}
    $
    \forall i \in [1,n]. t_i' =
    \begin{cases}
      t_i & \text{ if } \Kap(iso, t_i) \\
      \VPA{\Paused}{t_i} & \text{ otherwise}
    \end{cases}
    $
  \item \llabel{gammadef} $\Gamma = x_1 : t_1', \dots, x_n : t_n'$
  \item \llabel{gammapdef} $\Gamma' = \Gamma, w : k~\Cell[t']$
  \item $\Gammas' = \Gamma :: \Gamma_{n+1} :: \Gammas^*$
  \end{enumerate}
  By \rn{region-enter-ok}
  \begin{enumerate}[resume*]
  \item $\RCfg = \RCfglong{(r, S, F_1)::\RS; \Hop; (r', S') * \Hcl; \Hfr}$
  \item $w$ fresh
  \item $\forall i \in [1,n]. x_i \text{ fresh}$
  \item $\forall i \in [1,n]. \Get(F, \Use_i) = ((k_i, \iota_i), F_{i+1})$
  \item \llabel{valvpa} $\forall i \in [1,n] (k_i', \iota_i) =
    \begin{cases}
      (k_i, \iota_i) & \If k_i = \Iso \\
      (\VPA{\Paused}{k_i}, \iota_i)) & \Ow
    \end{cases}$
  \item \llabel{fdef} $F = [x_i \mapsto (k_i', \iota) \suchthat i \in [1, n]]$
  \item $F_{n+1}(y) = (k_{\mathsf{here}}, \iota)$
  \item \llabel{cfgload} $\CfgLoad((r, S, F_{n+1})::RS, \Hop, \iota) = o[f \mapsto (k_{\mathsf{bridge}}, \iota')]$
  \item $\iota' \in \dom(S')$
  \item $\iota''$ fresh
  \item $F' = F, w \mapsto (k, \iota'')$
  \item $RF = (r', [\iota'' \mapsto (\Cell, [\Valfield \mapsto (\Mut,
    \iota')])], F')$
  \item $\RCfg' = \RCfglong{RF::(r, S, F_{n+1})::\RS; (r', S') * \Hop; \Hcl; \Hfr}$
  \end{enumerate}
  By repeated application of lemma \ref{lemma:wf_get}
  \begin{enumerate}[resume*]
  \item \llabel{wffn1}$\Delta; \Gamma_{n+1} \vdash F_{n+1}$
  \item $\iota_i : CL \in \Delta$
  \item \llabel{deltasubtype} $k_i \CL_i <: t_i$
  \end{enumerate}

  For each $i$ we split into case on $k_i$:
  \begin{pfsubcases}
    \pfsubcase{$k = \Iso$} This, \lref[case]{deltasubtype}, \lref[case]{typevpa}, and \lref[pf]{obs:iso_subtype_implies_vpa_undefined} implies that
    \begin{enumerate}[label=B\arabic*]
    \item $\Kap(\Iso, t_i)$
    \end{enumerate}
    By \lref[case]{typevpa} and \lref[case]{valvpa},
    \begin{enumerate}[resume*]
    \item \llabel{a} $k_i' = k_i$
    \item \llabel{b} $t_i' = t_i$
    \end{enumerate}
    Thus by \lref[case]{deltasubtype}, \lref{a}, \lref{b}
    \begin{enumerate}[resume*]
    \item $k_i' CL <: t_i'$
    \end{enumerate}
    \pfsubcase{$k \neq \Iso$} Similarly to first first subcase, \lref[case]{deltasubtype},
    \lref[case]{typevpa}, \lref[pf]{obs:iso_subtype_implies_vpa_undefined}
    implies
    \begin{enumerate}[label=B\arabic*]
    \item \llabel{a} $k_i' = \VPA{\Paused}{k_i}$
    \item \llabel{b} $t_i' = \VPA{\Paused}{t_i}$
    \end{enumerate}
    Specifically $t_i'$ is well-defined, and thus, by \lref{a}, \lref{b},
    \lref[case]{deltasubtype}, \lref[pf]{obs:vpa_subtype}
    \begin{enumerate}[resume*]
    \item $k_i' CL <: t_i'$
    \end{enumerate}
  \end{pfsubcases}
  \pfsetscope{case}

  We conclude that
  \begin{enumerate}[label=C\arabic*]
  \item \llabel{primesubtypes} $\forall i \in [1, n]. k_i'~CL <: t_i'$
  \end{enumerate}
  Starting with the $\Delta; \NilGamma \vdash \NilF$, by repeated application of
  \rn{wf-F-cons}, \lref{gammadef}, \lref{fdef}, \lref[case]{primesubtypes}
  \begin{enumerate}[resume*]
  \item $\Delta; \Gamma \vdash F$
  \end{enumerate}
  By \lref[pf]{obs:expanding_delta}
  \begin{enumerate}[resume*]
  \item \llabel{deltapdef} $\Delta' = \Delta, \iota'' : \Cell[t']$
  \item \llabel{wffconsrec} $\Delta'; \Gamma \vdash F$
  \end{enumerate}
  By \rn{wf-F-cons}, \lref{wffconsrec}, $k~\Cell[t'] <: k~\Cell[t']$
  \begin{enumerate}[resume*]
  \item \llabel{wfrff} $\Delta'; \Gamma' \vdash F'$
  \end{enumerate}

  By \lref{cfgload} and well-formedness of $\RCfg$
  \begin{enumerate}[resume*]
  \item $o[f \mapsto (k_{\mathsf{bridge}}, \iota')] = (\CLTag, F^*)$
  \item $\iota : \CL \in \Delta$
  \item $\Delta; \FTypes(\CL) \vdash F^*$
  \item $\iota' : \CL' \in \Delta$
  \item \llabel{ref1} $k_{\mathsf{bridge}} \CL' <: t$
  \end{enumerate}
  By \lref{kapt} and \lref{ref1}
  \begin{enumerate}[resume*]
  \item $k_{\mathsf{bridge}} = \Iso$
  \end{enumerate}
  By \lref{ref1}, \lref[pf]{obs:make_mut_subtype},
  \begin{enumerate}[resume*]
  \item \llabel{subtypetp} $\Mut \CL' <: \Hmakemut(t) = t'$
  \end{enumerate}
  By \lref{deltapdef}, \rn{cltag-cell}, \lref{subtypetp}
  \begin{enumerate}[resume*]
  \item \llabel{bunch1}$\CellTag \subtag \Cell[t']$
  \item $\iota'' : \Cell[t'] \in \Delta'$
  \item $\Delta'; \Valfield : t' \vdash \Valfield \mapsto (\Mut, \iota')$
  \item \llabel{bunch2}$\Delta; \NilDelta \vdash \NilS$
  \end{enumerate}
  By \lref{bunch1} - \lref{bunch2}, \rn{wf-subheap-cons},
  \begin{enumerate}[resume*]
  \item \llabel{wfrfs} $\Delta; \iota'' : \Cell[t'] \vdash \iota'' \mapsto (\CellTag, \Valfield
    \mapsto (\Mut, \iota')$
  \end{enumerate}
  From \lref{wfrfs}, \lref{wfrff}, \lref{wffn1}, and \lref[pf]{wf_rcfg},
  it follows that there is $\Delta_\RS, \Psi_\RS$ such that
  \begin{enumerate}[resume*]
    \item $\Gamma' :: \Gammas_1; \Delta'; \Delta_\RS, \iota'' : Cell[t'];  r', \Psi_\RS \vdash \RF :: (r, S, F_{n+1}) :: \RS$
  \end{enumerate}
  By applying rules for well formedness, we end up with
  \begin{enumerate}[resume*]
    \item $\Gammas'; \Delta'; \Psi \vdash \RCfg'$
  \end{enumerate}

  Now, we move on to $\CapOK(\RCfg')$.

  We write $\Graph = G(\RCfg)$ and $\Graph' = G(\RCfg')$. Because of
  well-formedness of $\RCfg$ and $\RCfg'$, these are well-formed.

  We write
  \begin{itemize}
  \item $\rho = \rho(\Gammas, \RCfg) = r \rhosepe \rho^*$
  \item $\rho' = \rho(\Gammas', \RCfg') = r' \rhosep{y.f} r \rhosepe \rho^*$
  \end{itemize}

  We let $\Graph_i$ be defined as follows
  \begin{itemize}
  \item $\Graph_0 = \Graph + \ruut(r') + \temp(r', \iota'')$
  \item $\REF_i = \ruut(r) \refto{z_i, k_i} \loc_i = \REF_{\Graph_i}(\ruut(r), z_i)$
  \item $\REF_i' = \ruut(r') \refto{x_i, k_i'} \loc_i$
  \item
    $k_i' =
    \begin{cases}
      k_i & \If k_i = \iso \\
      \paused \sees k_i & \Ow
    \end{cases}
    $
  \item $\Graph_{i+1} = \Graph_i - \REF_i + \REF_i'$ if $\Use_i = \Drop z_i$.
  \item $\Graph_{i+1} = \Graph_i + \REF_i'$ if $\Use_i = z_i$.
  \end{itemize}
  Then, from lemma \ref{lemma:wd_graph_act},
  \begin{enumerate}[resume*]
  \item $\Graph' = \Graph_n + \ruut(r') \refto{w, k} \temp(r', \iota'') +
    \temp(r', \iota'') \refto{\Valfield, \mut} \heap(r', \iota')$
  \end{enumerate}

  We make some general \textbf{observations}:

  If we have $\Graph, \rho, \Cl, \Fr, r$ such that $\vdash \rho, \Cl \cup \{r\},
  \Fr$ and $r \not\in \Cl$
  then
  \begin{enumerate}[label=\textbf{OBS}\arabic*]
  \item If
    \begin{itemize}
    \item $\regionorder1(\rho, \Cl \cup \{r \}, \Fr, \Loc_1[r_1] \RefTo{f, k}
      \Loc_2[r_2])$
    \end{itemize}
    Then
    $\RegionOrder1(r :: \rho, \Cl, \Fr, \Loc_1[r_1] \RefTo{f, k} \Loc_2[r_2])$

    \textbf{Follows} by
    \begin{enumerate}
    \item if $k \neq \Iso$, moving $r$ from the closed set to region stack does
      not affect ordering of $r_1$ and $r_2$
    \item If $k = \Iso$, then either $r_2 = r'$ in which case $\topok(\RCfg)$
      implies $r_1 = r$, thus $r :: \rho \vdash r_1 < r_2$.

      Otherwise, $r_2 \neq r'$ and region order of $r_1$ and $r_2$ will be
      unaffected.
    \end{enumerate}

    This in turn implies
  \item If $\regionorder(\rho, \Cl \cup r, \Fr, \Graph)$, then $\regionorder(r :: \rho,
    \Cl, \Fr, \Graph)$.

  \item If
    \begin{itemize}
    \item $\entrypointsok(r :: \rho, \Graph)$
    \item $r' \not\in r :: \rho$
    \item $\ruut(r) \refto{y, k_y} \loc_y \in \Graph$
    \item $\loc_y \refto{f', k_{f}} \Heap(r', \_) \in \Graph$
    \end{itemize}
    Then $\entrypointsok(r' \rhosep{y.f} r :: \rho, \Graph)$.

    \textbf{Follows} from definition of $\entrypointsok$.
  \item
    $\topokgraph(\rho, \Fr, \Graph)$ implies $\topokgraph(r :: \rho, \Fr, \Graph)$.

    \textbf{Follows} from that the relation on region ids defined by $\rho
    \vdash r1 \leq r2$ is included in the relation defined by $r :: \rho \vdash
    r1 \leq r2$.
  \item\llabel[pf]{obs:enter_region_order_paused}
    If
    \begin{itemize}
    \item $\RegionOrder1(r :: \rho, \Cl, \Fr, \Loc_1[r_1] \RefTo{f, k}
      \Loc_2[r_2])$
    \item $r_1 \in \rho$
    \item $k \neq \Iso$
    \end{itemize}
    Then $\regionorder1(r :: \rho, \Cl, \Fr, \Loc[r] \refto{f', \Paused \sees k}
    \loc_2[r_2])$

    \textbf{Follows} from case analysis on $k$.
  \item\llabel[pf]{obs:enter_region_order_iso}
    If
    \begin{enumerate}
    \item \llabel{shit} $\RegionOrder(r' :: r :: \rho, \Cl, \Fr, \ruut(r) \RefTo{f, \iso}
      \Loc_2[r_2])$
    \item \llabel{piss}$r_2 \neq r'$
    \end{enumerate}
    Then $\RegionOrder(r' :: r :: \rho, \Cl, \Fr, \ruut(r') \RefTo{f', \iso}
    \Loc_2[r_2])$

    \textbf{Follow} by the following argument:
    (i) and capability $\iso$ implies that $r_2 \neq r$. Furthermore
    $r_2 \neq r'$ by (ii), so we do not have $r' :: r :: \rho \vdash r
    < r_2$. Also, $r \not\in \Fr$. Thus $r_2 \in Cl$, and thus the conclusion
    holds.
  \item\llabel[pf]{obs:enter_locationok_paused}
    If
    \begin{enumerate}
    \item $\locationok1(\loc_1 \refto{f, k} \loc_2)$
    \item $k \neq \iso$
    \end{enumerate}
    Then
    \begin{itemize}
      \item $\locationok1(\ruut(r) \refto{f, \paused \sees k} \loc_2)$
    \end{itemize}

    \textbf{Follows} from case analysis on $k$.
  \item\llabel[pf]{obs:enter_locationok_iso}
    Similary, if
    \begin{enumerate}
    \item $\locationok1(\loc_1 \refto{f, \iso} \loc_2)$
    \end{enumerate}
    Then
    \begin{itemize}
      \item $\locationok1(\ruut(r) \refto{f, \iso} \loc_2)$
    \end{itemize}
  \end{enumerate}

  Because of these observations, and since adding locations does not affect
  invariants we have
  \begin{enumerate}[label=E\arabic*]
  \item $\varunique(\Graph_0)$
  \item $\regionorder(\rho', \Cl', \Fr, \Graph_0)$
  \item $\locationok(\Graph_0)$
  \item $\deepfreeze(\Fr, \Graph_0)$
  \item $\entrypoints(\rho', \Graph_0)$ using well-formedness to get $(\ruut(r)
    \refto{y, k'} \loc_y) \in \Graph_0$, $(\loc_y \refto{f, \iso} \heap(r',
    \iota')) \in \Graph_0$ and that $\kw{open}(k')$ implies $\Rid(\loc_y) \in \rho$.
  \item $\topokgraph(\rho', \Fr, \Graph_0)$
  \end{enumerate}

  By an inductive argument on $i$, by cases on $k_i$:
  \begin{pfsubcases}
    \pfsubcase{$k = \var$} Implies $\Use_i = \Drop z_i$, and thus $\Graph_i =
    \Graph_{i-1} - \REF_i + \REF_i'$.
    \begin{itemize}
      \item $\varunique(\Graph_i)$ follows from \lref[pf]{obs:varuniq_swap}.
      \item $\regionorder(\rho', \Cl', \Fr, \Graph_i)$ follows from \lref[pf]{obs:enter_region_order_paused}
      \item $\locationok(\Graph_i)$ follows from
        \lref[pf]{obs:enter_locationok_paused}
      \item $\deepfreeze(\Fr, \Graph_i)$ follows from $r' \not\in \Fr$ and $\deepfreeze(\Graph_{i-1})$.
      \item $\entrypointsok(\rho', \Graph_i)$ since no entry points have changed,
      \item $\topokgraph(\rho', \Fr, \Graph_i)$ follows from $k_i = \paused$ and $\regionorder(\Graph_i)$
    \end{itemize}
    \pfsubcase{$k = \iso$} Implies $\Use_i = \Drop z_i$, and thus $\Graph_i =
    \Graph_{i-1} - \REF_i + \REF_i'$.
    \begin{itemize}
    \item $\varunique(\Graph_i)$ follows from \lref[pf]{obs:varuniq_neq_var}.
    \item $\regionorder(\rho', \Cl', \Fr, \Graph_i)$ follows from $\Rid(\loc_i)
      \neq r'$ \\ (from $\topokgraph(\rho', \Fr, \Graph_i)$) and
      \lref[pf]{obs:enter_region_order_iso}.
      \item $\locationok(\Graph_i)$ follows from \lref[pf]{obs:enter_locationok_iso}
    \item $\deepfreeze(\Graph_i)$ follows from $r' \not\in \Fr$ and $\deepfreeze(\Graph_{i-1})$.
    \item $\entrypointsok(\rho', \Graph_i)$ since no entry points have changed,
    \item $\topokgraph(\rho', \Fr, \Graph_i)$ follows from
      \lref[pf]{obs:topok_swap} and that $\Rid(\loc_i) \in \Cl$ (which
      implies that we do not have $\rho' \vdash \rid(\loc_i) \leq r$).
    \end{itemize}
  \end{pfsubcases}

  From a simple case analysis on $k$, and the observation that there is no
  reference $\REF$ in $\Graph_{n}$ such that $\REF$ points to $\temp(r',
  \iota'')$, we have
  \begin{itemize}
  \item $\varunique(\Graph')$
  \item $\regionorder(\rho', \Cl', \Fr, \Graph')$
  \item $\locationok(\Graph')$
  \item $\deepfreeze(\Fr, \Graph')$
  \item $\topokgraph(\rho', \Cl', \Fr, \Graph')$
  \end{itemize}
  Meaning
  \begin{itemize}
  \item $\capok(\RCfg')$
  \item $\topok(\RCfg')$
  \end{itemize}
\pfcase{\rn{wf-eff-swap}}
\llabel[pf]{case_eff_swap_var}

  We have
  \begin{enumerate}[label=A\arabic*]
  \item $\Eff = \swap{x, y.\Valfield, \Use}$
  \item $\Gammas = \Gamma :: \Gamma^*$
  \item \llabel{gammausegammap} $\Gamma \vdash \Use : t \dashv \Gamma'[y: \var~\Cell[t']]$
  \item \llabel{capnotvar} $\Kap(\{ \var \}^c, t)$
  \item $\Gammas' = \Gamma'[y : \var~\Cell[t]], x : t :: \Gamma^*$
  \end{enumerate}

  \begin{itemize}
  \item $\Graph = G(\RCfg)$
  \item $\rho = \rho(\RCfg)$
  \item $\Cl = \Closed(\RCfg)$
  \item $\Fr = \Frozen(\RCfg)$
  \end{itemize}

  There are two cases for the dynamic execution of $\Eff$:
  \rn{region-swap-heap} and \rn{region-swap-temp}. However, the
  former is impossible because of well-formedness and $\capok(\RCfg)$,
  specifically $\locationok(\Graph)$ and $\regionorder(\rho, \Cl, \Fr, \Graph)$,
which says that $y$ with $\var$ capability must point to the temporary store in
the top region frame.

  By \rn{region-swap-temp}
  \begin{enumerate}[resume*]
  \item $\RCfg = \RCfglong{(r, S, F)::RS; \Hop; \Hcl; \Hfr}$
  \item $x$ fresh
  \item $\Get(F, \Use) = ((k, \iota), F')$
  \item $F'(y) = (k_y, \iota_y)$
  \item $\load{S, \iota_y} = o[\Valfield \mapsto (k', \iota')]$
  \item $o' = o[\Valfield \mapsto (k, \iota)]$
  \item $\store(S, \iota, o') = S'$
  \item $F'' = F', x \mapsto (k', \iota')$
  \item $\RCfg' = \RCfglong{(r, S', F'')::RS; \Hop; \Hcl; \Hfr}$
  \end{enumerate}

  We note that
  \begin{enumerate}[resume*]
  \item $\rho' = \rho(\RCfg') = \rho$
  \item $\Cl' = \Closed(\RCfg') = \Cl$
  \item $\Fr' = \Closed(\RCfg') = \Fr$
  \end{enumerate}

  By well-formedness of $\RCfg$ we have
  \begin{enumerate}[resume*]
    \item \llabel{deltadeltass}$\Delta; \Delta_S \vdash S$
  \end{enumerate}

  Using lemma \ref{lemma:wf_get}, we know

  \begin{enumerate}[resume*]
  \item \llabel{deltagammapfp} $\Delta; \Gamma'[y : \var~\Cell[t']] \vdash F'$
  \end{enumerate}

  We argue that there is a $\Delta'$ such that $\Gammas'; \Delta'; \Psi \vdash
  \RCfg$.
  We know $\iota_y : \Cell[\_]$ is in $\Delta$ since $\RCfg$ is well typed.
  Let
  \begin{enumerate}[resume*]
    \item $\Delta = \Delta^*[\iota_y : \Cell[\_]]$
    \item $\Delta' = \Delta^*[\iota_y : \Cell[t]]$
  \end{enumerate}

  By induction on the structure of the derivation on \lref{deltagammapfp}
  \begin{enumerate}[resume*]
    \item \llabel{deltapgammapfpp} $\Delta'; \Gamma'[y : \var~\Cell[t]] \vdash F''$
  \end{enumerate}

  By induction on the structure of the derivation of \lref{deltadeltass}
  \begin{enumerate}[resume*]
  \item $\Delta_S = \Delta_S^*[\iota_y : \Cell[\_]]$
  \item $\Delta_S' = \Delta_S^*[\iota_y : \Cell[t]]$
  \item \llabel{deltadeltaspsp} $\Delta'; \Delta_S' \vdash S'$
  \end{enumerate}

  For any frame $(\_, \_, F_1) \in RS$ there is a corresponding $\Gamma_1 \in
  \Gammas^*$. By induction on structure of $\Delta; \Gamma_1 \vdash F_1$: if
  $x \mapsto (k_x, \iota_x) \in F_1$, and corresponding typing $x : t_x$, then
  $\iota_x \neq \iota_y$, since otherwise we would contradict
  $\varunique(\Graph)$. We conclude that $\iota_x : \CL \in \Delta \implies
  \iota_x : \CL \in \Delta'$ and thus $k_x \CL <: t_x$. Thus by \rn{wf-F-cons} we
  can conclude that
  \begin{enumerate}[resume*]
  \item $\Delta'; \Gamma_1 \vdash F_1$
  \end{enumerate}

  A analogous argument holds for any $(\CLTag, F_2)$ in $\Hop * \Hcl * \Hfr$,
  with rule \rn{wf-subheap-cons}.

  Thus all subheaps and frames are well-formed under $\Gammas'; \Delta'; \Psi$,
  allowing us to conclude that
  \begin{enumerate}[resume*]
    \item $\Gammas'; \Delta'; \Psi \vdash \RCfg'$
  \end{enumerate}

  Letting $\Graph = G(\RCfg)$, by lemma \ref{lemma:wd_graph_act}
  \begin{enumerate}[resume*]
  \item
    $
    \Graph' =
    \begin{cases}
      \Graph - (\ruut(r) \refto{z, k_z} \loc_z) - (\loc_y \refto{f,
      k_f} \loc_f) + (\loc_y \refto{f, k_z} \loc_z) & \If \Use = \Drop
                                                      z \\
      \Graph - (\loc_y \refto{f, k_f} \loc_f) + (\loc_y \refto{f, k_z}
      \loc_z) & \Ow
    \end{cases}
    $
  \end{enumerate}

  \begin{itemize}
  \item $\varunique(\Graph')$ holds because of \lref[pf]{obs:varuniq_neq_var}
    and \lref[pf]{obs:varuniq_swap}, because the only way to break the invariant
    is if $k_z = \var$. But this implies $\Use = \Drop z$ and thus we remove the
    old reference to $\loc_z$.
  \item $\regionorder(\rho, \Cl, \Fr, \Graph')$ holds because region
    $\rid(\ruut(r)) = \rid(\loc_y)$, i.e. the new reference will have the same
    relation between start and end region.
  \item $\locationok(\Graph')$ follows from a case analysis on $k_z$, notin $k_z \neq \var$ by \lref{gammausegammap} and \lref{capnotvar}.
  \item $\deepfreeze(\Fr, \Graph')$ holds since $r \not\in \Fr$.
  \item $\entrypointsok(\rho, \Graph')$ since no entrypoints have changed.
  \item $\topokgraph(\rho, \Fr, \Graph')$ holds by case analysis on $k_z$, using
    $\regionorder(\rho, \Cl, \Fr, \Graph')$ and \lref[pf]{obs:topok_swap}.
  \end{itemize}

\pfcase{\rn{wf-eff-swap-class}}
  We have
  \begin{enumerate}[label=A\arabic*]
  \item $\Eff = \swap{x, y.f, \Use}$
  \item $\Gammas = \Gamma :: \Gamma^*$
  \item \llabel{gammausegammap} $\Gamma \vdash \Use : t \dashv \Gamma'$
  \item \llabel{capnotvar} $\Kap(\{ \var \}^c, t)$
  \item $\Gamma(y) = k \CL$
  \item $k \in \{\mut, \tmp\}$
  \item $\Gammas' = \Gamma'[y : k \CL], x : t :: \Gamma^*$
  \end{enumerate}




  This case is similar to case \lref[pf]{case_eff_swap_var} above. One thing to
  note is that the case where $k = \mut$ together with $\vdash k \CL$ (from
  lemmas \ref{lemma:wf_gamma}, \ref{lemma:wf_type_gamma}) implies that the
  reference specified by $\Use$ will be either $\imm, \iso$ or $\mut$, (i.e. a
  reference to a closed, frozen or within the same region) which will
  allow us to prove $\capok$ and $\topok$.

\pfcase{\rn{wf-eff-exit}}

  We have
  \begin{enumerate}[label=A\arabic*]
  \item $\Eff = \exit{x, \Use, y.f, w.g}$
  \item $\Gammas = \Gamma' :: \Gamma[y : k \CL] :: \Gammas^*$
  \item $k \in \{ \var, \tmp, \mut, \paused \}$
  \item $\Gamma' \vdash \Use : t \dashv \Gamma''$
  \item $\Kap(\{ \iso, \imm \}, t)$
  \item $\Gamma''(w) = k' \CL'$
  \item $\FType(\CL', g) = t'$
  \item $\Kap(\mut, t')$
  \item $\Kap(\iso, \FType(\CL, f))$
  \item
    $
    k'' \CL'' =
    \begin{cases}
      \var \Cell[\Hmakeiso(t')] & \If k \CL = \var~\Cell[\_] \\
      k \CL & \Ow
    \end{cases}
    $
  \item $\Gammas' = \Gamma[y : k'' \CL''], x : t :: \Gammas^*$
  \end{enumerate}

  For the dynamic rule we have two cases: \rn{region-exit-temp} and
  \rn{region-exit-heap}.
  \begin{pfsubcases}
    \pfsubcase{\rn{region-exit-temp}}
    \llabel[case]{ret}
    \begin{enumerate}[label=B\arabic*]
    \item $\RCfg = \RCfglong{(r', S', F') :: (r, S, F) :: RS; (r', S_{op}) *
        \Hop; \Hcl; \Hfr}$
    \item $x$ fresh
    \item $\Get(F', \Use) = ((k_z, \iota_z), F'')$
    \item $F''(w) = (k_w, \iota')$
    \item $\sload(S', \iota') = o'[g \mapsto (k_g, \iota'')]$
    \item $F(y) = (k_y, \iota)$
    \item $\stackload((r, S, F) :: \RS, \iota) = o[f \mapsto(k_f, \iota_f)]$
    \item $\stackstore((r, S, F) :: \RS, \iota, o[f \mapsto (k_f, \iota'')]) =
      (r, S'', F) :: \RS'$
    \item $F''' = F, x \mapsto (k_z, \iota_z)$
    \item $\RCfg' = \RCfglong{(r, S'', F''') :: RS'; \Hop; (r', S_{op}) * \Hcl;
        \Hfr}$
    \end{enumerate}
    \pfsubcase{\rn{region-exit-heap}}
    \llabel[case]{reh}
    \begin{enumerate}[label=C\arabic*]
    \item $\RCfg = \RCfglong{(r', S', F') :: (r, S, F) :: RS; (r', S_{op}) *
        \Hop; \Hcl; \Hfr}$
    \item $x$ fresh
    \item $\Get(F', \Use) = ((k_z, \iota_z), F'')$
    \item $F''(w) = (k_w, \iota')$
    \item $\sload(S', \iota') = o'[g \mapsto (k_g, \iota'')]$
    \item $F(y) = (k_y, \iota)$
    \item $\heapload(\Hop, \iota) = o[f \mapsto(k_f, \iota_f)]$
    \item $\heapstore(\Hop, \iota, o[f \mapsto (k_f, \iota'')]) =
      \Hop'$
    \item $F''' = F, x \mapsto (k_z, \iota_z)$
    \item $\RCfg' = \RCfglong{(r, S, F''') :: RS; \Hop'; (r', S_{op}) * \Hcl;
        \Hfr}$
    \end{enumerate}
  \end{pfsubcases}
  \pfsetscope{case}

  We note
  \begin{enumerate}[label=D\arabic*]
  \item $\Graph = G(\RCfg)$
  \item $\rho = \rho(\RCfg) = r' :: r :: \rho^*$
  \item $\rho' = \rho(\RCfg') = r :: \rho^*$
  \item $\Fr = \Frozen(\RCfg) = \Frozen(\RCfg')$
  \item $\Cl = \Closed(\RCfg)$
  \item $\Cl = \Closed(\RCfg') = \Cl \cup \{ r' \}$
  \item $\Graph' = G(\RCfg')$
  \item $\varunique(\Graph)$
  \item $\regionorder(\rho, \Cl, \Fr, \Graph)$
  \item $\locationok(\Graph)$
  \item $\deepfreeze(\Graph)$
  \item $\entrypoints(\rho, \Graph)$
  \item $\topokgraph(\rho, \Fr, \Graph)$
  \end{enumerate}

  We note the similarity between \lref{ret} and \lref{reh}, and furthermore note that the
  they are similar to a $\swap$, with the extra action of popping a region
  frame.

  $\regionorder(\rho, \Cl, \Fr, \RCfg)$, $\topokgraph(\rho, \Graph)$,
  $\locationok(\Graph)$, and $\entrypointsok(\rho, \RCfg)$ implies that the
  only reference from another region into $r'$ is $\loc_y \refto{f, \iso}
  \heap(r', \_)$. Thus we can construct $\Delta'$ from $\Delta$ by removing all
  $\iota^* \in \Delta$ such that $\iota^* \in S'$. By these facts, we can
  conclude

  \begin{enumerate}[resume*]
  \item $\Gamma[y : k \CL] :: \Gammas^*; \Delta'; \Psi \vdash \RCfglong{(r, S,
    F)::RS; \Hop; (r', S_{op}) * \Hcl; \Hfr}$.
  \end{enumerate}

  From similar reasoning to the swap
  effects e.g. \lref[pf]{case_eff_swap_var} above.
  \begin{enumerate}[resume*]
  \item $\Gammas'; \Delta'; \Psi \vdash \RCfg'$
  \end{enumerate}

  By lemma \ref{lemma:wd_graph_act}, and well-formedness
  \begin{enumerate}[resume*]
  \item
    $
    \Graph' = \Graph
    \begin{aligned}[t]
      &- \ruut(r') - \{ \temp(r', \_) \suchthat \temp(r', \_) \in \Graph \} \\
      &- \{ \ruut(r') \refto{\_, \_} \_ \suchthat \ruut(r') \refto{\_, \_} \_ \in \Graph \} \\
      &- \{ \temp(r', \_) \refto{\_, \_} \_ \suchthat \temp(r', \_) \refto{\_, \_} \_ \in \Graph \} \\
      &+ \ruut(r) \refto{x, k_z} \loc_z \\
      &- \loc_y \refto{f, \iso} \heap(r', \_) + \loc_y \refto{f, \iso} \heap(r', \iota'')
    \end{aligned}
    $
  \item $\Use = z$ or $\Use = \Drop z$.
  \item $\REF_\Graph(\ruut(r'), z) = \ruut(r') \refto{z, k_z} \loc_z$
  \item $\REF_\Graph(\ruut(r), y) = \ruut(r) \refto{y, k_y} \loc_y$
  \item $\REF_\Graph(\loc_y, f) = \loc_y \refto{f, \iso} \heap(r', \_)$
  \item \llabel{kz} $k_z \in \{ \imm, \iso \}$
  \item $k_y \in \{ \mut, \paused, \var, \tmp \}$
  \item $\REF_\Graph(\ruut(r'), w) = \ruut(r') \refto{w, k_w} \temp(r',
    \iota')$
  \item $k_w \in \{ \tmp, \var \}$
  \item $\REF_\Graph(\temp(r', \iota'), g) = \temp(r', \iota') \refto{g, \mut}
    \heap(r', \iota'')$
  \end{enumerate}
  \begin{itemize}
  \item $\varunique(\Graph')$ by applying observation
    \lref[pf]{obs:subgraph_varuniq}, and then using well-formedness and \\ $\Kap(\iso,
    \FType(\CL, f))$ to get $k_f \neq \var$ and $\Kap(\{\iso, \imm\}, t)$ to get
    $k_z \neq \var$, and then applying observation
    \lref[pf]{obs:varuniq_neq_var}.
  \item $\regionorder(\rho', \Cl', \Fr, \Graph')$ by \lref{kz} implying
    $\Rid(\loc_z) \in \Cl \cup \Fr \subseteq \Cl' \cup \Fr$ thus
    $\regionorder1(\rho', \Cl', \Fr, \ruut(r) \refto{x, k_z} \loc_z)$. Furthermore
    $r' \in \Cl'$ and so\\ $\regionorder1(\rho', \Cl', \Fr, \loc_y \refto{f, \iso}
    \heap(r', \iota''))$
  \item $\locationok(\Graph')$ by case analysis on $k_z$.
  \item $\deepfreeze(\Fr, \Graph')$ since frozen regions are unchanged.
  \item $\entrypointsok(\rho', \Graph')$ by arguing that all region ids in
    $\rho$ are unique (by well-typedness of $\RCfg$) and thus by
    $\topokgraph(\rho, \Graph)$ and \lref[pf]{obs:entrypoint_objects_disjoint}
    all entrypoint are disjoint and therefore the swap at $\loc_y$, field $f$,
    does not affect any other entrypoint.
  \item $\topokgraph(\rho', \Fr, \Graph')$ by \lref[pf]{obs:topok_swap}.
  \end{itemize}

\pfcase{\rn{wf-eff-freeze}}
\llabel[pf]{case_eff_freeze}

  We have
  \begin{enumerate}[label=A\arabic*]
  \item $\Eff = \freeze(x, \Use)$
  \item $\Gammas = \Gamma :: \Gammas^*$
  \item $\Gamma \vdash \Use : t \dashv \Gamma'$
  \item $\Kap(\iso, t)$
  \item $\Gammas' = \Gamma', x : \Hmakeimm(t) :: \Gammas^*$
  \end{enumerate}

  By \rn{region-freeze}
  \begin{enumerate}[resume*]
  \item $\RCfg = \RCfglong{(r, S, F) :: RS; \Hop; (R * H) * \Hcl; \Hfr}$
  \item $x$ fresh
  \item $\Get(F, \Use) = ((k, \iota), F')$
  \item $\iota \in \dom(R.S)$
  \item $H = \reachableregions(\iota, \Hop * H * \Hcl)$
  \item $F'' = F', x \mapsto (\imm, \iota)$
  \item $\RCfg' = \RCfglong{(r, S, F'') :: RS; \Hop; \Hcl; (R * H) * \Hfr}$
  \end{enumerate}

  Well-typedness follows by induction on the well-formedness of the top region
  frame. We have
  \begin{enumerate}[resume*]
  \item $\Gammas'; \Delta; \Psi \vdash \RCfg'$
  \end{enumerate}

  We note
  \begin{enumerate}[resume*]
  \item $\rho = \rho(\Gammas, \RCfg) = \rho(\Gammas', \RCfg')$
  \item $\Cl = \Closed(\RCfg)$
  \item $\Fr = \Frozen(\RCfg)$
  \item $\Fr_{new} = \rids(R * H)$
  \item $\Cl' = \Cl - \Fr_{new} = \Closed(\RCfg')$
  \item $\Fr' = \Fr + \Fr_{new} = \Frozen(\RCfg')$
  \item $\Graph = G(\RCfg)$
  \item $\Graph' = G(\RCfg')$
  \item $\Use = \Drop z$
  \end{enumerate}

  By lemma \ref{lemma:wd_graph_act},
  \begin{enumerate}[resume*]
  \item
    $\Graph' =
    \begin{aligned}[t]
      \Graph - (\ruut(r) \refto{z, k_z} \loc_z[r']) + (\ruut(r) \refto{x, \imm} \loc_z[r'])
    \end{aligned}
    $
  \item $k_z = \iso$
  \end{enumerate}

  \begin{itemize}
    \item $\varunique(\Graph')$ since we do not handle $\var$ references.
    \item $\regionorder(\rho, \Cl', \Fr', \Graph')$ by noting that
      $\reachableregions$ says that all references between two regions
      reachable from $R$ will have both ends in $R * H$, and that $R$ is in $R *
      H$.
    \item $\locationok(\Graph')$ noting that $\loc_z[r'] = \heap(r', \_)$ by
      $\locationok(\Graph)$ and $k_z = \iso$.
    \item $\deepfreeze(\Fr', \Graph')$ by similar argument to $\reachableregions$.
    \item $\topokgraph(\rho, \Fr', \Graph')$ since $\Fr \subseteq \Fr'$ and $r'
      \in \Fr'$.
    \item $\entrypointsok(\rho, \Graph')$ since entrypoints are unaffected.
  \end{itemize}

\pfcase{\rn{wf-eff-merge}}

  We have
  \begin{enumerate}[label=A\arabic*]
  \item $\Eff = \freeze(x, \Use)$
  \item $\Gammas = \Gamma :: \Gammas^*$
  \item $\Gamma \vdash \Use : t \dashv \Gamma'$
  \item $\Kap(\iso, t)$
  \item $\Gammas' = \Gamma', x : \Hmakeimm(t) :: \Gammas^*$
  \end{enumerate}

  By \rn{region-merge}
  \begin{enumerate}[resume*]
  \item $\RCfg = \RCfglong{(r, S, F) :: RS; (r, S') * \Hop; R * \Hcl; \Hfr}$
  \item $x$ fresh
  \item $\Get(F, \Use) = ((k, \iota), F')$
  \item $\iota \in \dom(R.S)$
  \item $R' = (r, S' \uplus R.S)$
  \item $F'' = F', x \mapsto (\mut, \iota)$
  \item $\RCfg' = \RCfglong{(r, S, F'') :: RS; R * \Hop; \Hcl; \Hfr}$
  \end{enumerate}

  Well-typedness holds by induction on the well-formedness of the top frame and
  region configuration heaps.
  \begin{enumerate}[resume*]
  \item $\Gammas'; \Delta; \Psi \RCfg'$
  \end{enumerate}

  We note
  \begin{enumerate}[resume*]
  \item $\rho = \rho(\Gammas, \RCfg) = \rho(\Gammas', \RCfg')$
  \item $\Cl = \Closed(\RCfg)$
  \item $\Fr = \Frozen(\RCfg) = \Frozen(\RCfg')$
  \item $r' = R.r$
  \item $\Cl' = \Cl - r$
  \item $\Graph = G(\RCfg)$
  \item $\Graph' = G(\RCfg')$
  \item $\Use = \Drop z$
  \end{enumerate}

  By lemma \ref{lemma:wd_graph_act},
  \begin{enumerate}[resume*]
  \item $LOCS = \{ \Loc[r'] \suchthat \Loc[r'] \in \Graph \}$
  \item $LOCS' = \{ \Loc[r] \suchthat \Loc[r'] \in LOCS \}$
  \item $REFS = \{ \Loc \refto{f, k} \Loc' \suchthat \Loc \in LOCS \}$
  \item $REFS' = \left\{ \Loc[r] \refto{f, k} \Loc'[r_2'] \Suchthat \Loc[r'] \refto{f, k}
      \Loc'[r_2] \in REFS \AN r_2' = \left[\begin{array}{ll}
                                       r & \If r_2 = r' \\
                                       r_2 & \Ow
                                            \end{array}\right. \right\}$
  \item
    $
    \Graph' =
    \begin{aligned}[t]
      \Graph & - LOCS + LOCS' - REFS + REFS' \\
      &- (\ruut(r) \refto{z, \iso} \loc_z[r']) + (\ruut(r) \refto{x, \mut} \loc_z[r])
    \end{aligned}
    $
  \end{enumerate}

  Now, we can conclude that
  \begin{itemize}
  \item $\varunique(\Graph')$ since each $\REF \in REFS$ has been replaced with
    a corresponding one in $REFS'$.
  \item $\regionorder(\rho, \Cl', \Fr, \Graph')$ by case analysis on the
    capability of the references in $REFS'$ reference $\loc_1[r_1] \refto{f, k}
    \loc_2[r_2] \in REFS$, $k = \paused$ would violate $\regionorder(\rho, \Cl,
    \Fr, \Graph)$, since $r' \in \Cl$.
  \item $\locationok(\Graph')$ by same argument as $\varunique$.
  \item $\deepfreeze(\Fr, \Graph')$ since we have not changed frozen regions.
  \item $\topokgraph(\rho, \Fr, \Graph')$ since intra-region references have been
    converted to intra-region references, and iso-references (i.e. inter-region
    references) have been converted to corresponding iso-references.
  \item $\entrypointsok(\rho, \Graph')$ since we have not touched entry points.
  \end{itemize}

\pfcase{\rn{wf-eff-halloc-iso}}

  \begin{enumerate}[label=A\arabic*]
  \item $\Eff = \halloc{x, \iso, \#C, \Use_1, \dots, \Use_n}$
  \item $\Gammas = \Gamma_1 :: \Gammas^*$
  \item $x$ fresh
  \item $\vdash \iso~C$
  \item $\FTypes(C) = f_1 : t_1, \dots, f_n : t_n$
  \item $\forall i \in [1, n]. \Gamma_i \vdash \Use_i : t_i' \dashv \Gamma_{i+1} \AN
    \Kap({iso, imm}, t_i) \AN t_i' <: t_i$
  \item $\Gamma_{n+1}, x : \iso~C :: \Gammas^*$
  \end{enumerate}

  From \rn{region-alloc-heap-iso}
  \begin{enumerate}[resume*]
  \item $\RCfg = \RCfglong{(r, S, F_1) :: RS; \Hop; \Hcl; \Hfr}$
  \item $x$ fresh
  \item $\forall i \in [1, n]. \Get(F_i, \Use_i) = ((k_i, \iota_i), F_{i+1})$
  \item $\kw{fields}(C) = f_1, ..., f_n$
  \item $o = (\#C, [f_1 \mapsto (k_1, \iota_1), ..., f_n \mapsto (k_n,
    \iota_n)])$
  \item $\iota$ fresh
  \item $r'$ fresh
  \item $F' = F_{n+1}, x \mapsto (\iso, \iota)$
  \item $\RCfg = \RCfglong{(r, S, F_1) :: RS; \Hop; (r', [\iota \mapsto o]) *\Hcl; \Hfr}$
  \end{enumerate}

  Well-typedness follows similarly to case $\swap()$, where we use an inductive
  argument to account for the multiple $\Use_1, \dots, \Use_n$. $\capok$ and
  $\topok$ follows from the observation that all references originating from the
  created object is either $\iso$ or $\imm$, (i.e. $k_i \in \{ \iso, \imm \}$) implying that they point into closed or
  frozen regions respectively. The object itself is created in a closed (fresh
  region), to which we create an iso reference.

\pfcase{\rn{wf-eff-halloc-mut}}
  \begin{enumerate}[label=A\arabic*]
  \item $\Eff = \halloc{x, \mut, \#C, \Use_1, \dots, \Use_n}$
  \item $\Gammas = \Gamma_1 :: \Gammas^*$
  \item $x$ fresh
  \item $\vdash \mut~C$
  \item $\FTypes(C) = f_1 : t_1, \dots, f_n : t_n$
  \item $\forall i \in [1, n]. \Gamma_i \vdash \Use_i : t_i \dashv \Gamma_{i+1}$
  \item $\Gamma_{n+1}, x : \mut~C :: \Gammas^*$
  \end{enumerate}

  From \rn{region-alloc-heap-mut}
  \begin{enumerate}[resume*]
  \item $\RCfg = \RCfglong{(r, S, F_1) :: RS; (r, S') * \Hop; \Hcl; \Hfr}$
  \item $x$ fresh
  \item $\forall i \in [1, n]. \Get(F_i, \Use_i) = ((k_i, \iota_i), F_{i+1})$
  \item $\kw{fields}(C) = f_1, ..., f_n$
  \item $o = (\#C, [f_1 \mapsto (k_1, \iota_1), ..., f_n \mapsto (k_n,
    \iota_n)])$
  \item $\iota$ fresh
  \item $S'' = S',\iota \mapsto o$
  \item $F' = F_{n+1}, x \mapsto (\mut, \iota)$
  \item $\RCfg = \RCfglong{(r, S, F_1) :: RS; (r, S'') * \Hop; (r', [\iota \mapsto o]) *\Hcl; \Hfr}$
  \end{enumerate}

  Follows from a similar argument as the $\mut$ case for $\swap{}$, using an
  inductive argument to account for the multiple $\Use_i$s.

\pfcase{\rn{wf-eff-salloc}}
  \begin{enumerate}[label=A\arabic*]
  \item $\Eff = \halloc{x, k, \#C, \Use_1, \dots, \Use_n}$
  \item $\Gammas = \Gamma_1 :: \Gammas^*$
  \item $x$ fresh
  \item $k \in \{ \tmp, \var \}$
  \item $\vdash k~C$
  \item $\FTypes(C) = f_1 : t_1, \dots, f_n : t_n$
  \item $\forall i \in [1, n]. \Gamma_i \vdash \Use_i : t_i \dashv \Gamma_{i+1}$
  \item $\Gamma_{n+1}, x : k~C :: \Gammas^*$
  \end{enumerate}

  From \rn{region-alloc-temp}
  \begin{enumerate}[resume*]
  \item $\RCfg = \RCfglong{(r, S, F_1) :: RS; \Hop; \Hcl; \Hfr}$
  \item $x$ fresh
  \item $k \in \{ \tmp, \var \}$
  \item $\forall i \in [1, n]. \Get(F_i, \Use_i) = ((k_i, \iota_i), F_{i+1})$
  \item $\kw{fields}(C) = f_1, ..., f_n$
  \item $o = (\#C, [f_1 \mapsto (k_1, \iota_1), ..., f_n \mapsto (k_n,
    \iota_n)])$
  \item $\iota$ fresh
  \item $S' = S,\iota \mapsto o$
  \item $F' = F_{n+1}, x \mapsto (k, \iota)$
  \item $\RCfg = \RCfglong{(r, S', F_1) :: RS; \Hop; (r', [\iota \mapsto o]) *\Hcl; \Hfr}$
  \end{enumerate}

  Follows from a similar argument as the $\tmp$ or $\var$ case for $\swap{}$, using an
  inductive argument to account for the multiple $\Use_i$s.

\pfcase{\rn{wf-eff-load}}

  \begin{enumerate}[label=A\arabic*]
  \item $\Gammas = \Gamma :: \Gamma^*$
  \item $x$ fresh
  \item $\Gamma(y) = k \CL$
  \item $\FType(\CL, f) = t$
  \item $k \neq \iso$
  \item $\vdash k \sees t$
  \end{enumerate}

  By \rn{region-load}
  \begin{enumerate}[resume*]
  \item $\RCfg = \RCfglong{(r, S, F) :: \RS; \Hop; \Hcl; \Hfr}$
  \item $x$ fresh
  \item $F(y) = (k_y, \iota_y)$
  \item $\CfgLoad((r, S, F)::\RS; \Hop * \Hfr, \iota_y) = o[f \mapsto (k_f, \iota_f)$
  \item $F' = F, x \mapsto (k_y \sees k_f, \iota_f)$
  \end{enumerate}

  Well-typedness follows from a similar argument to the $swap()$ cases. $\capok$
  and $\topok$ follows from observation \lref[pf]{obs:graph_add_vpa_ref}.


\pfcase{\rn{wf-eff-bind}}
  By trivial inductive argument on $\many{x = Use}$.

\pfcase{\rn{wf-eff-badenter}}
  Trivial.
\pfcase{\rn{wf-eff-eps}}
  Trivial.

\end{pfcases}

\subsubsection{Observations for Proof}

\begin{enumerate}[label=O\arabic*]
\item \llabel[pf]{obs:subgraph_varuniq}\[\Graph \subseteq \Graph' \AN \VarUnique(\Graph') \implies \VarUnique(\Graph) \]
  \item \llabel[pf]{obs:varuniq_neq_var}
    If
    \begin{itemize}
    \item $\REF = \Loc_1 \RefTo{f, k} \Loc_2 \in \Graph$
    \item $k \neq \Var$
    \item $k' \neq \Var $
    \item $\VarUnique(\Graph )$
    \end{itemize}
    Then
    \begin{itemize}
    \item $\VarUnique(\Graph + (\_ \RefTo{f', k'} \Loc_2))$
    \item $\VarUnique(\Graph - \REF + (\_ \RefTo{f', k'} \Loc_2))$
    \end{itemize}

    \textbf{Follows} from $k \neq \var$ implying that there is no reference $\_
    \refto{g, k''} \loc_2$ with $k'' = \var$.

\item\llabel[pf]{obs:varuniq_swap} $\VarUnique(\Graph) \AN  (\Loc_1 \RefTo{f,
    \Var} \Loc_2) \in \Graph \implies \VarUnique(\Graph - (\Loc_1 \RefTo{f,
    \Var} \Loc_2) + (\Loc_3 \RefTo{f', k'} \Loc_2)$
\item \llabel[pf]{obs:topok_swap} If $\REF = (\Loc_1[r_1] \RefTo{f, k} \Loc_2[r_2]$),
  $\REF' = (\Loc_1'[r_1'] \RefTo{f', k'} \Loc_2[r_2])$, $\REF \in \Graph$, \\ $\rho
  \vdash r_2 \leq r_1 \implies \rho \vdash r_2 \leq r_1'$ and
  \[
    \topokgraph(\rho, \Fr, \Graph)
  \]
  then
  \[
    \topokgraph(\rho, \Fr, \Graph - \REF + \REF')
  \]

  \textbf{Follows} from a simple case analysis on the clauses of $\toppok$.
\item\llabel[pf]{obs:wf_rcfg_gammas_extend}

  If $\Gammas_1 \gammadefincl \Gammas_2$ and
  $\Gammas_1 \vdash \RCfg$, then $\Gammas_1 | \Gammas_2 \vdash \RCfg$.
\item\llabel[pf]{obs:wf_eff_result_subgamma} If $\Gammas_1 \gammadefincl \Gammas_2$,
  $\Gammas_1 \vdash \Eff \dashv \Gammas_1'$ and $\Gammas_2 \vdash \Eff \dashv
  \Gammas_2'$, then $\Gammas_1' \gammadefincl \Gammas_2'$.

  \textbf{Follows} by induction on $\Gammas_1 \vdash \Eff \dashv \Gammas_1'$.
\item\llabel[pf]{obs:wf_rcfg_split} If $\Gammas[x : t | t'] \vdash \RCfg$, then
  $\Gammas[x: t] \vdash \RCfg$ or $\Gammas[x : t'] \vdash \RCfg$

  \textbf{Follows} since the actual type of $x$ must be a subtype of $t|t'$, meaning that
  the actual type is a subtype of either $t$ or $t'$.
\item\llabel[pf]{obs:vpa_subtype} If $t <: t'$, then $\VPA{k}{t} <: \VPA{k}{t'}$
\item\llabel[pf]{obs:iso_subtype_implies_vpa_undefined} $t <: t'$ and
  $\Kap(iso, t)$ implies that $\VPA{\Paused}{t} = \FUndef$. \textbf{Follows} by
  definition of subtyping and view-point adaptation (\ref{tab:vpa}).
\item\llabel[pf]{obs:expanding_delta} $\Delta; \Gamma \vdash F$, $\iota \not\in
  \dom(\Delta)$ then $\Delta, \iota : CL; \Gamma \vdash F$. \textbf{Follows} from that
  $\iota \not\in \dom(\Delta)$ implies that $\iota \not\in \mathbf{rng}(F)$
\item\llabel[pf]{obs:make_mut_subtype} If $t <: t'$ then $\Hmakemut(t) <:
  \Hmakemut(t')$. \textbf{Follows} by induction on structure of $t$.


\item\llabel[pf]{obs:entrypoint_objects_disjoint}
  If
  \begin{itemize}
  \item $\entrypointsok(\rho, \Graph)$
  \item $\topokgraph(\rho, \Fr, \Graph)$
  \item $\rho$ contains unique region ids
  \item $\rho = \dots \rhosep{x.f} r' \dots \rhosep{y.g} r'' :: \dots \Nilrho$
  \end{itemize}
  Then
  \begin{itemize}
    \item $\REF_\Graph(\ruut(r'), x) = \ruut(r') \refto{x, k_x} \loc_x$
    \item $\REF_\Graph(\ruut(r''), y) = \ruut(r'') \refto{y, k_y} \loc_y$
    \item Either $\loc_x \neq \loc_y$ or $f \neq g$.
  \end{itemize}

  \textbf{Follows} from uniqueness constraints of $\topokgraph$ and that $\rho$ does not
  have repeated region ids.
\item\llabel[pf]{obs:top_iso_ref_closed}
  If
  $\regionorder1(r :: \rho, \Cl, \Fr, (\ruut(r) \refto{x, \iso} \loc_x[r']))$,
  then $r' \in \Cl$.

  \textbf{Follows} by definition of $\regionorder1$
\item\llabel[pf]{obs:graph_add_vpa_ref}
  If
  \begin{itemize}
  \item $\rho = r :: \rho^*$
  \item $\vdash \rho, \Cl, \Fr$
  \item $\varunique(\Graph)$
  \item $\regionorder(\rho, \Cl, \Fr, \Graph)$
  \item $\locationok(\Graph)$
  \item $\deepfreeze(\Fr, \Graph)$
  \item $\topokgraph(\rho, \Fr, \Graph)$
  \item $\entrypointsok(\rho, \Graph)$
  \item $\ruut(r) \refto{x, k_x} \loc_x[r_x] \in \Graph$
  \item $\loc_x[r_x] \refto{f, k_f} \loc_f \in \Graph$
  \item $r_x \in \rho$
  \item $\REF = \ruut(r) \refto{z, k_x \sees k_f} \loc_f$
  \end{itemize}
  Then
  \begin{itemize}
  \item $\varunique(\Graph + \REF)$
  \item $\regionorder(\rho, \Cl, \Fr, \Graph + \REF)$
  \item $\locationok(\Graph + \REF)$
  \item $\deepfreeze(\Fr, \Graph + \REF)$
  \item $\topokgraph(\rho, \Fr, \Graph + \REF)$
  \item $\entrypointsok(\rho, \Graph + \REF)$
  \end{itemize}

  \textbf{Follows} from a case analysis on $k_x, k_f$.
\end{enumerate}

\subsection{Theorem: Soundness of the Tandem Semantics}

A well formed configuration is either done, is has reached a failed state
(because of failing to enter an already opened region), or can take a step to a
well formed configuration.
\[
  \begin{aligned}
    \vdash \langle \SB de \FB~\RCfg\rangle \implies de = \Use \OR
    de = \mathbf{Failure} \OR \langle \SB de \FB~\RCfg \rangle \rightarrow
    \langle \SB de' \FB~\RCfg' \rangle \AN \vdash \langle \SB de' \FB~\RCfg' \rangle
  \end{aligned}
\]

\subsubsection{Proof of Theorem}

Follows from theorems \ref{theorm:cmd_progress}, \ref{theorm:cmd_preservation},
\ref{theorm:region_progress}, and \ref{theorm:region_preservation}.

\subsection{Lemma: Subtyping is Transitive}
\label{lemma:subtype_trans}

Subtyping is transitive, i.e. if $t_1 <: t_2$ and $t_2 <: t_3$, then $t_1 <: t_3$.

\subsubsection{Proof of Lemma}
\pf
By strong induction over the subtyping relations.

\TODO{real proof}

\subsection{Lemma: get is Total}
\label{lemma:get_total}

\[
  \begin{array}{l}
    \Delta; \Gamma \vdash F \AN \Gamma \vdash use : t \dashv \Gamma' \implies\\
    \quad\exists F', k, \iota. \Get(F, use) = ((k, \iota), F')
  \end{array}
\]

\subsubsection{Proof of Lemma}
\pf

By induction over the well-formedness judgment of $F$.

\TODO{real proof}

\subsection{Lemma: Well-Formedness of \Get}
\label{lemma:wf_get}

\[
  \begin{aligned}
    \forall &\Delta, \Gamma, \Gamma', use, F, F', k, \iota. \\
            &\Delta; \Gamma \vdash F \AN \\
            &\Gamma \vdash use : t \dashv \Gamma' \AN \\
            &\Get(F, use) = ((k, \iota), F') \\
            &\implies \\
            &\Delta; \Gamma' \vdash F' \AN \\
            &\iota : CL \in \Delta \AN \\
            &k~CL <: t
  \end{aligned}
\]

\subsubsection{Observations}
It is a simple exercise to prove the following observations
\begin{enumerate}[label=O\arabic*, align=left]
\item \label{obs:deltagammaf_order} If $\Delta; \Gamma \vdash F$, then if
  $\phi$ is a permutation of the order of the mappings in $\Gamma$ and $F$,
  we have $\Delta; \phi(\Gamma) \vdash \phi(F)$.
\item \label{obs:subtype_sub} If $t <: t'$, then $t[k/k'] <: t'[k/k']$ for all
  capabilities $k, k'$.
\end{enumerate}

\subsubsection{Proof of Lemma}
\pf

In order to prove the lemma we begin by stating the assumptions

\begin{enumerate}[label=\textbf{Main \arabic*}, align=left]
\item \llabel{deltagammaf} $\Delta; \Gamma \vdash F$
\item \llabel{gammausetgammap} $\Gamma \vdash \Use : t \dashv \Gamma'$
\item \llabel{getfuse} $\Get(F, \Use) = ((k , \iota), F')$.
\end{enumerate}

By structural induction on $\Delta; \Gamma \vdash F$:

\begin{enumerate}[label=\textbf{Case \arabic*}, align=left]
\pfcase{\rn{wf-F-nil}}:
  Impossible.
\pfcase{\rn{wf-F-cons}}:
  \begin{enumerate}[label=A\arabic*, align=left]
  \item $\Gamma = \Gamma_1, y : t_y$
  \item $F = F_1, y \mapsto (k_y, \iota_y)$
  \item $\Delta; \Gamma_1 \vdash F_1$
  \item $\iota_y : \CL_y \in \Delta$
  \item $k_y \CL_y <: t_y$
  \item $y \not\in \dom(F_1)$
  \end{enumerate}

  We now split on cases of \Use.
  \begin{enumerate}[label*=\textbf{.\arabic*}, align=left]
    \pfsubcase{$\Use = z = y$} \llabel[pf]{subcase:use_eq_z_eq_y}

    By lemma \ref{lemma:inversion} we we have
    \begin{enumerate}[label=B\arabic*, align=right]
    \item \llabel{a} $\Kap(\{\Mut, \Tmp, \Paused, \Imm\}, t_y)$
    \item \llabel{b} $\Gamma' = \Gamma_1, y : t_y$
    \item \llabel{c} $t_y <: t$
    \end{enumerate}
    By definition of  $\Get$
    \begin{enumerate}[resume*]
    \item $\Get(y, (F_1, y \mapsto (k_y, \iota_y))) = ((k_y, \iota_y), (F_1, y
      \mapsto (k_y, \iota_y)))$
    \end{enumerate}
    We can thus identify the following
    \begin{enumerate}[resume*]
    \item $k = k_y$
    \item $\iota = \iota_y$
    \item $\CL = \CL_y$
    \item $F' = F_1, y \mapsto (k_y, \iota_y)$
    \end{enumerate}

    Together with these facts, and with observations
    \ref{obs:deltagammaf_order}, \ref{obs:subtype_sub}, and lemma
    \ref{lemma:subtype_trans}, it follows that
    \begin{itemize}[label=-]
    \item $\Delta; \Gamma' \vdash F'$
    \item $\iota : \CL \in \Delta$, and
    \item $k \CL <: t$.
    \end{itemize}

    \pfsubcase{$\Use = z \neq y$} \llabel[pf]{subcase:use_eq_z_neq_y}

    Follows from definition of $\Get$ and application of induction hypothesis
    together with observation \ref{obs:deltagammaf_order}.

    \pfsubcase{$\Use = \kw{drop}~z$, $z = y$}

    This case is similar to \lref[pf]{subcase:use_eq_z_eq_y} above .

    \pfsubcase{$\Use = \kw{drop}~z$, $z \neq y$}

    Similar to \lref[pf]{subcase:use_eq_z_neq_y}.
  \end{enumerate}

  \pfcase{\rn{wf-F-cons-undef}}

  We have the assumptions
  \begin{enumerate}[label=A\arabic*, align=left]
  \item $\Gamma = \Gamma_1, y : \Undef$
  \item $F = F_1, y \mapsto \Undef$
  \item $\Delta; \Gamma_1 \vdash F_1$
  \item $\iota_y : \CL_y \in \Delta$
  \item $k_y \CL_y <: t_y$
  \item $y \not\in \dom(F_1)$
  \end{enumerate}

  Because $y$ is undefined in both $\Gamma$ and $F$, looking at the definition
  of $\Get$ together with \lref[pf]{gammausetgammap}, the only possible cases
  are $\Use = \kw{drop}~z$ or $\Use = z$, where $z \neq y$. We can thus apply
  the induction hypothesis similar to \lref[pf]{subcase:use_eq_z_neq_y} above.
\end{enumerate}

\subsection{Lemma: Well-Formed cfg\_load}
\label{lemma:wf_cfg_load}

\[
  \begin{array}{l}
    \RCfg = \RCfglong{RS; H_{op}; H_{cl}; H_{fr}} \AN
    \Gammas; \Delta; \Psi \vdash \RCfg \AN
    \iota : CL \in \Delta
    \implies \\
    \quad\CfgLoad(RS, H_{op} * H_{cl} * H_{fr}) = (CL, F) \AN
         \Delta; \FTypes{CL} \vdash F
  \end{array}
\]

\subsubsection{Proof of Lemma}
\pf

By induction over the well-formedness rules for $\RS$ and $H$.
Note that \emph{every} $\iota$ in $\Delta$ maps to \emph{exactly
  one} object in the configuration.

\TODO{real proof}

\subsection{Lemma: Well-Formed Environment}
\label{lemma:wf_gamma}

If $\Gamma \vdash b : t \dashv \Gamma'$, then $\vdash \Gamma$ and $\vdash
\Gamma'$.

\subsubsection{Proof of lemma}
By structural induction on $\Gamma \vdash b : t \dashv \Gamma'$.

\subsection{Lemma: Well-Formed Type Under Environment}
\label{lemma:wf_type_gamma}

If $\vdash \Gamma$ and $\Gamma(x) = t$, then $\vdash t$.

\subsubsection{Proof of lemma}
By structural induction on $\vdash \Gamma$.

\subsection{Inversion Lemma}
\label{lemma:inversion}

Because of subtyping, we cannot invert typing judgments directly
to get useful information. Instead we prove an inversion lemma
that gives us the premises of the corresponding rule, with
subtyping information regarding the result type.

Because some of the type rules are only defined when variables
have types of the shape $k~CL$, we define a helper function for
calculating the type of a field lookup for type unions (this is
the result of typing \c{*x.f} when $x$ has type $t$):

\[
\mathbf{fresult}(t, f) =
\begin{cases}
  \VPA{k}{\mathbf{ftype}(CL, f)} & \text{if } t = k~CL\\
  \mathbf{fresult}(t_1, f)~|~\mathbf{fresult}(t_2, f) & \text{if } t = t_1~|~t_2
\end{cases}
\]

When doing strong updates of ref cells, we need to turn a union of
types into a union of \Cell{} types:

\[
  \begin{array}{rcl}
    \mathit{make\_cell}(k~CL) & = & \Var{}~\Cell[k~CL]\\
    \mathit{make\_cell}(t_1~|~t_2) & = & \mathit{make\_cell}(t_1) ~|~ \mathit{make\_cell}(t_2)
  \end{array}
\]


When freezing a type, we need to change \Iso{} capabilities to \Imm{}:
\[
  \begin{array}{rcl}
    \mathit{make\_imm}(k~CL) & = &
    \begin{cases}
      \Imm~CL & \text{if } k = \Iso\\
      k~CL    & \text{otherwise}
    \end{cases}\\
    \mathit{make\_imm}(t_1~|~t_2) & = & \mathit{make\_imm}(t_1) ~|~ \mathit{make\_imm}(t_2)
  \end{array}
\]

The inversion lemma itself is stated as follows:

\[
  \begin{array}{rcl}
    \Gamma \vdash x : t \dashv \Gamma' & \iff &
           \vdash \Gamma \AN \Gamma = \Gamma_1[x : t'] \AN t' <: t \AN
                                                \Kap(\{\Iso, \Var\}^c, t') \AN\\
    &    & \Gamma' = \Gamma_1[x : t']\\
    \\

    \Gamma \vdash \kw{drop}~x : t \dashv \Gamma' & \iff &
           \vdash \Gamma \AN \Gamma = \Gamma_1[x : t'] \AN t' <: t \AN \Gamma' = \Gamma_1[x : \mathbf{undef}]\\
    \\

    \Gamma \vdash \c{*}x.f : t \dashv \Gamma' & \implies &
           \Gamma(x) = t_x \AN \mathbf{fresult}(t_x, f) = t' \AN t' <: t\\
    &    & \Kap(\{\Iso\}^c, t_x) \AN \forall k~CL\in t_x. \vdash \VPA{k}{t'}\\
    \\

    \Gamma \vdash \c{*}x : t \dashv \Gamma' & \implies &
           \Gamma(x) = t_x \AN \mathbf{fresult}(t_x, \c{val}) = t' \AN t' <: t\\
    &    & \Kap(\Var{}, t_x) \AN \vdash \VPA{\Var{}}{t'}\\
    \\

    \Gamma \vdash x.f~\c{:=}~\Use : t \dashv \Gamma' & \implies &
            \Gamma \vdash \Use : t' \dashv \Gamma' \AN \Gamma'(x) = t_x \AN\\
    &    & \Kap(\{\Mut{}, \Tmp\}, t_x) \AN \mathbf{fresult}(t_x, f) = t' \AN t' <: t\\
    \\

    \Gamma \vdash x \c{:=}~\Use : t \dashv \Gamma' & \implies &
           \Gamma \vdash \Use : t' \dashv \Gamma''[x : t_x] \AN \Kap(\Var{}, t_x) \AN\\
    &    & \mathbf{fresult}(t_x, \c{val}) = t_f \AN t_f <: t \AN\\
    &    & \Gamma' = \Gamma''[x : \mathit{make\_cell}[t']] \\
    \\

    \Gamma_1 \vdash \fnc(\Use_1,...,\Use_n) : t \dashv \Gamma_{n+1} & \implies &
           \mathbf{fnctype}(\fnc) = t_1,...,t_n \rightarrow t' \AN t' <: t \AN\\
     &   & \forall i \in [1,n]. \Gamma_i \vdash \Use_i : t_i \dashv \Gamma_{i+1}\\
    \\

    \Gamma \vdash \kw{var}~\Use : t \dashv \Gamma' & \implies &
           \Gamma \vdash \Use : t' \dashv \Gamma' \AN \Var{}~\Cell[t'] <: t\\
    \\

    \Gamma_1 \vdash \kw{new}~k~C(\many{\Use}) : t \dashv \Gamma_{n+1} & \implies &
           \vdash \Gamma_1 \AN \vdash k~C \AN k~C <: t \AN k\in\{\Mut, \Tmp, \Iso\} \AN \\
    &    & \mathbf{ftypes}(C)= f_1:t_1,...f_n:t_n \AN \many{\Use} = \Use_1, ..., \Use_n \AN\\
    &    & \forall i \in [1,n]. (\Gamma_i \vdash \Use_i : t'_i \dashv \Gamma_{i+1} \AN t'_i <: t_i)\\
    &    & \quad k=\Iso \implies \Kap(\{\Iso, \Imm\}, t'_i)\\
    \\

    \Gamma \vdash \kw{freeze}~\Use : t \dashv \Gamma' & \implies &
           \Gamma \vdash \Use : t' \dashv \Gamma' \AN \Kap(iso, t') \AN
           \mathit{make\_imm}(t') <: t)\\
    \\

    \Gamma \vdash \kw{merge}~\Use : t \dashv \Gamma' & \implies &
           \Gamma \vdash \Use : t' \dashv \Gamma' \AN \Kap(iso, t') \AN
           \mathit{make\_mut}(t') <: t\\
    \\
    \text{Continued on next page...}
  \end{array}
\]\[
  \begin{array}{l}
        \Gamma_1 \vdash \kw{enter}~x.f~[\many{y=\Use}] \SB z~\c{=>}~e \FB : t \dashv \Gamma_{n+1} \implies\\
        \quad\forall i \in [1,n]. \Gamma_i \vdash \Use_i : t_i \dashv \Gamma_{i+1} \AN \Gamma_{n+1}(x) = t_x \AN\\
        \quad\Kap(\{\Mut, \Tmp, \Var, \Paused\}, t_x) \AN \\
        \quad\mathbf{fresult}(t_x, f)) = t_f \AN \Kap(\Iso, t_f) \AN\\
        \quad\Gamma = y_1 : t'_1, ..., y_n : t'_n \\
        \quad\quad\text{ where } t'_i =
               \begin{cases}
                 t_i & \text{if } \Kap(\Iso, t_i)\\
                 \VPA{\Paused}{t_i} & \text{otherwise}
               \end{cases} \AN\\
        \quad t_f' = \mathit{make\_mut}(t_f) \AN \Kap(\{\Iso, \Imm\}, t') \AN t' <: t \AN\\
        \quad \Gamma, z : \Tmp~\Cell[t_f'] \vdash e : t' \dashv \Gamma', z : \Tmp~\Cell[t_f']\\
    \\

    \Gamma_1 \vdash \kw{enter}~x~[\many{y=\Use}] \SB z~\c{=>}~e \FB : t \dashv \Gamma' \implies\\
    \quad \forall i \in [1,n]. \Gamma_i \vdash \Use_i : t_i \dashv \Gamma_{i+1} \AN \Gamma[x : t_x] = \Gamma_{n+1} \AN\\
    \quad \Kap(\Var, t_x) \AN \\
    \quad \mathbf{fresult}(t_x, \c{val})) = t_f \AN \Kap(\Iso, t_f) \AN\\
    \quad \Gamma_y = y_1 : t'_1, ..., y_n : t'_n \\
    \quad \quad\text{ where } t'_i =
               \begin{cases}
                 t_i & \text{if } \Kap(\Iso, t_i)\\
                 \VPA{\Paused}{t_i} & \text{otherwise}
               \end{cases} \AN\\
    \quad t_z = \mathit{make\_cell}(\mathit{make\_mut}(t_f)) \AN \\
    \quad t_z' = \mathit{make\_cell}(t_f') \AN\\
    \quad \Gamma_y, z : t_z \vdash e : t' \dashv \Gamma_y', z : t_z' \AN \Kap(\{\Iso, \Imm\}, t') \AN\\
    \quad \Kap(\Mut, t_f') \AN t' <: t \AN\\
    \quad \Gamma' = \Gamma[x : t_x'] \AN t_x' = \mathit{make\_cell}(\mathit{make\_iso}(t_f'))\\
  \end{array}
\]\[
  \begin{array}{l}
    \Gammas \vdash \kw{entered}~x.f~y.\c{val} \SB de \FB : t \vdash \Gamma' \implies\\
    \quad \Gammas = \many{\Gamma}' :: \Gamma_1 \underset{x.f}{::} \Gamma_0[x : t_x] \AN \\
    \quad \Kap(\{\Mut, \Tmp, \Var, \Paused\}, t_x) \AN \mathbf{fresult}(t_x, f) = t_f \AN\\
    \quad \Kap(\Iso{}, t) \AN \many{\Gamma}' :: \Gamma_1 \vdash de : t' \dashv \Gamma_1' \AN \Kap(\{\Iso{}, \Imm{}\}, t') \AN\\
    \quad t' <: t \AN \Gamma_1'(y) = t_y \AN \mathbf{fresult}(t_y, \c{val}) = t'' \AN\\
    \quad \Kap(\Mut{}, t'') \AN \Kap(\{\Tmp{}, \Var{}\}, t_y) \AN \Gamma' = \Gamma_0[x : t_x'] \AN\\
    \quad t_x' =
    \begin{cases}
      \mathit{make\_cell}(\mathit{make\_iso}(t'')) & \text{if } t_x = \mathit{make\_cell}(\_)\\
      t_x & \text{otherwise}
    \end{cases}\\

  \end{array}
\]

\subsubsection{Proof of Lemma}
\pf

We first prove the $\Rightarrow$-cases by induction over the
typing judgments. The only two interesting subcases are using the
rules \rn{cmd-ty-sub} and \rn{cmd-ty-split}, which are similar for
all the cases. We show the cases for $x$, $\kw{drop}~x$ and
function call here; the remaining cases are similar. \TODO{More if
  there is time}

\begin{pfcases}
  \pfcase{$\Gamma \vdash x : t \dashv \Gamma'$}
  \begin{pfsubcases}
    \pfsubcase{\rn{cmd-ty-use-keep}}
    \begin{itemize}
    \item Proof follows immediately from reflexivity of subtyping
    \end{itemize}

    \pfsubcase{\rn{cmd-ty-sub}}
    \begin{itemize}
    \item[A1] $\Gamma \vdash x : t' \dashv \Gamma'$
    \item[A2] $t' <: t$
    \end{itemize}
    \begin{itemize}
    \item By the induction hypothesis, we have $\Gamma(x) = t''$
      and $t'' <: t'$,
    \item By transitivity of subtyping, we get $t'' <: t$.
    \item The remaining obligations follow from the induction
      hypothesis.
    \end{itemize}

    \pfsubcase{\rn{cmd-ty-split}}
    \begin{itemize}
    \item[A1] $t = t_{x} | t'_{x}$
    \item[A2] $\Gamma = \Gamma_1[y : t_{y1} ~|~ t_{y2}]$
    \item[A3] $\Gamma_1[y : t_{y1}] \vdash x : t_{x} \dashv \Gamma_2$
    \item[A4] $\Gamma_1[y : t_{y2}] \vdash x : t'_{x} \dashv \Gamma_2'$
    \item[A5] $\Gamma' = \Gamma_2 ~|~ \Gamma_2'$
    \end{itemize}
    \begin{itemize}
    \item if $x \neq y$ the proof follows from the induction
      hypothesis.
    \item If $x = y$, by the induction hypothesis we have
      $t_{y1} <: t_{x}$ and
      $t_{y2} <: t'_{x}$, that \Iso{} is not
      in the capabilities of $t_{y1}$ or $t_{y2}$, and that
      $\Gamma_2(x) = t_{y1}$ and
      $\Gamma_2'(x) = t_{y2}$.
    \item From the definition of $\Gamma_2 ~|~ \Gamma_2'$, we get
      that $\Gamma'(x) = t_{y1}~|~t_{y2}$.
    \item From the subtyping rules, we get that
      $t_{y1}~|~t_{y2} <: t_{x}~|~t'_{x}$.
    \end{itemize}
  \end{pfsubcases}

  \pfcase{$\Gamma \vdash \kw{drop}~x : t \dashv \Gamma'$}
  \begin{pfsubcases}
    \pfsubcase{\rn{cmd-ty-drop}}
    \begin{itemize}
    \item Proof follows immediately
    \end{itemize}

    \pfsubcase{\rn{cmd-ty-sub}}
    \begin{itemize}
    \item[A1] $\Gamma \vdash \kw{drop}~x : t' \dashv \Gamma'$
    \item[A2] $t' <: t$
    \end{itemize}
    \begin{itemize}
    \item By the induction hypothesis, we have $\Gamma(x) = t''$
      and $t'' <: t'$,
    \item By transitivity of subtyping, we get $t'' <: t$.
    \item The remaining obligations follow from the induction
      hypothesis.
    \end{itemize}

    \pfsubcase{\rn{cmd-ty-split}}
    \begin{itemize}
    \item[A1] $t = t_{x} | t'_{x}$
    \item[A2] $\Gamma = \Gamma_1[y : t_{y1} ~|~ t_{y2}]$
    \item[A3] $\Gamma_1[y : t_{y1}] \vdash \kw{drop}~x : t_{x} \dashv \Gamma_2$
    \item[A4] $\Gamma_1[y : t_{y2}] \vdash \kw{drop}~x : t'_{x} \dashv \Gamma_2'$
    \item[A5] $\Gamma' = \Gamma_2 ~|~ \Gamma_2'$
    \end{itemize}
    \begin{itemize}
    \item if $x \neq y$ the proof follows from the induction
      hypothesis.
    \item If $x = y$, by the induction hypothesis we have
      $t_{y1} <: t_{x}$ and $t_{y2} <: t'_{x}$, and that
      $\Gamma_2(x) = \mathbf{undef}$ and
      $\Gamma_2'(x) = \mathbf{undef}$.
    \item From the definition of $\Gamma_2 ~|~ \Gamma_2'$, we get
      that $\Gamma'(x) = \mathbf{undef}$.
    \item From the subtyping rules, we get that
      $t_{y1}~|~t_{y2} <: t_{x}~|~t'_{x}$.
    \end{itemize}
  \end{pfsubcases}

  \pfcase{$\Gamma \vdash \fnc(use_1,...,use_n) : t \dashv \Gamma'$}
  \begin{pfsubcases}
    \pfsubcase{\rn{cmd-ty-call}}
    \begin{itemize}
    \item Proof follows immediately from reflexivity of subtyping
    \end{itemize}

    \pfsubcase{\rn{cmd-ty-sub}}
    \begin{itemize}
    \item[A1] $\Gamma \vdash b : t' \dashv \Gamma'$
    \item[A2] $t' <: t$
    \end{itemize}
    \begin{itemize}
    \item By the induction hypothesis, we have
      $\mathbf{fnctype}(\fnc) = t_1,...,t_n \rightarrow
      t''$ and $t'' <: t'$.
    \item By transitivity of subtyping, we get
      $t'' <: t$.
    \item The rest of the obligations follow from the induction
      hypothesis.
    \end{itemize}

    \pfsubcase{\rn{cmd-ty-split}}
    \begin{itemize}
    \item[A1] $t = t_{b1} | t_{b1}'$
    \item[A2] $\Gamma = \Gamma_1[x : t_{x1} ~|~ t_{x2}]$
    \item[A3] $\Gamma_1[x : t_{x1}] \vdash b : t_{b1} \dashv \Gamma_2$
    \item[A4] $\Gamma_1[x : t_{x2}] \vdash b : t_{b1}' \dashv \Gamma_2'$
    \item[A5] $\Gamma' = \Gamma_2 ~|~ \Gamma_2'$
    \end{itemize}
    \begin{itemize}
    \item By the induction hypothesis, we have
      $\mathbf{fnctype}(\fnc) = t_1,...,t_n \rightarrow
      t'$, and $t' <: t_{b1}$ and $t' <: t_{b2}$.
    \item By the subtyping rules, $t' <: t_{b1} | t_{b1}'$.
    \item The arguments are typed using the same reasoning as in the
      cases for $b = use$, but generalized to a sequence. We note
      that a \kw{drop} of the same variable cannot appear twice in
      the same argument list.
    \end{itemize}
  \end{pfsubcases}
\end{pfcases}

We then prove the $\Leftarrow$-cases of $x$ and $\kw{drop}~x$ by
induction on the shape of $t'$.

\begin{pfcases}
  \pfcase{$x$}
  \begin{pfsubcases}
    \pfsubcase{$t' = k~CL$}\\
    $\Gamma_1[x : k~CL] \vdash x : t \dashv \Gamma_1[x : k~CL]$ by
    \rn{cmd-ty-sub} with \rn{cmd-ty-use-keep}.

    \pfsubcase{$t' = t'_1~|~t'_2$}
    \begin{itemize}
    \item By the induction hypothesis, we have
      $\Gamma_1[x : t'_1] \vdash x : t \dashv \Gamma_1[x : t'_1]$
      and
      $\Gamma_1[x : t'_2] \vdash x : t \dashv \Gamma_1[x : t'_2]$.
    \item By \rn{cmd-ty-split} we have
      $\Gamma_1[x : t'_1~|~t'_2] \vdash x : t~|~t \dashv \Gamma_1[x : t'_1~|~t'_2]$.
    \item Since $t~|~t <: t$, by \rn{cmd-ty-sub} we have
      $\Gamma_1[x : t'_1~|~t'_2] \vdash x : t \dashv \Gamma_1[x : t'_1~|~t'_2]$.
    \end{itemize}
  \end{pfsubcases}

  \pfcase{$\kw{drop}~x$}\\
  No induction is needed. We have
  $\Gamma[x : t'] \vdash x : t \dashv \Gamma'[x : \Undef]$ by
  \rn{cmd-ty-sub} with \rn{cmd-ty-use-drop}.
\end{pfcases}

\subsection{Graph Actions}

To describe the action of an effect $\Eff$ on the object graph $\Graph$, we
make some basic definitions of what an \emph{graph action} is. The main idea is
that each effect has some basic action consisting of removal and addition of
locations and references.

Given a configuration $\Gammas \vdash \RCfg$ and $\Gammas \vdash \Eff \dashv
\Gammas'$, the graph $G(\RCfg)$ is well defined and well formed. By progress,
WLOG we can assume that $\RCfg \RegStep{\Eff} \RCfg'$. For such an $\Eff$ we can
calculate a delta $\delta$ such that $G(\RCfg') = G(\RCfg) + \delta$ (this
addition is defined below). $\delta$, called a graph action, is a pair of set
actions, one for locations and one for references. A set action is a pair
of sets, one for removal and one for addition:
\[
  \begin{array}{lcll}
    \delta & \in & (\delta_\AllLocs, \delta_\AllRefs) & \text{Graph
                                                             action:
                                                             action for
                                                             locations/references} \\
    \delta_E & \in & \Pow(E) \times \Pow(E) & \text{Set action: pair of
                                              \emph{remove} and \emph{add} set}
  \end{array}
\]

For convenience we define
\[
  \Nildelta = (\emptyset, \emptyset).
\]

Addition between a set $E$ and a $\delta = (E_\mathsf{r}, E_\mathsf{a})$ is
defined as
\[
  \begin{array}{lcl}
  E + (E_{\mathsf{r}}, E_{\mathsf{a}}) &=& (E \setminus E_{\mathsf{r}}) \cupdisj E_{\mathsf{a}}
  \end{array}
\]

Given a graph action $\delta = (\delta_l, \delta_r)$ and a graph $\Graph =
(\LocSet, \RefSet)$, we define

\[
  \begin{array}{lcl}
    \Graph + \delta &=& (\LocSet + \delta_l, \RefSet + \delta_r) \\
  \end{array}
\]

Furthermore for convenience, we define the addition of two actions $delta_1 =
(E^{\mathsf{r}}_1, E^{\mathsf{a}}_1)$ and $\delta_2 = (E^{\mathsf{r}}_2, E^{\mathsf{a}}_2)$ as
\[ \delta_1 + \delta_2 = (E^{\mathsf{r}}_1 \cupdisj E^{\mathsf{r}}_2,
  E^{\mathsf{a}}_1 \cupdisj E^{\mathsf{a}}_2 ) \]

When not ambiguous, we can write
\[
  \begin{array}{lcl}
    \Graph + \LocSet &\equiv& \Graph + ((\emptyset, \LocSet), \Nildelta) \\
    \Graph - \LocSet &\equiv& \Graph + ((\LocSet, \emptyset), \Nildelta) \\
    \Graph + \Loc &\equiv& \Graph + ((\emptyset, \{\Loc\}), \Nildelta) \\
    \Graph - \Loc &\equiv& \Graph + ((\{\Loc\}, \emptyset), \Nildelta) \\
    \Graph + \RefSet &\equiv& \Graph + (\Nildelta, (\emptyset, \RefSet)) \\
    \Graph - \RefSet &\equiv& \Graph + (\Nildelta, (\RefSet, \emptyset)) \\
    \Graph + \REF &\equiv& \Graph + (\Nildelta, (\emptyset, \{\REF\})) \\
    \Graph - \REF &\equiv& \Graph + (\Nildelta, (\{\REF\}, \emptyset))
  \end{array}
\]
Both $+$ and $-$ associate to the left.

\subsection{Lemma: Well-Defined Graph Actions}
\label{lemma:wd_graph_act}

If
\begin{itemize}
\item $\Gammas \vdash \RCfg$
\item $\Gammas \vdash \Eff \dashv \Gammas'$
\item $\RCfg \RegStep{\Eff} \RCfg'$
\end{itemize}
then there is a unique $\delta$, such that
\[
  G(\RCfg') = G(\RCfg) + \delta
\]
Furthermore, this $\delta$ can be computed using only $G(\RCfg), \rho(\Gammas, \RCfg),
\Eff$, and access to any fresh $\iota$ and $r$ used in the step.

\subsubsection{Proof of Lemma}
\pf

We prove this by defining a partial function, taking a $\rho = \rho(\Gammas, \RCfg)$,
$\Graph = G(\RCfg)$, and $\Eff$:
\[
  \delta^*(\rho, \Graph, \Eff) = (\delta_{LOC}, \delta_{REF})
\]

In the cases below, we make statements of the form
\[
  \REF_\Graph(\Loc, f) = (\Loc \RefTo{f, k_f} \Loc_f)
\]
These equalities are well-defined because of the well formedness of $\RCfg$ and
$\Eff$. Furthermore, uniqueness of $\delta$ follows from that $\REF_\Graph$ is a
partial function.

\begin{pfcases}
\pfcase{$\enter{w, k, y.f, \many{x = \Use}}$}
\llabel[pf]{caseenter}
  \[
    \delta^*(r :: \rho, \Graph, \enter{w, k, y.f, \many{z = \Use}}) = (\delta_{LOC}, \delta_{REF})
  \]
  Looking at the rule \rn{region-enter-ok}, the $\enter{}$ effect
  adds two new locations: a new $\Root(r')$ for the region $r'$ on the region
  stack, and a $\Temp(r', \iota'')$, representing the $\Cell$ holding the entry
  (with $\iota''$ a fresh object id). We do not remove any graph locations. I.e.
  \[
    \delta_{LOC} = (\{\}, \{ \Root(r'), \Temp(r', \iota'') \} ).
  \]
  As for the references, looking at the rule again, we have the following:
  \begin{itemize}
  \item For each $z_i = \kw{drop}~x_i \in \many{z = \Use}$, we remove the
    reference
    \[\REF_\Graph(\Root(r), x_i) = (\Root(r) \RefTo{x_i, k_i} \Loc_{x_i}) \]
    and add the reference
    \[\Root(r') \RefTo{z_i, k} \Loc_{x_i} \]
    We also note that well-formedness of the effect and configuration,
    $\REF_\Graph(\Root(r), x_i)$ is well defined, and $k_i = \Iso$.

    For such an $i$, we let
    \[
      \delta_{REF}^i = \begin{cases}
                         (\{\Root(r) \RefTo{x_i, k_i} \Loc_{x_i} \}, \{ \Root(r)
                         \RefTo{z_i, k_i} \Loc_{x_i} \} ) & \If k_i = \iso \\
                         (\{\Root(r) \RefTo{x_i, k_i} \Loc_{x_i} \}, \{ \Root(r)
                         \RefTo{z_i, \paused \sees k_i} \Loc_{x_i} \} )& \Ow
                       \end{cases}
    \]
  \item For each $z_i = x_i \in \many{z = \Use}$, if
    \[\REF_\Graph(\Root(r), x_i) = (\Root(r) \RefTo{x_i, k_i} \Loc_{x_i}), \]
    we add the reference
    \[\Root(r') \RefTo{z_i, \VPA{\Paused}{k_i}} \Loc_{x_i}. \]
    Well-formedness implies that $\REF_\Graph(\Root(r), x_i)$ and
    $\VPA{\Paused}{k_i}$ is well defined.

    For such an $i$, we let
    \[
      \delta_{REF}^i = (\emptyset, \{ \Root(r) \RefTo{z_i, k_i} \Loc_{x_i} \} ).
    \]
  \item We finally add a reference from $\Root(r')$ to $\Temp(r', \iota'')$
    \[ \Root(r') \RefTo{w, k} \Temp(r', \iota'') \]
    (note that b.c. of well-formedness, $k \in \{ \Var, \Temp \}$)
    and from $\Temp(r', \iota'')$ to $\Loc_f$:
    \[ \Temp(r', \iota'') \RefTo{\Valfield, \Mut} \Loc_f \]
    where
    \[
      \begin{array}{lcl}
        \REF_\Graph(\Root(r), y) &=& (\Root(r) \RefTo{y, k_y}  \Loc_y) \\
        \REF_\Graph(\Loc_y, f)   &=& (\Loc_y   \RefTo{f, \Iso} \Loc_f).
      \end{array}
    \]
    These are well defined because of well-formedness.

    We let
    \[
      \delta_{REF}^{\mathsf{entry}} = \left(\emptyset, \left\{(\Root(r') \RefTo{w, k}
          \Temp(r', \iota'')), (\Temp(r', \iota'') \RefTo{\Valfield, \Mut}
          \Loc_f) \right\} \right)
    \]
  \end{itemize}

  From the above we conclude that
  \[
    \delta_{REF} = \sum_{i = 1}^{\Len(\many{z = \Use})}{\delta_{REF}^i} + \delta_{REF}^\mathsf{entry}
  \]
\pfcase{$\badenter$}

  The rule \rn{region-enter-fail} has $\RCfg = \RCfg'$ and we conclude that
  \[
    \delta^*(\rho, \Graph, \badenter) = (\Nildelta, \Nildelta)
  \]
  i.e. $\delta_{LOC} = \Nildelta$ and $\delta_{REF} = \Nildelta$.

\pfcase{$\exit{x, \Use, y.f, z.f'}$}
  \[
    \delta^*(r :: \rho, \Graph, \exit{x, \Use, y.f, z.f'}) = (\delta_{LOC}, \delta_{REF})
  \]
  We must now consider two cases: \rn{region-exit-temp} and
  \rn{region-exit-heap}. By examining both rules, we see that the top region
  frame is popped. We can thus immediately conclude that $\Root(r')$ and
  $\Temp$-locations pertaining to the top region $r'$ are removed, i.e.
  \[
    \begin{aligned}
      \delta_{LOC} = (&\{ \Root(r') \} \cupdisj \\ & \{ \Temp(r', \_) ~|~
                                                          \Temp(r', \_) \in \Graph \}, \emptyset )
    \end{aligned}
  \]

  For references, since we remove the top region frame all outgoing references
  are removed. This corresponds to:
  \[
    \begin{aligned}
    \delta_{REF}^1 = ( &\left\{ \Root(r') \RefTo{\_, \_} \_ ~|~ \Root(r') \RefTo{\_, \_} \_ \in
      \Graph  \right \} \cupdisj \\
                      & \left\{ \Temp(r', \_) \RefTo{\_, \_} \_ ~|~ \Temp(r', \_) \RefTo{\_, \_} \_ \in
      \Graph  \right \}, \emptyset )
    \end{aligned}
  \]

  The return value specified by $\Use$ is picked up from the frame being
  popped. Assuming $\Use = w$ or $\Use = \kw{drop}~w$, this means that we add
  a reference:
  \[
    \delta_{REF}^2 = (\emptyset, \{\Root(r) \RefTo{x, k_w} \Loc_w \})
  \]
  where
  \[
    \REF_\Graph(\Root(r'), w) = (\Root(r') \RefTo{w, k_w} \Loc_w)
  \]
  Note that $\REF_\Graph(\Root(r'), z)$ is removed as part of $\delta_{REF}^1$ above.

  Lastly, we move the new entrypoint referred to by $z.f'$ in the top frame,
  into $y.f$ referred to in the second top frame:
  \[
    \delta_{REF}^3 = (\{ \Loc_y \RefTo{f, k} \Loc_f \}, \{\Loc_y \RefTo{f,
      k} \Loc_{f'} \})
  \]
  where
  \[
    \begin{array}{lcl}
      \REF_\Graph(\Root(r), y) &=& (\Root(r) \RefTo{y, k_y} \Loc_y) \\
      \REF_\Graph(\Loc_y, f)   &=& (\Loc_y \RefTo{f, k} \Loc_f)  \\
      \REF_\Graph(\Root(r'), z)&=& (\Root(r') \RefTo{z, k_z} \Loc_{f'})
    \end{array}
  \]

  In total we get:
  \[
    \delta_{REF} = \delta_{REF}^1 + \delta_{REF}^2 + \delta_{REF}^3
  \]

  Note that all of this holds for both \rn{region-exit-temp} and
  \rn{region-exit-heap}. The only difference between the two cases is the
  format of $\Loc_y$.
\pfcase{$\load{x, y.f}$}
  \[
    \delta^*(r::\rho, \Graph, \load{x, y.f}) = (\delta_{LOC}, \delta_{REF})
  \]
  It is clear from \rn{region-load} that
  \[
    \delta_{LOC} = \Nildelta.
  \]

  Furthermore, we see that the only action on references is adding a
  reference from $\Root(r)$ to $\Loc_f$, the location referred to by $y.f$:
  \[
    \delta_{REF} = (\emptyset, \{ \Root(r) \RefTo{x, \VPA{k}{k'}} \Loc_f \})
  \]
  where
  \[\begin{array}{lcl}
      \REF(\Root(r), y) &=& (\Root(r)\RefTo{y, k} \Loc_y) \\
      \REF(\Loc_y, f)   &=& (\Loc_y  \RefTo{f, k'}\Loc_f)
    \end{array}\]

\pfcase{$\swap{x, y.f, \Use}$}

  The two rules for swap are \rn{region-swap-temp} and \rn{region-swap-heap},
  which are very similar. We note that $\swap{}$ has no action on the
  locations of the graph:
  \[
    \delta^*(r::\rho, \Graph, \swap{x, y.f, \Use}) = (\Nildelta, \delta_{REF})
  \]

  We have $\Use = z$ or $\Use = \kw{drop}~z$. If we let
  \[
    \REF_\Graph(\Root(r), z) = (\Root(r) \RefTo{z, k_z} \Loc_z)
  \]
  we have two cases:
  \[
    \delta_{REF}^{\mathsf{drop}} =
    \begin{cases}
      (\{ \Root(r) \RefTo{z, k_z} \Loc_z \}, \emptyset) & \text{ if } \Use =
      \kw{drop}~z \\
      \Nildelta & \text{ otherwise}
    \end{cases}
  \]

  The action of the actual swap on the graph is
  \[
    \delta_{REF}^\mathsf{switch} = (\{ \Loc_y \RefTo{f, k_y} \Loc_f \}, \{\Loc_y
    \RefTo{f, k_z} \Loc_z \})
  \]

  In total we have
  \[
    \delta_{REF} = \delta_{REF}^\mathsf{drop} + \delta_{REF}^\mathsf{switch}.
  \]

  Because of the similarity, this holds in both \rn{region-swap-tmp} and
  \rn{region-swap-heap}. The only difference will be the format of $\Loc_y$.

\pfcase{$\halloc{x, k, \#C, \many{\Use}}$}
\llabel[pf]{casehalloc}
  \[
    \delta^*(r::\rho, \Graph, \halloc{x, \#C, k, \many{\Use}}) = (\delta_{LOC}, \delta_{REF}).
  \]
  By well-formedness, $k = \Mut$ or $k = \Iso$, which corresponds to the two
  cases \rn{region-halloc-mut} and \rn{region-halloc-iso} respectively. In both cases
  the set of locations will be expanded with a new heap location with fresh
  object id $\iota$. Its region id will depend on $k$:
  \[
    r^* =
    \begin{cases}
      r & \text{ if } k = \Mut \\
      r'  & \text{ if } k = \Iso, \text{ where } r' \text{ is fresh}
    \end{cases}
  \]
  \[
    \delta_{LOC} = (\emptyset, \{ \Heap(r^*, \iota) \})
  \]

  Similar to the enter case (\lref[pf]{caseenter}) above, we have a sequence of $\Use$
  to consider. These correspond to the fields of the new object being created.
  Letting $\Fnames(C) = f_1, \dots, f_n$:
  \begin{itemize}
  \item For each $\kw{drop}~z_i \in \many{\Use}$,
    \[
      \REF_\Graph(\Root(r), z_i) = (\Root(r) \RefTo{z_i, k_i} \Loc_i)
    \]
    and
    \[
      \delta_{REF}^i = (\{ \Root(r) \RefTo{z_i, k_i} \Loc_i \},
                        \{ \Heap(r^*, \iota) \RefTo{f_i, k_i} \Loc_i \})
    \]
  \item For each $z_i \in \many{\Use}$,
    \[
      \REF_\Graph(\Root(r), z_i) = (\Root(r) \RefTo{z_i, k_i} \Loc_i)
    \]
    and
    \[
      \delta_{REF}^i = (\emptyset, \{ \Heap(r^*, \iota) \RefTo{f_i, k_i} \Loc_i \})
    \]
  \end{itemize}
  This is all well defined because of well-formedness. In total we have
  \[
    \delta_{REF} = (\emptyset, \{ \Root(r) \RefTo{x, k} \Heap(r^*, \iota) \}) + \sum_{i = 1}^{\Len(\many{\Use})} \delta_{REF}^i
  \]
\pfcase{$\salloc{x, k, \CLTag, \many{\Use}}$}
\llabel[pf]{casesalloc}
  \[
    \delta^*(r::\rho, \Graph, \salloc{x, k, C, \many{\Use}}) = (\delta_{LOC}, \delta_{REF}).
  \]
  By similar reasoning to \lref[pf]{casehalloc}, for a fresh $\iota$
  \[
    \delta_{LOC} = (\emptyset, \{ \Temp(r, \iota) \})
  \]
  Letting $\Fnames(C) = f_1, \dots, f_n$:
  \begin{itemize}
  \item For each $\kw{drop}~z_i \in \many{\Use}$,
    \[
      \REF_\Graph(\Root(r), z_i) = (\Root(r) \RefTo{z_i, k_i} \Loc_i)
    \]
    and
    \[
      \delta_{REF}^i = (\{ \Root(r) \RefTo{z_i, k_i} \Loc_i \},
      \{ \Temp(r, \iota) \RefTo{f_i, k_i} \Loc_i \})
    \]
  \item For each $z_i \in \many{\Use}$,
    \[
      \REF_\Graph(\Root(r), z_i) = (\Root(r) \RefTo{z_i, k_i} \Loc_i)
    \]
    and
    \[
      \delta_{REF}^i = (\emptyset, \{ \Temp(r, \iota) \RefTo{f_i, k_i} \Loc_i \})
    \]
  \end{itemize}

  In total:
  \[
    \delta_{REF} = (\emptyset, \{ \Root(r) \RefTo{x, k} \Temp(r, \iota) \}) + \sum_{i = 1}^{\Len(\many{\Use})} \delta_{REF}^i
  \]
  We note that $k \in \{ \Tmp, \Var \}$ by well-formedness.

\pfcase{$\freeze{x, \Use}$}

  By inspecting \rn{region-freeze},
  \[
    \delta^*(r :: rn, \Graph, \freeze{x, \Use}) = (\Nildelta, \delta_{REF})
  \]
  i.e. $\delta_{LOC} = \Nildelta$.

  By well-formedness, $\Use = \kw{drop}~z$ and
  \[
    \REF_\Graph(\Root(r), z) = (\Root \RefTo{z, \Iso} \Loc_z)
  \]
  From rule \rn{region-freeze} we have that this reference is consumed and
  converted into a immutable reference:
  \[
    \delta_{REF} = (\{\Root(r) \RefTo{z, \Iso} \Loc_z\}, \{\Root(r) \RefTo{x,
      \Imm} \Loc_z\})
  \]
\pfcase{$\merge{x, \Use}$}
  \[
    \delta^*(r :: rn, \Graph, \merge{x, \Use}) = (\delta_{LOC}, \delta_{REF})
  \]
  By well-formedness, $\Use = \kw{drop}~z$. By \rn{region-merge},
  \[
    \REF_\Graph(\Root(r), z) = (\Root(r) \RefTo{z, \Iso} \Heap(r', \iota'))
  \]
  The $\delta$s for merging a region is somewhat tricky. In the configuration
  semantics, this ammounts to merging the subheap of the region $r'$ into the the
  subheap of the currently active region $r$. This means that all locations
  $loc[r']$ will be converted to $loc[r]$.

  \[
    D_{LOC} = \left\{ \Loc[r'] \Suchthat \Loc[r'] \in \Graph \right\}
  \]

  We write
  \[
    \delta_{LOC} = (D_{LOC}, D_{LOC}[r'/r])
  \]
  where $D_{LOC}[r'/r]$ is the set of elements of $D_{LOC}$ but with $r'$
  substituted for $r$.

  For references we have the analogue
  \[
    D_{REF} = \left\{ (\Loc_1[r_1] \RefTo{f, k} \Loc_2[r_2]) \Suchthat
    (\Loc_1[r_1] \RefTo{f, k} \Loc_2[r_2]) \in \Graph \AN
    (r_1 = r' \OR r_2 = r') \right\}
  \]
  We have
  \[
    \delta_{REF} = (D_{REF}, D_{REF}[r'/r])
  \]

\pfcase{$\cast{x, \Use, k~CL}$}
\llabel[pf]{casecast}

  $\Use$ can either be $z$ or $\kw{drop}~z$. From the rule \rn{region-cast} it is
  obvious that
  \[
    \delta^*(r :: \rho, \Graph, \cast{x, \Use, k~CL}) = (\Nildelta, \delta_{REF})
  \]
  where
  \[
    \delta_{REF} = \delta_{REF}^\mathsf{drop} + (\emptyset, \{ \Root(r) \RefTo{x,
      k_z} \Loc_z \}
  \]
  where
  \[
    \begin{array}{lcl}
      \REF_\Graph(\Root(r), z) &=& (\Root(r) \RefTo{z, k_z} \Loc_z)\\
      \delta_{REF}^\mathsf{drop} &=& \begin{cases}
                                       (\{\Root(r) \RefTo{z, k_z} \Loc_z \},
                                       \emptyset) & \text{ if } \Use = \kw{drop}~z
                                       \\
                                       \Nildelta & \text{ otherwise}
                                     \end{cases}
    \end{array}
  \]

\pfcase{$\nocast{x, \Use, k~CL}$}

By similar reasoning to \lref[pf]{casecast}, using \rn{region-nocast} instead:
  \[
    \delta^*(r :: \rho, \Graph, \nocast{x, \Use, k~CL}) = (\Nildelta, \delta_{REF})
  \]
  where
  \[
    \delta_{REF} = \delta_{REF}^\mathsf{drop} + (\emptyset, \{ \Root(r) \RefTo{x,
      k_z} \Loc_z \}
  \]
  where
  \[
    \begin{array}{lcl}
      \REF_\Graph(\Root(r), z) &=& (\Root(r) \RefTo{z, k_z} \Loc_z)\\
      \delta_{REF}^\mathsf{drop} &=& \begin{cases}
                                      (\{\Root(r) \RefTo{z, k_z} \Loc_z \},
                                      \emptyset) & \text{ if } \Use = \kw{drop}~z
                                      \\
                                      \Nildelta & \text{ otherwise}
                                    \end{cases}
    \end{array}
  \]

\pfcase{$\bind{\many{x = \Use}}$}

  Looking at rule \rn{region-bind}, similarly to $\halloc{}$ and $\salloc{}$
  cases (\lref[pf]{casehalloc} and \lref[pf]{casesalloc}), we have a sequence of
  $\Use$s. Here we instead create references from $\Root(r)$. We get
  \[
    \delta^*(r::\rho, \Graph, \bind{\many{x = \Use}}) = (\Nildelta, \delta_{REF}).
  \]
  \begin{itemize}
  \item For each $x_i = \kw{drop}~z_i \in \many{x = \Use}$,
    \[
      \REF_\Graph(\Root(r), z_i) = (\Root(r) \RefTo{z_i, k_i} \Loc_i)
    \]
    and
    \[
      \delta_{REF}^i = (\{ \Root(r) \RefTo{z_i, k_i} \Loc_i \},
      \{ \Root(r) \RefTo{x_i, k_i} \Loc_i \})
    \]
  \item For each $x_i = z_i \in \many{x = \Use}$,
    \[
      \REF_\Graph(\Root(r), z_i) = (\Root(r) \RefTo{z_i, k_i} \Loc_i)
    \]
    and
    \[
      \delta_{REF}^i = (\emptyset, \{ \Root(r) \RefTo{x_i, k_i} \Loc_i \})
    \]
  \end{itemize}

  In total:
  \[
    \delta_{REF} = \sum_{i = 1}^{\Len(\many{\Use})} \delta_{REF}^i
  \]

\end{pfcases}

\end{document}